\documentclass[a4paper,11pt]{article}
\pdfoutput=1 

\usepackage{jheppub} 

\usepackage[T1]{fontenc}
\usepackage{multirow,bbold,slashed,hhline}

\usepackage[utf8]{inputenc}
\usepackage{hyperref}
\usepackage{booktabs}
\usepackage[T1]{fontenc} 
\usepackage{siunitx} 
\usepackage[table,xcdraw,dvipsnames]{xcolor}
\usepackage{graphicx}
\usepackage{dcolumn}
\usepackage{bm}
\usepackage[normalem]{ulem}
\usepackage{braket} 
\usepackage{amsmath}
\usepackage{cancel}
\usepackage{multicol}
\usepackage[capitalise, english]{cleveref}
\usepackage{arydshln}
\usepackage{lscape}
\usepackage{pdflscape}
\usepackage{makecell}
\usepackage[export]{adjustbox}
\usepackage{tikz}
\usepackage[compat=1.1.0]{tikz-feynman}
\usepackage[compat=1.1.0]{tikz-feynhand}
\usepackage{ytableau}
\usepackage[shortlabels]{enumitem}
\usepackage{dsfont}
\usepackage{listings,newtxtt} 
\usepackage{subcaption}

\newcommand{\nc}{\newcommand}
\nc{\vp}{H}
\nc{\tvp}{\widetilde{H}}
\nc{\D}{\mbox{$\not\!\!D$}}
\nc{\Db}{\mbox{${\raisebox{2mm}{\boldmath ${}^\leftarrow$}\hspace{-4mm} D}$}}
\nc{\Dfb}{\mbox{$\raisebox{2mm}{\boldmath ${}^\leftrightarrow$}\hspace{-4mm} D$}}
\nc{\vpj }{\mbox{${\vp^\dag i\,\raisebox{2mm}{\boldmath ${}^\leftrightarrow$}\hspace{-4mm} D_\mu\,\vp}$}}
\nc{\vpjt}{\mbox{${\vp^\dag i\,\raisebox{2mm}{\boldmath ${}^\leftrightarrow$}\hspace{-4mm} D_\mu^{\,I}\,\vp}$}}

\newcommand{\eps}{{\epsilon}}

\DeclareFontFamily{OT1}{mathc}{}
\DeclareFontShape{OT1}{mathc}{m}{it}{<-> mathc10}{}
\DeclareMathAlphabet{\mathcal}{OT1}{mathc}{m}{it}

\allowdisplaybreaks
 
\definecolor{niceblue}{rgb}{0.0,0.1,0.7}
\definecolor{nicered}{rgb}{0.7,0.1,0.1} 
\definecolor{nicegreen}{rgb}{0.1,0.5,0.1}
\hypersetup{colorlinks,citecolor=niceblue,linkcolor=niceblue,urlcolor=niceblue}

\def \bm#1{\mbox{\boldmath$#1$\unboldmath}}
\def \beq{\begin{equation}}
\def \eeq{\end{equation}}
\def \bea{\begin{eqnarray}}
\def \eea{\end{eqnarray}}

\def \LLone{{\color{nicered} L}}
\def \LLtwo{{\color{nicered} L^2}}
\def \NLL{{\color{nicegreen} L}}
\def \LLcolor{{\color{nicered} red}}
\def \NLLcolor{{\color{nicegreen} green}}

\title{Electroweak precision observables \\ at NLL order in the SMEFT}

\author[a]{Ulrich~A.~Haisch}
\author[b]{and Ben~A.~Stefanek}

\affiliation[a]{Max Planck Institute for Physics, \\ Boltzmannstr.~8, 85748 Garching, Germany}
\affiliation[b]{Instituto de F{\`i}sica Corpuscular (IFIC), Consejo Superior de Investigaciones Cient{\`i}ficas (CSIC) and Universitat de Val{\'e}ncia (UV), 46980 Valencia, Spain}

\emailAdd{haisch@mpp.mpg.de}
\emailAdd{bstefan@ific.uv.es}

\preprint{MPP-2026-136}

\abstract{Electroweak precision observables~(EWPO) remain among the most powerful indirect probes of physics beyond the Standard Model~(BSM), and future $e^+ e^-$ colliders are expected to substantially enhance their sensitivity. Within the framework of the Standard Model Effective Field Theory~(SMEFT), we show how~EWPO predictions can be systematically improved to next-to-leading-logarithmic~(NLL) accuracy by combining fixed-order tree-level and one-loop matching with two- and one-loop renormalization-group evolution. We derive NLL expressions that encode the full dependence on all~210~independent Wilson coefficients and illustrate their impact through fits to both current and projected collider measurements. This work provides a basis for global SMEFT analyses at~NLL~precision, thereby maximizing the indirect reach to BSM~physics through~EWPO measurements. The~phenomenological relevance of this framework is illustrated explicitly in both the custodial Randall–Sundrum model and a custodial composite Higgs~model.}

\begin{document} 
\maketitle
\flushbottom

\section{Introduction} 
\label{sec:introduction}

The legacy of the Large Electron~Positron~Collider~(LEP) and the Stanford~Linear~Collider~(SLC) is that they transformed electroweak~(EW) precision physics into a high-precision science and firmly established the Standard Model~(SM) as a quantitatively predictive theory. Operating at the $Z$ pole and at higher energies, LEP and SLC performed extraordinarily precise measurements of $Z$- and $W$-boson properties, fermion couplings, and asymmetries. These measurements fixed EW parameters at the permille level, confirmed the gauge structure of the SM, and established the existence of three light neutrino~species~\cite{ALEPH:2005ab}.

Decades later, EW precision observables~(EWPO) continue to define the precision frontier, providing some of the most stringent indirect probes of physics beyond the~SM~(BSM). Their precision is expected to improve significantly at future $e^+e^-$ colliders, such as the proposed Future~Circular~Collider~(FCC-ee)~\cite{FCC:2018evy} and the Circular~Electron~Positron~Collider~(CEPC)~\cite{CEPCStudyGroup:2018ghi}. In light of this, substantial theoretical effort has been devoted to providing precision predictions for~EWPO within the Standard~Model~Effective~Field~Theory~(SMEFT)~\cite{Buchmuller:1985jz,Grzadkowski:2010es,Brivio:2017vri,Isidori:2023pyp}. These predictions have reached next-to-leading~order~(NLO) accuracy in fixed-order perturbation theory~\cite{Dawson:2019clf,Dawson:2022bxd,Bellafronte:2023amz,Biekotter:2025nln,Bellafronte:2026jic} for the complete set of dimension-six operators, with results available for various EW input schemes~\cite{Biekotter:2023xle,Biekotter:2023vbh}. Next-to-next-to-leading order~(NNLO) fixed-order corrections to the~EWPO for a small subset of dimension-six operators have also been studied~\cite{Haisch:2024wnw,DiNoi:2025uhu}.

Combining the NLO fixed-order results~\cite{Dawson:2019clf,Dawson:2022bxd,Bellafronte:2023amz,Biekotter:2025nln,Bellafronte:2026jic} with the one-loop beta functions of the dimension-six SMEFT operators~\cite{Jenkins:2013zja,Jenkins:2013wua,Alonso:2013hga} allows one to compute the leading-logarithmic~(LL) corrections associated with the renormalization-group~(RG) running from the scale $\Lambda$, which characterizes the onset of BSM~physics, down to the EW scale of order the on-shell $Z$-boson mass~$m_Z$, relevant for $Z$-pole observables. Introducing
\beq \label{eq:defL}
L = \ln \frac{\Lambda}{m_Z} \,,
\eeq
and focusing on Quantum~Chromodynamics~(QCD) for the moment, the LL~corrections take the form
\beq \label{eq:LL}
\alpha_s^n L^n \,, \quad n = 1, 2, \dots \,,
\eeq
where $\alpha_s = g_s^2/(4 \pi)$, with $g_s$ denoting the $SU(3)_C$ gauge coupling.

In this article, we show how the two-loop beta functions of the complete set of dimension-six SMEFT operators, recently made available in the instant classic~\cite{Born:2026xkr},\footnote{Other contributions to the two-loop beta functions in the SMEFT have been reported in~\cite{Gorbahn:2016uoy,Bern:2020ikv,Jin:2020pwh,Haisch:2022nwz,DiNoi:2023ygk,Jenkins:2023bls,DiNoi:2024ajj,Born:2024mgz,Naterop:2024cfx,Duhr:2025zqw,Haisch:2025lvd,Zhang:2025ywe,Assi:2025fsm,Naterop:2025cwg,DiNoi:2025arz,Haisch:2025vqj,Duhr:2025yor,DiNoi:2025tka,Banik:2025wpi,Henriksson:2025vyi,Chala:2025crd,Guedes:2025sax,Duhr:2026btw}.} can be leveraged to systematically improve~EWPO predictions to next-to-leading-logarithmic~(NLL) accuracy. In the QCD case, this corresponds to contributions of the~form
\beq \label{eq:NLL}
\alpha_s^n L^{n-1} \,, \quad n = 1, 2, \dots \,,
\eeq
which can now be consistently incorporated into~EWPO fits to the SMEFT dimension-six Wilson~coefficients. With FCC-ee and CEPC expected to push~EWPO measurements to subpermille precision, achieving NLL accuracy becomes essential to fully exploit the anticipated experimental gains. This point has been highlighted and studied concurrently in~\cite{Mantani:2026fao,Born:2026tgm}, which provide the first phenomenological applications of the two-loop beta functions of~\cite{Born:2026xkr}.

The remainder of this paper is organized as follows. In Section~\ref{sec:anatomy}, we present a series of pedagogical examples that illustrate how logarithmically enhanced contributions beyond~LL~order arise in~EWPO. Particular emphasis is placed on dimension-six operators that contribute only at loop level in fixed-order calculations. The Wilson coefficients of these operators are often only weakly constrained by existing measurements, making~EWPO one of the most powerful probes of their effects. In Section~\ref{sec:matching}, we demonstrate that NLL-accurate predictions for~EWPO can be obtained without incorporating two-loop matching corrections, since such contributions are formally of next-to-next-to-leading-logarithmic~(NNLL) order according to the logarithmic counting introduced in~\eqref{eq:LL} and~\eqref{eq:NLL}. Numerical NLL expressions for the full set of~EWPO, including their dependence on the relevant Wilson coefficients, are derived in Section~\ref{sec:NLLEWPO}. Section~\ref{sec:Zpolefits} presents global fits to current LEP and SLC measurements, together with projections for future $e^+e^-$ collider facilities, and quantifies the resulting sensitivity to the corresponding SMEFT Wilson coefficients. In Section~\ref{sec:BSM}, these model-independent constraints are applied to two representative BSM~scenarios. Finally, Section~\ref{sec:conclusions} summarizes our main results and discusses possible directions for future work. Additional technical details and complementary material are collected in a series of~appendices.

\section{Anatomy of~EWPO} 
\label{sec:anatomy}

In this section, we examine the anatomy of~EWPO contributions in the SMEFT, classifying the dimension-six operators according to whether they first enter at tree level, one-loop or two-loop order. Particular emphasis is placed on operators that contribute to~EWPO for the first time at the two-loop level. Within this class, we further distinguish between operators that generate LL contributions through RG~mixing into the tree-level operators and those whose leading effect on~EWPO appears only at NLL or beyond.

\subsection{Tree-level contributions} 
\label{sec:tree}

To set the stage, we recall that tree-level matching at the EW scale generates sensitivity to only a restricted subset of 23 dimension-six operators~\cite{Dawson:2019clf,Dawson:2022bxd,Bellafronte:2023amz,Biekotter:2025nln}, or equivalently their associated Wilson~coefficients. In the Warsaw basis~\cite{Grzadkowski:2010es}, these operators are
\beq \label{eq:treeoperators}
\left \{ Q_{HD}, \, Q_{HW\!B}, \, Q_{H\ell}^{(1) \hspace{0.25mm} pp}, \, Q_{H\ell}^{(2) \hspace{0.25mm} pp}, \, Q_{He}^{\hspace{0.25mm} pp}, \, Q_{Hq}^{(1) \hspace{0.25mm} pp}, \, Q_{Hq}^{(3) \hspace{0.25mm} pp}, \, Q_{Hu}^{\hspace{0.25mm} rr}, \, Q_{Hd}^{\hspace{0.25mm} pp}, \, Q_{\ell\ell}^{1221} \right \} \,,
\eeq
where the flavor indices $p = 1,2,3$ and $r = 1,2$ run over diagonal components and are not summed. The explicit definitions of these operators are given in~Tables~\ref{tab:no4ferm} and~\ref{tab:4ferm}, while the underlying field-theoretic building blocks are collected in Appendix~\ref{app:definitions} for completeness.

BSM~physics inducing $\mathcal{O}(1/\mathrm{TeV}^2)$ corrections to the Wilson~coefficients of the operators in~\eqref{eq:treeoperators} is generally already tightly constrained by existing $Z$-pole data. Barring highly fine-tuned cancellations, LEP and SLC measurements therefore imply a parametric suppression of these Wilson~coefficients, for instance by a loop factor of~$1/(16\pi^2)$. In~concrete BSM models, such suppression can arise from symmetries: a prominent example is the custodial~$SU(2)$ symmetry, which protects the operator $Q_{HD}$ from phenomenologically unacceptable corrections. However, accidental symmetries arising in the individual sectors of the SM --- including custodial~$SU(2)$ and the flavor~$U(3)^5$ symmetry --- are not preserved by RG~evolution in the SMEFT. In particular, the top-quark Yukawa coupling breaks both custodial and flavor symmetries, while the EW gauge couplings break custodial symmetry (via~$U(1)_Y$ hypercharge) while preserving flavor universality. As a result, a theory that is custodial symmetric and flavor universal at the high scale~$\Lambda$ generically flows to an effective field theory~(EFT) at the EW scale in which custodial symmetry is violated and flavor non-universality as well as flavor-changing neutral currents are induced. In what follows, we illustrate this general mechanism through a few simple yet instructive one- and two-loop examples in the~EWPO~context.

\begin{table}[t!] 
\centering
\renewcommand{\arraystretch}{1.5}
\begin{tabular}{|c|c|}
\hline 
\multicolumn{2}{|c|}{$\vp^6$ \text{and} $\vp^4 D^2$} \\
\hline
\cellcolor{gray!20} $Q_\vp$ & $(\vp^\dag\vp)^3$ \\
$Q_{\vp\Box}$ & $(\vp^\dag \vp)\hspace{0.25mm}\raisebox{-.5mm}{$\Box$}\hspace{0.25mm}(\vp^\dag \vp)$ \\
$Q_{\vp D}$ & $|\vp^\dag D_\mu\vp|^2$ \\
\hline 
\end{tabular}

\vspace{4mm}

\begin{tabular}{|c|c|c|c|}
\hline
\multicolumn{2}{|c|}{$\psi^2\vp^2 D$} & \multicolumn{2}{c|}{$X^2\vp^2$ \text{and} $X^3$} \\
\hline
$Q_{\vp \ell}^{(1)}$ & $(\vpj)(\bar \ell \gamma^\mu \ell)_{pp}$ & $Q_{\vp B}$ & $\vp^\dag \vp\, B_{\mu\nu} B^{\mu\nu}$ \\
$Q_{\vp \ell}^{(3)}$ & $(\vpjt)(\bar \ell \tau^I \gamma^\mu \ell)_{pp}$ & $Q_{\vp W}$ & $\vp^\dag \vp\, W^I_{\mu\nu} W^{I\mu\nu}$ \\
$Q_{\vp e}$ & $(\vpj)(\bar e \gamma^\mu e)_{pp}$ & $Q_{\vp W\!B}$ & $\vp^\dag \tau^I \vp\, W^I_{\mu\nu} B^{\mu\nu}$ \\
$Q_{\vp q}^{(1)}$ & $(\vpj)(\bar q \gamma^\mu q)_{pp}$ & \cellcolor{green!20} $Q_{\vp G}$ & $\vp^\dag \vp\, G^A_{\mu\nu} G^{A\mu\nu}$ \\
$Q_{\vp q}^{(3)}$ & $(\vpjt)(\bar q \tau^I \gamma^\mu q)_{pp}$ & $Q_W$ & $\eps^{IJK} W_\mu^{I\nu} W_\nu^{J\rho} W_\rho^{K\mu}$ \\
$Q_{\vp u}$ & $(\vpj)(\bar u \gamma^\mu u)_{pp}$ & \cellcolor{green!20} $Q_G$ & $f^{ABC} G_\mu^{A\nu} G_\nu^{B\rho} G_\rho^{C\mu}$ \\
$Q_{\vp d}$ & $(\vpj)(\bar d \gamma^\mu d)_{pp}$ & & \\
\hline 
\end{tabular}
\vspace{4mm}
\caption{Non-four-fermion self-conjugate and CP-conserving dimension-six SMEFT operators. The repeated flavor index $p = 1,2,3$ labels diagonal flavor components and is not summed over. This class comprises a total of 30 independent dimension-six operators. Operators highlighted in green first contribute to the~EWPO through $\NLL$ effects arising at two-loop order. The operator highlighted in gray corresponds to the unique case in which the leading-logarithmic RG~corrections first appear only at four-loop order.}
\label{tab:no4ferm}
\end{table}

\subsection{One-loop contributions} 
\label{sec:one}

The complete one-loop EW~matching corrections in the SMEFT have been computed in the articles~\cite{Dawson:2019clf,Dawson:2022bxd,Bellafronte:2023amz,Biekotter:2025nln,Bellafronte:2026jic}, retaining the full dependence on the $U(1)_Y$, $SU(2)_L$, and $SU(3)_C$ gauge couplings $g_Y$, $g_L$, and $g_s$, the Higgs quartic coupling $\lambda$ entering the SM potential term~$-\lambda/2 \, (H^\dagger H)^2$, and the top-quark Yukawa coupling $y_t = \sqrt{2} \hspace{0.25mm} m_t/v$, with $m_t$ the top-quark mass and $v$ the Higgs vacuum expectation value~(VEV). Light-fermion Yukawa couplings are phenomenologically negligible and are omitted throughout, and CP-violating effects are likewise neglected. Tables~\ref{tab:no4ferm} and~\ref{tab:4ferm} collect all dimension-six operators that mix into the tree-level set~\eqref{eq:treeoperators} through RG~running proportional to $g_Y$, $g_L$, $g_s$, or~$\lambda$, while operators whose mixing requires $y_t$ are listed separately in~Table~\ref{tab:uXoperators}. 

The allowed RG mixing patterns can be understood from the requirement that BSM interference effects must match the analytic and chiral structure of the CP-even, self-conjugate tree-level SM neutral-current gauge interactions. Before including Yukawa interactions, the relevant SMEFT operators are those invariant under the Cartan subgroup $U(1)^{15}$ of the~$U(3)^5$ flavor symmetry of the SM gauge sector. This selects the 207 CP-even operators listed in Tables~\ref{tab:no4ferm} and~\ref{tab:4ferm}. Switching on the Yukawa interactions in the excellent approximation where only $y_t$ is non-zero breaks $U(1)^{15} \to U(1)^{14}$ and additionally permits precisely the four chirality-flipping operators listed in~Table~\ref{tab:uXoperators}. A $U(1)^{14}$ symmetry imposed on the~SMEFT therefore identifies the complete set of operators that can generate sizable contributions to EWPO. It was recently pointed out in~\cite{Greljo:2025mwj} that TeV-scale BSM scenarios require a high-quality $SU(2)_q$ symmetry, implying that the symmetry controlling low-scale~BSM effects in EWPO should be at least $SU(2)_q\times U(1)^{13}$. Under this symmetry, 180~CP-even operators are invariant. In this work, we retain the full set of 211 operators listed in~Tables~\ref{tab:no4ferm},~\ref{tab:4ferm}, and~\ref{tab:uXoperators}, corresponding to the more general $U(1)^{14}$ symmetry.\footnote{The counting of 210 operators quoted in the abstract excludes the Higgs self-interaction operator $Q_H$, which is allowed by the symmetry but contributes to EWPO only at NNLL order.} We now briefly recall several operator classes for which~EWPO provide relevant one-loop constraints.

\paragraph{Higgs kinetic operator:} 

The dimension-six operator modifying the Higgs kinetic term~is
\beq \label{eq:higgskinetic}
Q_{\vp\Box} \,,
\eeq
with its Warsaw-basis definition given in~Table~\ref{tab:no4ferm}. In this case,~EWPO receive LL corrections through one-loop RG~mixing of $Q_{\vp\Box}$ into $Q_{\vp D}$~\cite{Alonso:2013hga}, driven by the hypercharge coupling~$g_Y$ that breaks custodial symmetry. This has notable phenomenological implications~\cite{Durieux:2022hbu,Stefanek:2024kds,Allwicher:2024sso,Maura:2024zxz,Maura:2025rcv,terHoeve:2025omu,Ding:2026qto}, especially in BSM models where the leading dimension-six effect arises through the Higgs self-interaction operator~$Q_{H}$. While the operator~$Q_{\vp\Box}$ induces a universal tree-level shift in all single-Higgs observables, the improved~EWPO measurements anticipated at FCC-ee and CEPC may yield stronger or complementary constraints, depending on the ultimate precision achievable at these future $e^+ e^-$ facilities.

\begin{table}[t!]
\centering
\renewcommand{\arraystretch}{1.5}
\scalebox{0.92}{
\begin{tabular}{|c|c|c|c|c|c|}
\hline
\multicolumn{2}{|c|}{$(\bar LL)(\bar LL)$} & 
\multicolumn{2}{c|}{$(\bar RR)(\bar RR)$} &
\multicolumn{2}{c|}{$(\bar LL)(\bar RR)$} \\
\hline

$Q_{\ell\ell}$ & $(\bar \ell \gamma_\mu \ell)(\bar \ell \gamma^\mu \ell)_{\substack{pprr \\ prrp}}$ & $Q_{ee}$ & $(\bar e \gamma_\mu e)(\bar e \gamma^\mu e)_{pprr}$ & $Q_{\ell e}$ & $(\bar \ell \gamma_\mu \ell)(\bar e \gamma^\mu e)_{pprr}$ \\
$Q_{qq}^{(1)}$ & $(\bar q \gamma_\mu q)(\bar q \gamma^\mu q)_{\substack{pprr \\ prrp}}$ &
$Q_{uu}$ & $(\bar u \gamma_\mu u)(\bar u \gamma^\mu u)_{\substack{pprr \\ prrp}}$ & $Q_{\ell u}$ & $(\bar \ell \gamma_\mu \ell)(\bar u \gamma^\mu u)_{pprr}$ \\
$Q_{qq}^{(3)}$ & $(\bar q \gamma_\mu \tau^I q)(\bar q \gamma^\mu \tau^I q)_{\substack{pprr \\ prrp}}$ & $Q_{dd}$ & $(\bar d \gamma_\mu d)(\bar d \gamma^\mu d)_{\substack{pprr \\ prrp}}$ & $Q_{\ell d}$ & $(\bar \ell \gamma_\mu \ell)(\bar d \gamma^\mu d)_{pprr}$ \\
$Q_{\ell q}^{(1)}$ & $(\bar \ell \gamma_\mu\ell)(\bar q \gamma^\mu q)_{pprr}$ & $Q_{eu}$ & $(\bar e \gamma_\mu e)(\bar u \gamma^\mu u)_{pprr}$ & $Q_{qe}$ & $(\bar q \gamma_\mu q)(\bar e \gamma^\mu e)_{pprr}$ \\
$Q_{\ell q}^{(3)}$ & $(\bar \ell \gamma_\mu \tau^I \ell)(\bar q \gamma^\mu \tau^I q)_{pprr}$ & $Q_{ed}$ & $(\bar e \gamma_\mu e)(\bar d\gamma^\mu d)_{pprr}$ & $Q_{qu}^{(1)}$ & $(\bar q \gamma_\mu q)(\bar u \gamma^\mu u)_{pprr}$ \\ 
& & $Q_{ud}^{(1)}$ & $(\bar u \gamma_\mu u)(\bar d \gamma^\mu d)_{pprr}$ & $Q_{qu}^{(8)}$ & $(\bar q \gamma_\mu T^A q)(\bar u \gamma^\mu T^A u)_{pprr}$ \\ 
& & $Q_{ud}^{(8)}$ & $(\bar u \gamma_\mu T^A u)(\bar d \gamma^\mu T^A d)_{pprr}$ & $Q_{qd}^{(1)}$ & $(\bar q \gamma_\mu q)(\bar d \gamma^\mu d)_{pprr}$ \\
& & & & $Q_{qd}^{(8)}$ & $(\bar q \gamma_\mu T^A q)(\bar d \gamma^\mu T^A d)_{pprr}$ \\
\hline
\end{tabular}
}
\vspace{4mm}
\caption{Self-conjugate four-fermion operators, where the repeated flavor indices $p,r = 1,2,3$ are not summed over. This class contains a total of 177 independent dimension-six operators.}
\label{tab:4ferm}
\end{table}

\paragraph{Top-quark gauge-boson operators:} 

The Warsaw-basis dimension-six operators coupling a top-quark current to an EW gauge boson are
\beq \label{eq:2topEWoperators}
\left \{ Q_{Hq}^{(1) \hspace{0.25mm} 33}, \, Q_{Hq}^{(3) \hspace{0.25mm} 33}, \, Q_{Hu}^{\hspace{0.25mm} 33} \right \} \,,
\eeq
as defined in~Table~\ref{tab:no4ferm}. The importance of LL RG~running in generating~EWPO sensitivity to these operators was first recognized in the studies~\cite{Zhang:2012cd,Brod:2014hsa}. Since anomalous top-quark~EW~gauge-boson couplings remain presently only weakly bounded by direct Large~Hadron~Collider~(LHC) searches in processes such as $pp \to t\bar{t}Z$, $pp \to tW$, and related top-quark production channels~\cite{Rontsch:2014cca,CMS:2015uvn,BessidskaiaBylund:2016jvp,Degrande:2018fog,CMS:2019too,Ravina:2021kpr,Faham:2021zet,MammenAbraham:2022yxp,ATLAS:2025yww,Cornet-Gomez:2025jot},~EWPO provide complementary and often more stringent indirect bounds on the Wilson coefficients of the operators in~\eqref{eq:2topEWoperators}. These operators therefore feature prominently in global analyses of anomalous top-quark~EW~interactions, such as~\cite{Dawson:2019clf,Bellafronte:2023amz,Durieux:2019rbz,Aoude:2020dwv,Bissmann:2020mfi,Bruggisser:2022rhb,Grunwald:2023nli,Garosi:2023yxg,Allwicher:2023aql,Allwicher:2023shc,Stefanek:2024kds}, which typically also incorporate flavor constraints.

\paragraph{Triple-gauge coupling operator:} 

The dimension-six operator
\beq \label{eq:weaktriplefieldstrengthoperator}
Q_W \,,
\eeq
is the fully antisymmetric contraction of three $SU(2)_L$ gauge field-strength tensors, with its explicit definition given in~Table~\ref{tab:no4ferm}. Indirect constraints on its Wilson coefficient arise at~LL through one-loop RG~running and have been derived from~EWPO~\cite{Dawson:2019clf,Biekotter:2025nln,Haisch:2025jqr} and from the $h \to \gamma\gamma$ decay~\cite{Haisch:2025jqr,Dedes:2018seb}. At present, the~EWPO constraints on $Q_W$ are less stringent than those from Higgs physics, with the latter providing complementary indirect sensitivity to LHC diboson measurements~\cite{Ellis:2020unq,ATLAS:2020nzk,Ethier:2021ydt,CMS:2021cxr,Celada:2024mcf}.

\paragraph{Two-lepton two-third-generation quark operators:} 

The dimension-six Warsaw-basis operators involving a leptonic and a third-generation quark current are
\beq \label{eq:2top2eoperators}
\left \{ Q_{\ell q}^{(1) \hspace{0.25mm} pp33}, \, Q_{\ell q}^{(3) \hspace{0.25mm} pp33}, \, Q_{\ell u}^{pp33}, \, Q_{\ell d}^{pp33}, \, Q_{q e}^{33pp}, \, Q_{e u}^{pp33}, \, Q_{e d}^{pp33} \right \} \,,
\eeq
where the lepton flavor index $p = 1,2,3$ is not summed, and their explicit definitions can be found in~Table~\ref{tab:4ferm}. Since the operators in~\eqref{eq:2top2eoperators} remain only weakly constrained by LHC~measurements of $pp \to t\bar{t}\ell^+\ell^-$~\cite{CMS:2023xyc,ATLAS:2025yww} and $pp \to \ell^+\ell^-$~\cite{Allwicher:2022gkm}, recent theoretical studies~\cite{deBlas:2015aea,Garosi:2023yxg,Allwicher:2023shc,Greljo:2024ytg,Maura:2025rcv,Jiang:2025frv,Cao:2025xnp,Allwicher:2025mvd,Allwicher:2025bub} have explored their indirect sensitivity through~EWPO, the processes $e^+e^- \to Zh$, $e^+e^- \to t\bar{t}$, and $e^+e^- \to b\bar b$ at future $e^+e^-$ colliders, and flavor observables, with a complete fixed-order NLO analysis of the~EWPO constraints provided~in~\cite{Bellafronte:2025ubi}.

\paragraph{Top-quark EW dipole operators:}

The top-quark EW dipole operators in the Warsaw basis are
\beq \label{eq:weaktopdipoles}
\left \{ Q_{u B}^{33}, \, Q_{u W}^{33} \right \} \,,
\eeq
with their definitions given in~Table~\ref{tab:uXoperators}. While these operators can be probed directly in top-quark processes~\cite{Schulze:2016qas,Brivio:2019ius,Ethier:2021bye,Aoude:2022deh,deBlas:2025hbr}, the resulting constraints remain presently weak, rendering indirect probes via~EWPO measurements~\cite{Greiner:2011tt}, the $h \to \gamma\gamma$ decay~\cite{Hartmann:2015oia,Hartmann:2015aia,Dedes:2018seb}, and flavor observables~\cite{Bissmann:2019gfc,Aoude:2020dwv,Bruggisser:2021duo,Allwicher:2023shc,Garosi:2023yxg} complementary and often competitive. A complete fixed-order NLO analysis of the~EWPO constraints on the operators in~\eqref{eq:weaktopdipoles} has been performed in the article~\cite{Dawson:2019clf}.

\subsection{Two-loop contributions} 
\label{sec:two}

While the one-loop structure of~EWPO in the SMEFT is by now well understood, considerably less is known about their behavior at two-loop order, with the first dedicated analyses appearing only recently~\cite{Haisch:2024wnw,DiNoi:2025uhu,Mantani:2026fao,Born:2026tgm}. In the following, we present a complete and systematic study of all dimension-six operators that become accessible to~EWPO constraints for the first time through two-loop effects. Following the discussion in~Section~\ref{sec:one}, our analysis retains contributions proportional to $g_Y$, $g_L$, $g_s$, $\lambda$, and $y_t$, thereby capturing all phenomenologically relevant sources of two-loop mixing into the tree-level operator set~\eqref{eq:treeoperators}. Particular emphasis is placed on operators whose leading contributions to~EWPO arise only through genuine two-loop NLL effects. These operators are highlighted in green in~Tables~\ref{tab:no4ferm},~\ref{tab:4ferm}, and~\ref{tab:uXoperators}.

\paragraph{Third-generation four-quark operators:} 

The third-generation four-quark operators relevant for the following discussion are given in the Warsaw basis as
\beq \label{eq:4topoperators}
\left \{ Q_{qq}^{(1)\hspace{0.25mm}3333}, \, Q_{qq}^{(3)\hspace{0.25mm}3333}, \, Q_{\psi_A \psi_B}^{(1)\hspace{0.25mm}3333}, \, Q_{\psi_A \psi_B}^{(8) \hspace{0.25mm} 3333}, \, Q_{uu}^{3333}, \, Q_{dd}^{3333} \right \} \,,
\eeq
with $\psi_A \psi_B = qu, qd, ud$, and their explicit definitions provided in~Table~\ref{tab:4ferm}. At present, the dimension-six operators in~\eqref{eq:4topoperators} are only weakly constrained by LHC data, as their leading tree-level contributions arise mainly in rare multi-top final states such as $pp \to t\bar t t\bar t$ and $pp \to t\bar t b\bar b$~\cite{Hartland:2019bjb,Ethier:2021bye,Aoude:2022deh,ATLAS:2023ajo,CMS:2023ftu,Degrande:2024mbg,DiNoi:2025uhu,ATLAS:2018fwl,CMS:2019rvj,CMS:2019eih,ATLAS:2020hpj,Banelli:2020iau,ATLAS:2021kqb,CMS:2023zdh,CMS-PAS-TOP-24-008,CMS:2025ugn,Subba:2026opu}. This limited direct sensitivity has driven considerable interest in indirect probes of third-generation four-quark operators in recent years~\cite{deBlas:2015aea,Gauld:2015lmb,Hartmann:2016pil,Boughezal:2019xpp,Brivio:2019ius,Degrande:2020evl,Dawson:2022bxd,DiNoi:2023ygk,Allwicher:2023aql,Allwicher:2023shc,Heinrich:2023rsd,Stefanek:2024kds,Haisch:2024wnw,DiNoi:2025uhu,Haisch:2025vqj,Freiheit:2025orv}. In particular, their effects on~EWPO have been investigated in~\cite{Hartmann:2016pil,Boughezal:2019xpp,Dawson:2022bxd,Allwicher:2023aql,Allwicher:2023shc,Stefanek:2024kds,Haisch:2024wnw,DiNoi:2025uhu}, and in this work we extend these studies to~NLL~accuracy.

\begin{table}[t!] 
\centering
\renewcommand{\arraystretch}{1.5}

\begin{tabular}{|c|c|}
\hline 
\multicolumn{2}{|c|}{$\psi^2 \vp^3$ \text{and} $\psi^2 X \vp$} \\
\hline
\cellcolor{green!20}$Q_{u\vp}^{33}$ & $(\vp^\dag\vp) (\bar{q}_3 u_3 \tvp)$ \\
$Q_{u B}^{33}$ & $ (\bar{q}_3 \sigma^{\mu\nu} u_3 \tvp) B_{\mu\nu}$ \\
$Q_{u W}^{33}$ & $ (\bar{q}_3 \sigma^{\mu\nu}\tau^I u_3 \tvp) W_{\mu\nu}^I$ \\
$Q_{u G}^{33}$ & $(\bar{q}_3 \sigma^{\mu\nu}T^A u_3 \tvp) G_{\mu\nu}^A$ \\
\hline 
\end{tabular}
\vspace{4mm}
\caption{Non-Hermitian dimension-six operators that can mix into the~EWPO in the presence of a non-vanishing top-quark Yukawa coupling $y_t$. Operators highlighted in green first contribute to the~EWPO through $\NLL$ effects arising at two-loop order.}
\label{tab:uXoperators}
\end{table}

A particularly interesting example is the purely right-handed four-top operator $Q_{uu}^{3333}$, which can receive sizable tree-level matching contributions at the high scale in BSM~scenarios where the top quark is partially composite~\cite{Kaplan:1991dc}. In such frameworks, the right-handed top quark may couple to new strong dynamics with a strength comparable to that of the Higgs sector~\cite{Giudice:2007fh,Pomarol:2008bh}, as realized for instance in Randall–Sundrum (RS) models~\cite{Agashe:2006wa,Casagrande:2008hr} and composite Higgs (CH) models~\cite{Banelli:2020iau,Stefanek:2024kds}, which we revisit in~Section~\ref{sec:BSM}. In scenarios where~$Q_{uu}^{3333}$ dominates the new-physics effects,~EWPO receive contributions up to two-loop order from three sources: tree-level matching at the EW scale involving the two-loop RG-evolved Wilson coefficients of the operators in~\eqref{eq:treeoperators}, one-loop matching involving the one-loop RG-evolved Wilson coefficients of the same operators, and two-loop matching effects proportional to the tree-level coefficient of $Q_{uu}^{3333}$.

Focusing on terms involving only powers of $\alpha_t = y_t^2/(4\pi)$, all logarithmically enhanced contributions associated with $Q_{uu}^{3333}$ can be derived analytically. Using the recently computed beta functions of~\cite{Born:2026xkr} together with the general EFT framework of~\cite{Buras:2018gto,Haisch:2025lvd} for extracting RG-induced logarithmic enhancements, we obtain, for example
\beq \label{eq:2loop4top}
\begin{split} 
\frac{C_{HD} (m_Z)}{C_{uu}^{3333} (\Lambda)} & = \frac{12 \hspace{0.125mm} \alpha_t^2}{\pi^2} \hspace{0.25mm} \LLtwo - \frac{3 \hspace{0.125mm} \alpha_t^2}{\pi^2} \hspace{0.25mm} \NLL \,, \\[1mm] 
\frac{C_{Hq}^{(1) \hspace{0.25mm} 33} (m_Z)}{C_{uu}^{3333} (\Lambda)} & = \frac{\alpha_t^2}{\pi^2} \hspace{0.25mm} \LLtwo \,, \\[1mm] 
\frac{C_{Hq}^{(3) \hspace{0.25mm} 33} (m_Z)}{C_{uu}^{3333} (\Lambda)} & = - \frac{\alpha_t^2}{4 \pi^2} \hspace{0.25mm} \NLL \,,
\end{split}
\eeq
which correspond to the two-loop evolved Wilson coefficients generated from a non-zero initial condition $C_{uu}^{3333}(\Lambda)$. The LL and NLL contributions are highlighted in~\LLcolor~and~\NLLcolor, respectively, and this notation is used consistently throughout the article. Note that the $\LLtwo$ terms in~\eqref{eq:2loop4top} originate from two-step mixing in which RG~evolution first induces one-loop mixing of $Q_{uu}^{3333}$ into $Q_{Hu}^{33}$ or $Q_{Hu}^{33}$ and $Q_{qu}^{(1)\hspace{0.25mm}3333}$, after which these operators mix at one loop into $Q_{HD}$ or $Q_{Hq}^{(1) \hspace{0.25mm} 33}$, respectively. In the case of $Q_{HD}$, these LL contributions have already been computed in~\cite{Allwicher:2023aql,Allwicher:2023shc,Stefanek:2024kds}.\footnote{The phenomenological relevance of logarithmically enhanced effects arising from chains of operator mixing has been emphasized in several earlier works, including~\cite{Hisano:2012cc,Brod:2013cka,Cirigliano:2016njn,Cirigliano:2016nyn,Panico:2018hal,Brod:2018pli,Bauer:2020jbp,Ardu:2021koz,Brod:2023wsh,Biekotter:2023mpd,Garosi:2023yxg,Haisch:2025jqr,Allwicher:2023aql,Allwicher:2023shc,Stefanek:2024kds}.} By contrast, the $\NLL$ terms in~\eqref{eq:2loop4top} correspond to genuine two-loop effects and therefore require the two-loop beta functions of~\cite{Born:2026xkr} or equivalent computations~\cite{Haisch:2024wnw}. These contributions have also been recently calculated in~\cite{Born:2026tgm}. 

The~only~LL contributions proportional to $\alpha_t$ arising already at one-loop order from RG~evolution with~$C_{uu}^{3333}(\Lambda) \neq 0$ are given by
\beq \label{eq:1loop4top}
\begin{split}
\frac{C_{Hu}^{33} (m_Z)}{C_{uu}^{3333} (\Lambda)} = \frac{4 \hspace{0.125mm} \alpha_t}{\pi} \hspace{0.25mm} \LLone \,, \\[1mm]
\frac{C_{qu}^{(1)\hspace{0.25mm}3333} (m_Z)}{C_{uu}^{3333} (\Lambda)} = \frac{2 \hspace{0.125mm} \alpha_t}{3 \pi} \hspace{0.25mm} \LLone \,,
\end{split}
\eeq
These expression follow directly from the one-loop SMEFT beta functions, first derived in the articles~\cite{Jenkins:2013zja,Jenkins:2013wua,Alonso:2013hga}.

While explicit results were presented above for the operator $Q_{uu}^{3333}$, it is instructive to briefly discuss the logarithmic structure associated with the remaining operators in~\eqref{eq:4topoperators}. For $Q_{qq}^{(1)\hspace{0.25mm}3333}$ and $Q_{qq}^{(3)\hspace{0.25mm}3333}$, the two-loop logarithms proportional to powers of $\alpha_t$ follow the same pattern as in~\eqref{eq:2loop4top} for $C_{HD}(m_Z)$ and $C_{Hq}^{(3) \hspace{0.25mm} 33}(m_Z)$, while $C_{Hq}^{(1) \hspace{0.25mm} 33}(m_Z)$ receives additional~$\LLtwo$ contributions. At one-loop order, cf.~\eqref{eq:1loop4top}, the $\LLone$ corrections arise not in~$C_{Hu}^{33}(m_Z)$, but instead in $C_{Hq}^{(1) \hspace{0.25mm} 33}(m_Z)$ and $C_{Hq}^{(3) \hspace{0.25mm} 33}(m_Z)$. In addition, one-loop RG~evolution generates further logarithmically enhanced contributions through self-mixing of the initial operator. For the operators $Q_{qu}^{(1)\hspace{0.25mm}3333}$ and $Q_{qu}^{(8) \hspace{0.25mm} 3333}$, RG~evolution induces at two-loop order a~$\NLL$~contribution to $C_{HW\!B}(m_Z)$ proportional to $g_Y g_L \hspace{0.125mm} \alpha_t$, corresponding to a genuine two-loop effect. In addition, for $Q_{qu}^{(1)\hspace{0.25mm}3333}$ the coefficients $C_{Hq}^{(1) \hspace{0.25mm} 33}(m_Z)$, $C_{Hu}^{33}(m_Z)$, $C_{qq}^{(1)\hspace{0.25mm}3333}(m_Z)$, and $C_{qu}^{(1)\hspace{0.25mm}3333}(m_Z)$ acquire one-loop $\LLone$ corrections. By contrast, for $Q_{qu}^{(8) \hspace{0.25mm} 3333}$ the coefficients~$C_{qq}^{(1)\hspace{0.25mm}3333}(m_Z)$, $C_{qq}^{(3)\hspace{0.25mm}3333}(m_Z)$, and $C_{qu}^{(1)\hspace{0.25mm}3333}(m_Z)$ receive $\LLone$ contributions at one-loop order. Among the remaining operators, only $Q_{qd}^{(1)\hspace{0.25mm}3333}$ and $Q_{ud}^{(1)\hspace{0.25mm}3333}$ generate $\LLtwo$ and $\NLL$ corrections at two-loop order together with $\LLone$ one-loop contributions proportional to powers of $\alpha_t$. In~these cases, the induced coefficients include $C_{Hd}^{33}(m_Z)$, $C_{qd}^{(1)\hspace{0.25mm}3333}(m_Z)$, and $C_{ud}^{(1)\hspace{0.25mm}3333}(m_Z)$. Finally, once effects proportional to $\alpha_Y = g_Y^2/(4 \pi)$ and $\alpha_s$ are included, the operators $Q_{qd}^{(8) \hspace{0.25mm} 3333}$, $Q_{ud}^{(8) \hspace{0.25mm} 3333}$, and $Q_{dd}^{3333}$ also induce $\LLtwo$ and $\NLL$ corrections at two-loop order, along with~$\LLone$ one-loop contributions. Taken together with~\eqref{eq:2loop4top} and~\eqref{eq:1loop4top}, these observations imply that, once terms proportional to $g_Y$, $g_L$, $g_s$, $\lambda$, and $y_t$ are retained,~EWPO become sensitive to the full set of third-generation four-quark operators in~\eqref{eq:4topoperators} through~$\LLtwo$ and $\NLL$ two-loop effects as well as $\LLone$ one-loop corrections.

\paragraph{Top-quark chromomagnetic dipole operator:} 

In the Warsaw basis, the top-quark chromomagnetic dipole operator is called
\beq \label{eq:topgluonicdipole}
Q_{u G}^{33} \,.
\eeq
Among the dipole operators listed in~Table~\ref{tab:uXoperators}, this operator is the most extensively studied, as it can be constrained in both top-quark production and decay processes~\cite{Zhang:2010dr,Degrande:2010kt,BuarqueFranzosi:2015jrv,Brivio:2019ius,Degrande:2020evl,Hartland:2019bjb,Ethier:2021bye,Aoude:2022deh,CMS:2025ugn} as well as through Higgs-related observables~\cite{Deutschmann:2017qum,Grazzini:2018eyk,Battaglia:2021nys,DiNoi:2023onw,DiNoi:2023ygk,DiNoi:2024ajj,Maltoni:2024dpn,CampilloAveleira:2024rnp,DiNoi:2025tka,DiNoi:2025arz,Haisch:2025lvd,Haisch:2025vqj}. 

Some of the non-vanishing Wilson coefficients generated through two-loop RG~evolution from a non-zero initial condition $C_{uG}^{33}(\Lambda)$ are given by
\beq \label{eq:twoloopgluontopdipole}
\begin{split}
\frac{C_{HD} (m_Z)}{C_{uG}^{33} (\Lambda)} & = \frac{g_s \hspace{0.125mm} y_t}{16 \pi^3} \left ( 48 \hspace{0.25mm} \alpha_t - \frac{32}{3} \hspace{0.25mm} \alpha_Y \right ) \NLL \,, \\[1mm]
\frac{C_{HW\!B} (m_Z)}{C_{uG}^{33} (\Lambda)} & = \frac{g_Y g_L \hspace{0.125mm} g_s \hspace{0.125mm} y_t}{16 \pi^4} \left ( \frac{5}{3} \hspace{0.5mm} \LLtwo + \frac{2}{3} \hspace{0.5mm} \NLL \right ) \,, \\[1mm]
\frac{C_{Hq}^{(1) \hspace{0.25mm} 33} (m_Z)}{C_{uG}^{33} (\Lambda)} & = \frac{g_s \hspace{0.125mm} y_t}{16 \pi^3} \left ( \frac{32}{3} \hspace{0.25mm} \alpha_t + \frac{8}{9} \hspace{0.25mm} \alpha_s - \frac{1}{2} \hspace{0.25mm} \alpha_L - \frac{91}{54} \hspace{0.25mm} \alpha_Y \right ) \NLL \,, \\[1mm]
\frac{C_{Hq}^{(3) \hspace{0.25mm} 33} (m_Z)}{C_{uG}^{33} (\Lambda)} & = \frac{g_s \hspace{0.125mm} y_t}{16 \pi^3} \left ( \frac{4}{3} \hspace{0.25mm} \alpha_t - \frac{80}{9} \hspace{0.25mm} \alpha_s - \frac{77}{18} \hspace{0.25mm} \alpha_L - \frac{31}{54} \hspace{0.25mm} \alpha_Y \right ) \NLL \,. 
\end{split}
\eeq
The $\NLL$ terms arise from the two-loop beta functions computed in~\cite{Born:2026xkr}, while the $\LLtwo$ contribution to $C_{HW\!B}(m_Z)$ originates from a two-step one-loop mixing chain in which $Q_{uG}^{33}$ first mixes into $Q_{uB}^{33}$ and $Q_{uW}^{33}$, which subsequently mix at one loop into $Q_{HW\!B}$. Several additional $\psi^2 H^2 D$ operators listed in~Table~\ref{tab:no4ferm} likewise receive two-loop $\NLL$ corrections.

The only $\LLone$ contributions generated already at one-loop order through RG~evolution from $C_{uG}^{33}(\Lambda) \neq 0$ are
\beq \label{eq:1loop4top}
\begin{split}
\frac{C_{uB}^{33} (m_Z)}{C_{uG}^{33} (\Lambda)} = -\frac{5 \hspace{0.125mm} g_Y g_s}{18 \pi^2} \hspace{0.25mm} \LLone \,, \\[1mm]
\frac{C_{uW}^{33} (m_Z)}{C_{uG}^{33} (\Lambda)} = -\frac{g_L \hspace{0.125mm} g_s}{6 \pi^2} \hspace{0.25mm} \LLone \,.
\end{split}
\eeq
These expressions follow directly from the one-loop SMEFT beta functions first derived in the articles~\cite{Jenkins:2013zja,Jenkins:2013wua,Alonso:2013hga}. As in the case of the third-generation four-quark operators discussed above,~EWPO become sensitive to the top-quark chromomagnetic dipole operator through a combination of $\LLtwo$ and $\NLL$ two-loop effects together with $\LLone$ one-loop corrections.

\paragraph{Triple-gluon coupling operator:} 

The triple-gluon coupling operator 
\beq \label{eq:triplefieldstrengthoperators}
Q_G \,,
\eeq
corresponds to the fully antisymmetric contraction of three $SU(3)_C$ field-strength tensors and is defined in the Warsaw basis as shown in~Table~\ref{tab:no4ferm}. Phenomenologically, it can be probed through dijet measurements~\cite{Ghosh:2014wxa,Krauss:2016ely,Hirschi:2018etq,Goldouzian:2020wdq,Bardhan:2020vcl,Ellis:2020unq} as well as via Higgs production in gluon fusion $gg \to h$~\cite{Haisch:2025lvd}.

Assuming a non-vanishing initial condition $C_G(\Lambda)$, the two-loop evolved Wilson coefficients for operators involving third-generation quark~fields~are
\beq \label{eq:twolooptriplegluon}
\begin{split} 
\frac{C_{Hq}^{(1) \hspace{0.25mm} 33} (m_Z)}{C_G (\Lambda)} & = \frac{g_s \hspace{0.125mm} \alpha_s}{16 \pi^2} \left ( 4 \alpha_t + \frac{40}{9} \hspace{0.25mm} \alpha_Y \right ) \NLL \,, \\[1mm] 
\frac{C_{Hq}^{(3) \hspace{0.25mm} 33} (m_Z)}{C_G (\Lambda)} & = \frac{g_s \hspace{0.125mm} \alpha_s}{16 \pi^2} \left ( - 4 \alpha_t + \frac{40}{3} \hspace{0.25mm} \alpha_L \right ) \NLL \,, \\[1mm]
\frac{C_{Hd}^{33} (m_Z)}{C_G (\Lambda)} & = -\frac{5 \hspace{0.125mm} g_s \hspace{0.125mm} \alpha_Y \hspace{0.125mm} \alpha_s}{9 \pi^2} \hspace{0.5mm} \NLL \,.
\end{split}
\eeq
In addition, the complete set of $\psi^2 H^2 D$ operators in~Table~\ref{tab:no4ferm} involving only first- and second-generation quark fields also receives two-loop $\NLL$ corrections. By contrast, one-loop RG~evolution from $C_G(\Lambda) \neq 0$ does not generate any non-vanishing contributions to the~EWPO. The expressions in~\eqref{eq:twolooptriplegluon} therefore contain only genuine two-loop $\NLL$ terms, obtained from the two-loop beta functions computed in~\cite{Born:2026xkr}, with no accompanying $\LLtwo$ contributions. The~triple-gluon operator $Q_G$ thus provides an example in which the~EWPO sensitivity arises for the first time only through two-loop $\NLL$ effects and is therefore highlighted in green in~Table~\ref{tab:no4ferm}. Unfortunately, it remains only weakly constrained by EWPO, since the leading $\alpha_t$ contributions cancel in the linear combination $C_{Hq}^{(1)\hspace{0.25mm}33}(m_Z)+C_{Hq}^{(3)\hspace{0.25mm}33}(m_Z)$, which determines the shift in the $Z$-boson coupling to left-handed bottom quarks.

\paragraph{Top-quark Yukawa-type operator:} 

The Yukawa-type operator involving top quarks that is relevant for our analysis is, in the Warsaw basis, given by 
\beq \label{eq:topyukawaoperator}
Q_{u H}^{33} \,,
\eeq
and defined in~Table~\ref{tab:uXoperators}. The operator $Q_{uH}^{33}$ can be constrained directly in associated Higgs production with top quarks $pp \to t \bar t h$~\cite{Hartland:2019bjb,Ethier:2021bye,CMS:2025ugn,DiNoi:2023onw,Maltoni:2024dpn,Ellis:2020unq,Bartocci:2023nvp,Bevilacqua:2026udk}, as well as indirectly through loop-induced Higgs observables~\cite{Hartmann:2015oia,Hartmann:2015aia,Deutschmann:2017qum,Grazzini:2018eyk,Dedes:2018seb,Battaglia:2021nys,Heinrich:2023rsd,DiNoi:2024ajj,Maltoni:2024dpn,CampilloAveleira:2024rnp,DiNoi:2025tka,Azatov:2013xha,Grojean:2013nya,Azatov:2014jga,Buschmann:2014sia,ATL-PHYS-PUB-2023-012,Ghosh:2025fma}.

The two-loop evolved Wilson coefficients entering the~EWPO take the form
\beq \label{eq:twolooptopyukawa}
\begin{split} 
\frac{C_{HD} (m_Z)}{C_{uH}^{33} (\Lambda)} & = \frac{9 \hspace{0.125mm} y_t}{32 \pi^3} \left ( \alpha_t - \alpha_Y \right ) \NLL \,, \\[1mm] 
\frac{C_{HW\!B} (m_Z)}{C_{uH}^{33} (\Lambda)} & = \frac{g_Y g_L \hspace{0.125mm} y_t}{512 \pi^4} \hspace{0.5mm} \NLL \,, \\[1mm]
\frac{C_{Hq}^{(1) \hspace{0.25mm} 33} (m_Z)}{C_{uH}^{33} (\Lambda)} & = \frac{3 \hspace{0.125mm} y_t}{128 \pi^3} \left ( \alpha_t - 2 \alpha_Y + \alpha_\lambda \right ) \NLL \,, \\[1mm] 
\frac{C_{Hq}^{(3) \hspace{0.25mm} 33} (m_Z)}{C_{uH}^{33} (\Lambda)} & = \frac{y_t}{128 \pi^3} \left ( 2 \alpha_Y - 3 \alpha_\lambda \right ) \NLL \,,
\end{split}
\eeq
which originate from a non-zero initial condition $C_{uH}^{33} (\Lambda)$. Here, we defined $\alpha_\lambda = \lambda/(4\pi)$. In contrast, no one-loop–induced contributions to the~EWPO are generated. This implies that, for the top-quark Yukawa-type operator $Q_{uH}^{33}$, the~EWPO become sensitive only once the effects of the two-loop beta functions~\cite{Born:2026xkr} are included in the RG~analysis. This feature has recently also been emphasized in~\cite{Born:2026tgm}. As this corresponds to a NLL effect, the operator is highlighted in green in~Table~\ref{tab:uXoperators}.

\paragraph{Higgs-gluon operator:} 

The dimension-six operator in the Warsaw basis encoding Higgs-gluon interactions is denoted by
\beq \label{eq:Higgsgluonoperator}
Q_{HG} \,,
\eeq
and is defined in~Table~\ref{tab:no4ferm}. The operator $Q_{HG}$ is already strongly constrained by Higgs measurements at the LHC~\cite{Deutschmann:2017qum,Grazzini:2018eyk,Battaglia:2021nys,DiNoi:2024ajj,Maltoni:2024dpn,CampilloAveleira:2024rnp,DiNoi:2025tka,Azatov:2013xha,Grojean:2013nya,Azatov:2014jga,Buschmann:2014sia,ATL-PHYS-PUB-2023-012,Ghosh:2025fma,Gauld:2016kuu,Capozi:2019xsi,Haisch:2022nwz,Heinrich:2022idm,Brivio:2025sib,Ellis:2020unq}.

The Higgs–gluon operator $Q_{HG}$ first contributes to~EWPO at the two-loop level. For $C_{HG}(\Lambda) \neq 0$, the induced Wilson coefficients of operators involving only third-generation quark~fields~are
\beq \label{eq:twohiggsgluon}
\begin{split} 
\frac{C_{Hq}^{(1) \hspace{0.25mm} 33} (m_Z)}{C_{HG}(\Lambda)} & = \frac{\alpha_s}{16 \pi^2} \left ( \frac{40}{3} \hspace{0.5mm} \alpha_t - \frac{8}{3} \hspace{0.5mm} \alpha_Y \right ) \NLL \,, \\[1mm] 
\frac{C_{Hq}^{(3) \hspace{0.25mm} 33} (m_Z)}{C_{HG}(\Lambda)} & = \frac{\alpha_s}{16 \pi^2} \left ( -\frac{40}{3} \hspace{0.5mm} \alpha_t - 8 \hspace{0.125mm} \alpha_L \right ) \NLL \,, \\[1mm]
\frac{C_{Hd}^{33} (m_Z)}{C_{HG}(\Lambda)} & = \frac{\alpha_Y \hspace{0.125mm} \alpha_s}{3 \pi^2} \hspace{0.25mm} \NLL \,.
\end{split}
\eeq
The $\psi^2 H^2 D$ operators collected in~Table~\ref{tab:no4ferm} that involve first- and second-generation quark fields receive two-loop $\NLL$ corrections as well. These $\NLL$ contributions originate from the two-loop beta functions computed in~\cite{Born:2026xkr}. Consequently, the operator $Q_{HG}$ first contributes to the~EWPO at two-loop order via $\NLL$ effects and is therefore highlighted in green in~Table~\ref{tab:no4ferm}. Similarly to $Q_G$, the operator $Q_{HG}$ remains only weakly constrained by EWPO, since its leading $\alpha_t$ contributions do not induce a shift in the $Z$-boson coupling to left-handed bottom~quarks.

\paragraph{Higgs self-interaction operator:} 

The modifications of the Higgs self-interactions that are not associated with Higgs wave-function renormalization are described in the Warsaw basis by the operator
\beq \label{eq:Higgssexoperator}
Q_{H} \,,
\eeq
with its explicit definition given in~Table~\ref{tab:no4ferm}. At the LHC, the operator~$Q_{H}$ can be probed directly through double-Higgs production~\cite{Carvalho:2015ttv,Bizon:2018syu,Borowka:2018pxx,Heinrich:2019bkc,Capozi:2019xsi,Heinrich:2022idm,Bagnaschi:2023rbx,Bizon:2024juq,Li:2024iio,Brivio:2025sib,Ding:2026qto,CMS:2026nuu}, while complementary indirect constraints arise from loop-induced effects in single-Higgs production and decay observables~\cite{Gorbahn:2016uoy,Degrassi:2016wml,Bizon:2016wgr,Maltoni:2017ims,Gorbahn:2019lwq,Haisch:2021hvy,Gao:2023bll,Haisch:2024nzv,Ghosh:2025fma,CMS:2026nuu}. The sensitivity of future $e^+e^-$ colliders to the Higgs self-interaction operator~$Q_{H}$, in processes such as $e^+e^- \to Zh$, has also been investigated~\cite{McCullough:2013rea,DiVita:2017vrr,Maltoni:2018ttu,Rindani:2018ubx,Rao:2021eer,Maura:2025rcv,terHoeve:2025omu}, with promising projected constraints.

The leading logarithmic corrections induced by RG~evolution from a non-vanishing initial condition $C_H(\Lambda)$ to the Wilson coefficients entering the~EWPO take~the~form
\beq \label{eq:fourloopHiggsself}
\begin{split} 
\frac{C_{HD} (m_Z)}{C_{H}(\Lambda)} & = \frac{27 \hspace{0.125mm} \alpha_t}{128 \pi^5} \left ( \alpha_t - \alpha_Y \right ) \left ( 2 \alpha_t - \alpha_\lambda \right ) L^2 \,, \\[1mm]
\frac{C_{HW\!B} (m_Z)}{C_{H} (\Lambda)} & = \frac{3 \hspace{0.125mm} g_Y g_L \alpha_t}{2048 \pi^6} \left ( 2 \alpha_t - \alpha_\lambda \right ) \hspace{0.5mm} L^2 \,, \\[1mm] 
\frac{C_{Hq}^{(1) \hspace{0.25mm} 33} (m_Z)}{C_{H} (\Lambda)} & = \frac{9 \hspace{0.125mm} \alpha_t}{512 \pi^5} \left ( \alpha_t - 2 \alpha_Y + \alpha_\lambda \right ) \left ( 2 \alpha_t - \alpha_\lambda \right ) L^2 \,, \\[1mm] 
\frac{C_{Hq}^{(3) \hspace{0.25mm} 33} (m_Z)}{C_{H} (\Lambda)} & = \frac{3 \hspace{0.125mm} \alpha_t}{512 \pi^5} \left ( 2 \alpha_Y - 3 \alpha_\lambda \right ) \left ( 2 \alpha_t - \alpha_\lambda \right ) L^2 \,.
\end{split}
\eeq 
These effects first arise at the four-loop level through a two-step mixing chain and can be derived using the general formalism presented in~\cite{Buras:2018gto,Haisch:2025lvd}. In particular, RG~evolution induces a two-loop mixing of $Q_{H}$ into the top-quark Yukawa-type operator~$Q_{uH}^{33}$ in~\eqref{eq:topyukawaoperator}, which subsequently mixes at two-loop order into $Q_{HD}$, $Q_{HW\!B}$, $Q_{Hq}^{(1) \hspace{0.25mm} 33}$, and $Q_{Hq}^{(3) \hspace{0.25mm} 33}$, as shown in~\eqref{eq:twolooptopyukawa}. In the logarithmic counting defined in~\eqref{eq:LL} and~\eqref{eq:NLL}, these contributions therefore correspond to NNLL effects. At the same order, the~EWPO also receive fixed-order contributions from two-loop matching at the EW scale. In the case of the Peskin-Takeuchi parameters~\cite{Peskin:1991sw}, these effects have been computed in~\cite{Degrassi:2017ucl,Kribs:2017znd}. The operator $Q_H$ is therefore the only operator highlighted in gray in~Tables~\ref{tab:no4ferm}, \ref{tab:4ferm}, and~\ref{tab:uXoperators}, reflecting the fact that it remains inaccessible to~EWPO at NLL accuracy.

\section{NLL methodology for~EWPO}
\label{sec:matching}

In the previous section, we identified the LL and NLL contributions generated by the two-loop SMEFT RG~evolution between the scales $\Lambda$ and $m_Z$. Logarithmically enhanced effects, however, arise not only from RG running but also from the matching conditions at the EW~scale. It is therefore important to determine which loop order in the matching calculation is required to achieve NLL accuracy in the logarithmic counting of~\eqref{eq:LL} and~\eqref{eq:NLL}. While one might naively expect that two-loop matching corrections are required, we will show below, using a simple yet phenomenologically relevant example, that this is not the case. In fact, one-loop matching corrections are sufficient to obtain NLL-accurate predictions, whereas two-loop matching effects enter only at NNLL order. This observation is particularly useful since complete one-loop matching conditions are available for the full set of~EWPO~\cite{Dawson:2019clf,Dawson:2022bxd,Bellafronte:2023amz,Biekotter:2025nln,Bellafronte:2026jic}, while two-loop matching calculations currently exist only for a limited subset of dimension-six operators~\cite{Haisch:2024wnw,Degrassi:2017ucl,Kribs:2017znd}.

To illustrate this general statement, we consider the contribution of the purely right-handed four-top operator $Q_{uu}^{3333}$ to the $W$-boson mass, or equivalently to the Peskin-Takeuchi parameter $T$. The SMEFT prediction for $T$ can be decomposed into three separate contributions
\beq \label{eq:Tdefinition}
T = T^{(0)}(\mu_R) + T^{(1)}(\mu_R) + T^{(2)}(\mu_R) \,,
\eeq
where $T^{(0)}$, $T^{(1)}$, and $T^{(2)}$ denote the tree-level, one-loop, and two-loop matching contributions, respectively. These quantities are evaluated at a renormalization scale $\mu_R$ chosen to be of the order of the EW scale. By analyzing the logarithmic structure of these terms, we will show that the one-loop matching corrections already contain all ingredients required for NLL accuracy, while two-loop matching effects contribute only beyond NLL order.

Restricting the discussion to contributions involving only powers of the top-quark Yukawa coupling, the matching corrections entering~\eqref{eq:Tdefinition} can be written as
\beq \label{eq:Tmatching}
\begin{split}
T^{(0)} (\mu_R) & = -\frac{v^2}{2 \alpha} \hspace{0.5mm} C_{HD}(\mu_R) \,, \\[1mm]
T^{(1)} (\mu_R) & = -\frac{v^2}{2 \alpha} \hspace{0.5mm} \frac{6 \alpha_t}{\pi} \hspace{0.5mm} 
\ln \left ( \frac{\mu_R}{m_t} \right ) C_{Hu}^{33} (\mu_R) \,, \\[1mm]
T^{(2)} (\mu_R) & = -\frac{v^2}{2 \alpha} \hspace{0.5mm} \frac{12 \alpha_t^2}{\pi^2} \hspace{0.5mm}
\left [ \ln^2 \left ( \frac{\mu_R}{m_t} \right ) - \frac{1}{4} \hspace{0.25mm} 
\ln \left ( \frac{\mu_R}{m_t} \right ) \right ] C_{uu}^{3333} (\mu_R) \,,
\end{split}
\eeq
where the two-loop matching contribution, recently computed in~\cite{Haisch:2024wnw}, is taken directly from that reference. The Wilson coefficients at the low scale $\mu_R$ are related to the high-scale coefficient $C_{uu}^{3333}(\Lambda)$ through RG~evolution. Keeping only the logarithmic terms relevant for the present analysis, one finds
\beq \label{eq:TRGE}
\begin{split}
C_{HD} (\mu_R) & = \frac{12 \alpha_t^2}{\pi^2} \hspace{0.5mm}
\left [ \ln^2 \left ( \frac{\Lambda}{\mu_R} \right ) - \frac{1}{4} \hspace{0.25mm} 
\ln \left ( \frac{\Lambda}{\mu_R} \right ) \right ] C_{uu}^{3333} (\Lambda) \,, \\[1mm]
C_{Hu}^{33} (\mu_R) & = \frac{4 \alpha_t}{\pi} \hspace{0.5mm} 
\ln \left ( \frac{\Lambda}{\mu_R} \right ) C_{uu}^{3333} (\Lambda) \,, \\[1mm]
C_{uu}^{3333} (\mu_R) & = C_{uu}^{3333} (\Lambda) \,,
\end{split}
\eeq
where the first and second relations correspond to the results given in~\eqref{eq:2loop4top} and~\eqref{eq:1loop4top}, respectively.

Substituting~\eqref{eq:TRGE} into~\eqref{eq:Tmatching}, one obtains the final result
\beq \label{eq:Tfinal}
T = -\frac{v^2}{2 \alpha} \hspace{0.5mm} \frac{12 \alpha_t^2}{\pi^2} \hspace{0.5mm}
\left [ \ln^2 \left ( \frac{\Lambda}{m_t} \right ) - \frac{1}{4} \hspace{0.25mm} 
\ln \left ( \frac{\Lambda}{m_t} \right ) \right ] C_{uu}^{(3333)} (\Lambda) \,.
\eeq
Importantly, the final expression is independent of the renormalization scale $\mu_R$, providing a non-trivial consistency check of the matching calculation. For a more involved two-loop example, see also~\cite{Haisch:2025lvd}.

The NLL approximation of~\eqref{eq:Tdefinition} is obtained by retaining the tree-level and one-loop matching contributions evaluated at the EW scale
\beq \label{eq:TNLL}
T_{\rm NLL} = T^{(0)} (m_Z) + T^{(1)} (m_Z) \,.
\eeq
The difference between the full result~\eqref{eq:Tdefinition} and the NLL prediction~\eqref{eq:TNLL} is therefore
\beq \label{eq:TNNLL}
T - T_{\rm NLL} = T^{(2)} (m_Z) = -\frac{v^2}{2 \alpha} \frac{12 \alpha_t^2}{\pi^2} \hspace{0.5mm}
\left [ \ln^2 \left ( \frac{m_Z}{m_t} \right ) - \frac{1}{4} \hspace{0.25mm} 
\ln \left ( \frac{m_Z}{m_t} \right ) \right ] C_{uu}^{(3333)} (\Lambda) \,.
\eeq
Unlike the full result in~\eqref{eq:Tfinal}, this difference does not contain logarithms enhanced by the hierarchy between the BSM and EW scales. Since $m_Z$ and $m_t$ are parametrically of the same order, the logarithms in~\eqref{eq:TNNLL} are effectively $\mathcal O(1)$ and therefore contribute only to the~NNLL~remainder.

The above example therefore explicitly demonstrates that two-loop matching corrections are not required for NLL accuracy. More generally, the structure of the matching expansion implies that higher-order matching corrections cannot generate logarithms associated with the hierarchy between the BSM and EW scales, i.e. terms of the form~$\ln \left (\Lambda/m_Z \right )$, which are the logarithms relevant for the LL and NLL counting. Such logarithmic dependence can only be generated by the RG~evolution between the high scale and the EW scale. Consequently, the combination of one-loop matching conditions with LL and NLL~RG~evolution is sufficient to obtain NLL-accurate predictions for~EWPO. Finally, we emphasize that this argument is not specific to~EWPO, but applies more generally to any observable for which the one-loop SMEFT matching conditions are known. In such cases, the availability of the two-loop SMEFT beta functions~\cite{Born:2026xkr} allows one to systematically upgrade existing NLO predictions to NLL accuracy by consistently including the relevant~RG~evolution~effects.

\section{EWPO at NLL order} 
\label{sec:NLLEWPO}

Analytic NLO results for the complete set of~EWPO in five different EW input schemes were presented in~\cite{Biekotter:2025nln}. Building on these results, we extract the tree-level and one-loop matching conditions and combine them with the corresponding one- and two-loop RG-evolved Wilson coefficients to derive NLL-accurate expressions for the~EWPO. As explained in the previous section, the inclusion of these ingredients is sufficient to achieve NLL accuracy, since higher-order matching corrections contribute only beyond NLL order in the logarithmic counting of~\eqref{eq:LL} and~\eqref{eq:NLL}. Throughout this article, we employ the~LEP~scheme~\cite{Biekotter:2023xle,Biekotter:2023vbh}, which uses~$\alpha$, $G_F$, and $m_Z$ as input parameters. Here, $\alpha$ denotes the fine-structure constant, $G_F$~is the Fermi constant extracted from muon decay, and $m_Z$ corresponds to the on-shell $Z$-boson mass. In this scheme, the~EWPO associated with the $W$ boson are its on-shell mass $m_W$ together with its total, hadronic, and leptonic decay widths, $\Gamma_W$, $\Gamma_W^{\rm had}$, and $\Gamma_W^{\rm lep}$. For the $Z$ boson, the relevant observables comprise the total decay width $\Gamma_Z$, the hadronic cross section $\sigma_{\mathrm{had}}$ in the narrow-width approximation, as well as ratios, left-right asymmetries, and forward-backward asymmetries for decays into leptons and quarks, namely $R_\ell$, $R_q$, $A_\ell$, $A_q$, $A_{\rm FB}^\ell$, and $A_{\rm FB}^q$, with $\ell = e,\mu,\tau$ and $q = s,c,b$. We note that the leptonic effective weak mixing angle $\sin^2\theta_{\mathrm{eff}}^\ell$ is not treated as an independent~EWPO, since it is equivalent to $A_\ell$, as discussed for example in~\cite{Biekotter:2025nln}.

To obtain numerical expressions for the~EWPO in the LEP scheme, we use the following input parameters from the 2025 Particle Data Group~(PDG) update~\cite{ParticleDataGroup:2024cfk}
\beq \label{eq:numerics1}
\begin{gathered}
\alpha (m_Z) = \frac{1}{128.946} \,, \quad G_F = 1.1663787 \cdot 10^{-5} \, {\rm GeV}^{-2} \,, \quad m_Z = 91.1880 \, {\rm GeV} \,, \\[1mm]
\alpha_s (m_Z) = 0.1180 \,, \quad m_t = 172.6 \, {\rm GeV} \,, \quad m_h = 125.1 \, {\rm GeV} \,, \quad m_W = 80.359 \, {\rm GeV} \,,
\end{gathered}
\eeq
where the masses in the second line correspond to the SM on-shell values. The Higgs VEV and the sine of the weak mixing angle are then determined from
\beq \label{eq:LEPrelations}
v = \frac{1}{\sqrt[4]{2} \hspace{0.125mm} \sqrt{G_F}} = 246.22 \, {\rm GeV} \,, \qquad s_w^2 = \frac{1}{2} \left [ 1 - \sqrt{1 - \frac{2 \sqrt{2} \pi \hspace{0.125mm} \alpha}{G_F \hspace{0.125mm} m_Z^2}} \right ] = 0.231 \,.
\eeq
For the couplings entering the one- and two-loop SMEFT beta functions, we use
\beq \label{eq:numerics2}
g_Y = 0.37 \,, \qquad g_L = 0.65 \,, \qquad g_s = 1.22 \,, \qquad y_t = 0.97 \,, \qquad \lambda = 0.28 \,,
\eeq
while all other Yukawa couplings are set to zero. The values in~\eqref{eq:numerics2} correspond to the relevant SM parameters evaluated at the scale $m_Z$. Evaluating these couplings instead at the scale $\Lambda$ reproduces the same logarithmic structure up to NLL order in all formulas presented below, with any resulting numerical differences contributing only at NNLL order according to the logarithmic counting defined in~\eqref{eq:LL} and~\eqref{eq:NLL}.

The numerical formulas given below describe the $\LLtwo$, $\LLone$, and $\NLL$ dependence of the~EWPO on the dimensionless high-scale Wilson coefficients $c_i(\Lambda)$, defined through
\beq \label{eq:smallc}
C_i (\Lambda) = \frac{c_i (\Lambda)}{{\rm TeV}^2} \,.
\eeq
Explicit expressions are presented for the five BSM~scenarios discussed in~Section~\ref{sec:two}, while the ancillary electronic files accompanying the arXiv submission of this work extend these results to a total of 210 independent Wilson coefficients. This includes all operators listed in~Tables~\ref{tab:no4ferm},~\ref{tab:4ferm}, and~\ref{tab:uXoperators}, with the only exception being the Higgs self-interaction operator $Q_H$, whose contributions to the~EWPO first arise at the NNLL order.

\subsection{Purely right-handed four-top operator} 
\label{sec:semi4top}

As a first application, we consider the purely right-handed four-top operator $Q_{uu}^{3333}$, introduced as the first example in~Section~\ref{sec:two}. In this scenario, the complete set of~EWPO is predicted at NLL accuracy as follows:
\begin{align} \label{eq:semianalytic4top}
\frac{\Delta m_W}{m_W} &= \left( -1.38 \hspace{0.25mm} \LLtwo + 1.78 \hspace{0.25mm} \LLone + 0.345 \hspace{0.25mm} \NLL \right) \cdot 10^{-4} \, c_{uu}^{3333} (\Lambda) \,, \\[1mm]
\frac{\Delta \Gamma_W}{\Gamma_W} &= \left( -4.01 \hspace{0.25mm} \LLtwo + 5.16 \hspace{0.25mm} \LLone + 1.00 \hspace{0.25mm} \NLL \right) \cdot 10^{-4} \, c_{uu}^{3333} (\Lambda) \,, \\[1mm]
\frac{\Delta \Gamma_W^{\mathrm{had}}}{\Gamma_W^{\mathrm{had}}} &= \left( -3.97 \hspace{0.25mm} \LLtwo + 5.10 \hspace{0.25mm} \LLone + 0.991 \hspace{0.25mm} \NLL \right) \cdot 10^{-4} \, c_{uu}^{3333} (\Lambda) \,, \\[1mm]
\frac{\Delta \Gamma_W^{\mathrm{lep}}}{\Gamma_W^{\mathrm{lep}}} &= \left( -4.12 \hspace{0.25mm} \LLtwo + 5.29 \hspace{0.25mm} \LLone + 1.03 \hspace{0.25mm} \NLL \right) \cdot 10^{-4} \, c_{uu}^{3333} (\Lambda) \,, \\[1mm]
\frac{\Delta \Gamma_Z}{\Gamma_Z} &= \left( -2.50 \hspace{0.25mm} \LLtwo + 3.32 \hspace{0.25mm} \LLone + 0.614 \hspace{0.25mm} \NLL \right) \cdot 10^{-4} \, c_{uu}^{3333} (\Lambda) \,, \\[1mm]
\frac{\Delta \sigma_{\mathrm{had}}}{\sigma_{\mathrm{had}}} &= \left( -0.0244 \hspace{0.25mm} \LLtwo + 0.0246 \hspace{0.25mm} \LLone + 0.0124 \hspace{0.25mm} \NLL \right) \cdot 10^{-4} \, c_{uu}^{3333} (\Lambda) \,, \\[1mm]
\frac{\Delta R_\ell}{R_\ell} &= \left( -0.313 \hspace{0.25mm} \LLtwo + 0.420 \hspace{0.25mm} \LLone + 0.0627 \hspace{0.25mm} \NLL \right) \cdot 10^{-4} \, c_{uu}^{3333} (\Lambda) \,, \\[1mm]
\frac{\Delta A_\ell}{A_\ell} &= \left( -31.8 \hspace{0.25mm} \LLtwo + 41.0 \hspace{0.25mm} \LLone + 7.96 \hspace{0.25mm} \NLL \right) \cdot 10^{-4} \, c_{uu}^{3333} (\Lambda) \,, \\[1mm]
\frac{\Delta A_{\mathrm{FB}}^\ell}{A_{\mathrm{FB}}^\ell} &= \left( -65.5 \hspace{0.25mm} \LLtwo + 84.4 \hspace{0.25mm} \LLone + 16.4 \hspace{0.25mm} \NLL \right) \cdot 10^{-4} \, c_{uu}^{3333} (\Lambda) \,, \\[1mm]
\frac{\Delta R_s}{R_s} &= \left( -0.0583 \hspace{0.25mm} \LLtwo + 0.0575 \hspace{0.25mm} \LLone + 0.0305 \hspace{0.25mm} \NLL \right) \cdot 10^{-4} \, c_{uu}^{3333} (\Lambda) \,, \\[1mm]
\frac{\Delta A_s}{A_s} &= \left( -0.404 \hspace{0.25mm} \LLtwo + 0.520 \hspace{0.25mm} \LLone + 0.101 \hspace{0.25mm} \NLL \right) \cdot 10^{-4} \, c_{uu}^{3333} (\Lambda) \,, \\[1mm]
\frac{\Delta A_{\mathrm{FB}}^s}{A_{\mathrm{FB}}^s} &= \left( -32.4 \hspace{0.25mm} \LLtwo + 41.7 \hspace{0.25mm} \LLone + 8.10 \hspace{0.25mm} \NLL \right) \cdot 10^{-4} \, c_{uu}^{3333} (\Lambda) \,, \\[1mm]
\frac{\Delta R_c}{R_c} &= \left( -0.375 \hspace{0.25mm} \LLtwo + 0.514 \hspace{0.25mm} \LLone + 0.0628 \hspace{0.25mm} \NLL \right) \cdot 10^{-4} \, c_{uu}^{3333} (\Lambda) \,, \\[1mm]
\frac{\Delta A_c}{A_c} &= \left( -3.08 \hspace{0.25mm} \LLtwo + 3.84 \hspace{0.25mm} \LLone + 0.887 \hspace{0.25mm} \NLL \right) \cdot 10^{-4} \, c_{uu}^{3333} (\Lambda) \,, \\[1mm]
\frac{\Delta A_{\mathrm{FB}}^c}{A_{\mathrm{FB}}^c} &= \left( -35.1 \hspace{0.25mm} \LLtwo + 45.1 \hspace{0.25mm} \LLone + 8.89 \hspace{0.25mm} \NLL \right) \cdot 10^{-4} \, c_{uu}^{3333} (\Lambda) \,, \\[1mm]
\frac{\Delta R_b}{R_b} &= \left( 0.718 \hspace{0.25mm} \LLtwo - 0.939 \hspace{0.25mm} \LLone - 0.163 \hspace{0.25mm} \NLL \right) \cdot 10^{-4} \, c_{uu}^{3333} (\Lambda) \,, \\[1mm]
\frac{\Delta A_b}{A_b} &= \left( -0.352 \hspace{0.25mm} \LLtwo + 0.454 \hspace{0.25mm} \LLone + 0.0880 \hspace{0.25mm} \NLL \right) \cdot 10^{-4} \, c_{uu}^{3333} (\Lambda) \,, \\[1mm]
\frac{\Delta A_{\mathrm{FB}}^b}{A_{\mathrm{FB}}^b} &= \left( -32.2 \hspace{0.25mm} \LLtwo + 41.5 \hspace{0.25mm} \LLone + 8.06 \hspace{0.25mm} \NLL \right) \cdot 10^{-4} \, c_{uu}^{3333} (\Lambda) \,.
\end{align}
Here, the SMEFT-induced shifts in the~EWPO are normalized to their corresponding SM values. The SM predictions and associated uncertainties for the~EWPO are summarized in~Table~\ref{tab:EWPOummary}. To clearly illustrate the numerical impact of the LL and NLL contributions, we keep the~$\LLtwo$,~$\LLone$, and~$\NLL$ terms separate. Although no strict hierarchy exists among their coefficients, the $\LLtwo$ and $\LLone$ terms are typically larger in magnitude than the $\NLL$ contributions, thereby justifying the LL and NLL classification both formally and numerically. For~$Q_{uu}^{3333}$, the $\LLtwo$ contributions --- arising from two-step RG~mixing --- generally dominate the SMEFT shifts in the~EWPO. This feature is common to all third-generation four-quark operators listed in~\eqref{eq:4topoperators}. Finally, we note that the shifts $\Delta A_\ell$, $\Delta A_{\mathrm{FB}}^\ell$, $\Delta A_{\mathrm{FB}}^s$, $\Delta A_{\mathrm{FB}}^c$, and $\Delta A_{\mathrm{FB}}^b$ are particularly sensitive to $Q_{uu}^{3333}$ through top-quark Yukawa enhanced corrections to the $T$ parameter and the $Z$-boson couplings to bottom quarks. The same mechanism enhances the sensitivity to all operators in~\eqref{eq:4topoperators}. We emphasize that the analysis of~\cite{DiNoi:2025uhu} includes only oblique NNLO fixed-order corrections to the~EWPO induced by these operators and therefore neglects the corresponding two-loop non-oblique effects. As we demonstrate in~Appendix~\ref{app:4topEWPO}, a complete~EWPO analysis must incorporate both oblique and non-oblique observables in order to fully exploit the constraining power of precision $Z$-pole measurements.

\subsection{Top-quark chromomagnetic dipole operator} 
\label{sec:semitopdipole}

We now repeat the analysis for the top-quark chromomagnetic dipole operator $Q_{uG}^{33}$, introduced as the second example in~Section~\ref{sec:two}, obtaining the following NLL results:
\begin{align} \label{eq:semianalytictopdipole}
\frac{\Delta m_W}{m_W} &= \left( -0.139 \hspace{0.25mm} \LLtwo + 0.125 \hspace{0.25mm} \LLone -1.79 \hspace{0.25mm} \NLL \right) \cdot 10^{-4} \, c_{uG}^{3333} (\Lambda) \,, \\[1mm]
\frac{\Delta \Gamma_W}{\Gamma_W} &= \left( -0.405 \hspace{0.25mm} \LLtwo + 0.349 \hspace{0.25mm} \LLone - 5.21 \hspace{0.25mm} \NLL \right) \cdot 10^{-4} \, c_{uG}^{3333} (\Lambda) \,, \\[1mm]
\frac{\Delta \Gamma_W^{\mathrm{had}}}{\Gamma_W^{\mathrm{had}}} &= \left( -0.400 \hspace{0.25mm} \LLtwo + 0.344 \hspace{0.25mm} \LLone - 5.14 \hspace{0.25mm} \NLL \right) \cdot 10^{-4} \, c_{uG}^{3333} (\Lambda) \,, \\[1mm]
\frac{\Delta \Gamma_W^{\mathrm{lep}}}{\Gamma_W^{\mathrm{lep}}} &= \left( -0.415 \hspace{0.25mm} \LLtwo + 0.357 \hspace{0.25mm} \LLone - 5.34 \hspace{0.25mm} \NLL \right) \cdot 10^{-4} \, c_{uG}^{3333} (\Lambda) \,, \\[1mm]
\frac{\Delta \Gamma_Z}{\Gamma_Z} &= \left( -0.143 \hspace{0.25mm} \LLtwo + 0.229 \hspace{0.25mm} \LLone - 3.49\hspace{0.25mm} \NLL \right) \cdot 10^{-4} \, c_{uG}^{3333} (\Lambda) \,, \\[1mm]
\frac{\Delta \sigma_{\mathrm{had}}}{\sigma_{\mathrm{had}}} &= \left( 0.00956 \hspace{0.25mm} \LLtwo - 0.0389\hspace{0.25mm} \LLone + 0.141 \hspace{0.25mm} \NLL \right) \cdot 10^{-4} \, c_{uG}^{3333} (\Lambda) \,, \\[1mm]
\frac{\Delta R_\ell}{R_\ell} &= \left( -0.112 \hspace{0.25mm} \LLtwo + 0.211 \hspace{0.25mm} \LLone - 0.857 \hspace{0.25mm} \NLL \right) \cdot 10^{-4} \, c_{uG}^{3333} (\Lambda) \,, \\[1mm]
\frac{\Delta A_\ell}{A_\ell} &= \left( -7.49 \hspace{0.25mm} \LLtwo + 9.64 \hspace{0.25mm} \LLone - 42.9 \hspace{0.25mm} \NLL \right) \cdot 10^{-4} \, c_{uG}^{3333} (\Lambda) \,, \\[1mm]
\frac{\Delta A_{\mathrm{FB}}^\ell}{A_{\mathrm{FB}}^\ell} &= \left( -15.4 \hspace{0.25mm} \LLtwo + 19.8 \hspace{0.25mm} \LLone - 88.3\hspace{0.25mm} \NLL \right) \cdot 10^{-4} \, c_{uG}^{3333} (\Lambda) \,, \\[1mm]
\frac{\Delta R_s}{R_s} &= \left( 0.0254 \hspace{0.25mm} \LLtwo - 0.101 \hspace{0.25mm} \LLone - 3.30 \hspace{0.25mm} \NLL \right) \cdot 10^{-4} \, c_{uG}^{3333} (\Lambda) \,, \\[1mm]
\frac{\Delta A_s}{A_s} &= \left( -0.0951 \hspace{0.25mm} \LLtwo + 0.122 \hspace{0.25mm} \LLone + 0.421 \hspace{0.25mm} \NLL \right) \cdot 10^{-4} \, c_{uG}^{3333} (\Lambda) \,, \\[1mm]
\frac{\Delta A_{\mathrm{FB}}^s}{A_{\mathrm{FB}}^s} &= \left( -7.62 \hspace{0.25mm} \LLtwo + 9.81 \hspace{0.25mm} \LLone - 42.7 \hspace{0.25mm} \NLL \right) \cdot 10^{-4} \, c_{uG}^{3333} (\Lambda) \,, \\[1mm]
\frac{\Delta R_c}{R_c} &= \left( -0.0491 \hspace{0.25mm} \LLtwo -0.00487 \hspace{0.25mm} \LLone + 4.67 \hspace{0.25mm} \NLL \right) \cdot 10^{-4} \, c_{uG}^{3333} (\Lambda) \,, \\[1mm]
\frac{\Delta A_c}{A_c} &= \left( -0.724 \hspace{0.25mm} \LLtwo + 0.932 \hspace{0.25mm} \LLone - 6.19 \hspace{0.25mm} \NLL \right) \cdot 10^{-4} \, c_{uG}^{3333} (\Lambda) \,, \\[1mm]
\frac{\Delta A_{\mathrm{FB}}^c}{A_{\mathrm{FB}}^c} &= \left( -8.25 \hspace{0.25mm} \LLtwo + 10.6 \hspace{0.25mm} \LLone - 49.4 \hspace{0.25mm} \NLL \right) \cdot 10^{-4} \, c_{uG}^{3333} (\Lambda) \,, \\[1mm]
\frac{\Delta R_b}{R_b} &= \left( 0.0261 \hspace{0.25mm} \LLtwo + 0.215 \hspace{0.25mm} \LLone - 0.657 \hspace{0.25mm} \NLL \right) \cdot 10^{-4} \, c_{uG}^{3333} (\Lambda) \,, \\[1mm]
\frac{\Delta A_b}{A_b} &= \left( -0.0952 \hspace{0.25mm} \LLtwo + 0.144 \hspace{0.25mm} \LLone + 0.606 \hspace{0.25mm} \NLL \right) \cdot 10^{-4} \, c_{uG}^{3333} (\Lambda) \,, \\[1mm]
\frac{\Delta A_{\mathrm{FB}}^b}{A_{\mathrm{FB}}^b} &= \left( -7.60 \hspace{0.25mm} \LLtwo + 9.81 \hspace{0.25mm} \LLone - 42.3\hspace{0.25mm} \NLL \right) \cdot 10^{-4} \, c_{uG}^{3333} (\Lambda) \,.
\end{align}
Compared to the~EWPO results presented in~Section~\ref{sec:semi4top}, we find that, for all~EWPO, the coefficients of the $\NLL$ contributions are larger than those of the $\LLtwo$ and $\LLone$ terms. This~pattern can be traced to the sizable two-loop beta function proportional to $y_t^3$ that governs the mixing of $Q_{uG}^{33}$ into $Q_{HD}$, see the first relation in~\eqref{eq:twoloopgluontopdipole}. Furthermore, only $C_{HW\!B}(m_Z)$ receives a $\LLtwo$ correction, while $\LLone$ contributions arise exclusively in $C_{uB}^{33}(m_Z)$ and $C_{uW}^{33}(m_Z)$, as follows from~\eqref{eq:twoloopgluontopdipole} and~\eqref{eq:1loop4top}, respectively. The exceptions are $\Delta \sigma_{\rm had}$, $\Delta R_s$, $\Delta A_s$, $\Delta R_c$, $\Delta R_b$, and $\Delta A_b$, for which the dominant $\NLL$ effects originate from the $\psi^2 H^2 D$ operators listed in~Table~\ref{tab:no4ferm}. The sizable $\NLL$ terms indicate that, in obtaining precise~EWPO constraints on the top-quark chromomagnetic dipole operator, the two-loop beta functions computed in~\cite{Born:2026xkr} play a significantly more important role than in the case of $Q_{uu}^{3333}$ discussed above.

\subsection{Operators appearing at NLL level} 
\label{sec:semiNLL}

Finally, we consider the operators $Q_G$, $Q_{uH}^{33}$ and $Q_{HG}$, as discussed as the examples three to five in~Sections~\ref{sec:two}. In contrast to the BSM~scenarios considered so far in this section, these three operators give rise only to genuine two-loop effects proportional to $\NLL$. We obtain the following formulas:
\begin{align} \label{eq:semiNLL}
\frac{\Delta m_W}{m_W} &= -0.125 \hspace{0.25mm} \cdot 10^{-4} \, \NLL \, c_{uH}^{33} (\Lambda) \,, \\[1mm]
\frac{\Delta \Gamma_W}{\Gamma_W} &= \Big [ \hspace{0.25mm} 0.321 \, c_{G} (\Lambda) - 0.363 \, c_{uH}^{33} (\Lambda) - 0.158 \, c_{HG} (\Lambda) \hspace{0.25mm} \Big ] \hspace{0.25mm} \cdot 10^{-4} \, \NLL \,, \\[1mm]
\frac{\Delta \Gamma_W^{\mathrm{had}}}{\Gamma_W^{\mathrm{had}}} &= \Big [ \hspace{0.25mm} 0.476 \, c_{G} (\Lambda) - 0.358 \, c_{uH}^{33} (\Lambda) - 0.234 \, c_{HG} (\Lambda) \hspace{0.25mm} \Big ] \hspace{0.25mm} \cdot 10^{-4} \, \NLL \,, \\[1mm]
\frac{\Delta \Gamma_W^{\mathrm{lep}}}{\Gamma_W^{\mathrm{lep}}} &= -0.372 \hspace{0.25mm} \cdot 10^{-4} \, \NLL \, c_{uH}^{33} (\Lambda) \,, \\[1mm]
\frac{\Delta \Gamma_Z}{\Gamma_Z} &= \Big [ \hspace{0.25mm} 0.436 \, c_{G} (\Lambda) - 0.219 \, c_{uH}^{33} (\Lambda) - 0.215 \, c_{HG} (\Lambda) \hspace{0.25mm} \Big ] \hspace{0.25mm} \cdot 10^{-4} \, \NLL \,, \\[1mm]
\frac{\Delta \sigma_{\mathrm{had}}}{\sigma_{\mathrm{had}}} &= \Big [ \hspace{0.25mm} -0.252 \, c_{G} (\Lambda) - 0.00127 \, c_{uH}^{33} (\Lambda) + 0.124 \, c_{HG} (\Lambda) \hspace{0.25mm} \Big ] \hspace{0.25mm} \cdot 10^{-4} \, \NLL \,, \\[1mm]
\frac{\Delta R_\ell}{R_\ell} &= \Big [ \hspace{0.25mm} 0.629 \, c_{G} (\Lambda) - 0.0340 \, c_{uH}^{33} (\Lambda) - 0.309 \, c_{HG} (\Lambda) \hspace{0.25mm} \Big ] \hspace{0.25mm} \cdot 10^{-4} \, \NLL \,, \\[1mm]
\frac{\Delta A_\ell}{A_\ell} &= -3.16 \hspace{0.25mm} \cdot 10^{-4} \, \NLL \, c_{uH}^{33} (\Lambda) \,, \\[1mm]
\frac{\Delta A_{\mathrm{FB}}^\ell}{A_{\mathrm{FB}}^\ell} &= -6.51 \hspace{0.25mm} \cdot 10^{-4} \, \NLL \, c_{uH}^{33} (\Lambda) \,, \\[1mm]
\frac{\Delta R_s}{R_s} &= \Big [ \hspace{0.25mm} 3.21 \cdot 10^{-5} \, c_{G} (\Lambda) - 0.00284 \, c_{uH}^{33} (\Lambda) - 1.58 \cdot 10^{-5} \, c_{HG} (\Lambda) \hspace{0.25mm} \Big ] \hspace{0.25mm} \cdot 10^{-4} \, \NLL \,, \\[1mm]
\frac{\Delta A_s}{A_s} &= \Big [ \hspace{0.25mm} -1.20 \cdot 10^{-4} \, c_{G} (\Lambda) - 0.0401 \, c_{uH}^{33} (\Lambda) + 5.91 \cdot 10^{-5} \, c_{HG} (\Lambda) \hspace{0.25mm} \Big ] \hspace{0.25mm} \cdot 10^{-4} \, \NLL \,, \\[1mm]
\frac{\Delta A_{\mathrm{FB}}^s}{A_{\mathrm{FB}}^s} &= \Big [ \hspace{0.25mm} -1.24 \cdot 10^{-4} \, c_{G} (\Lambda) - 3.22 \, c_{uH}^{33} (\Lambda) + 6.11 \cdot 10^{-5} \, c_{HG} (\Lambda) \hspace{0.25mm} \Big ] \hspace{0.25mm} \cdot 10^{-4} \, \NLL \,, \\[1mm]
\frac{\Delta R_c}{R_c} &= \Big [ \hspace{0.25mm} -6.20 \cdot 10^{-5} \, c_{G} (\Lambda) - 0.0343 \, c_{uH}^{33} (\Lambda) + 3.05 \cdot 10^{-5} \, c_{HG} (\Lambda) \hspace{0.25mm} \Big ] \hspace{0.25mm} \cdot 10^{-4} \, \NLL \,, \\[1mm]
\frac{\Delta A_c}{A_c} &= \Big [ \hspace{0.25mm} -9.15 \cdot 10^{-4} \, c_{G} (\Lambda) - 0.306 \, c_{uH}^{33} (\Lambda) + 4.50 \cdot 10^{-3} \, c_{HG} (\Lambda) \hspace{0.25mm} \Big ] \hspace{0.25mm} \cdot 10^{-4} \, \NLL \,, \\[1mm]
\frac{\Delta A_{\mathrm{FB}}^c}{A_{\mathrm{FB}}^c} &= \Big [ \hspace{0.25mm} -9.42 \cdot 10^{-4} \, c_{G} (\Lambda) - 3.49 \, c_{uH}^{33} (\Lambda) + 4.63 \cdot 10^{-3} \, c_{HG} (\Lambda) \hspace{0.25mm} \Big ] \hspace{0.25mm} \cdot 10^{-4} \, \NLL \,, \\[1mm]
\frac{\Delta R_b}{R_b} &= \Big [ \hspace{0.25mm} 3.30 \cdot 10^{-5} \, c_{G} (\Lambda) + 0.0605 \, c_{uH}^{33} (\Lambda) - 1.62 \cdot 10^{-5} \, c_{HG} (\Lambda) \hspace{0.25mm} \Big ] \hspace{0.25mm} \cdot 10^{-4} \, \NLL \,, \\[1mm]
\frac{\Delta A_b}{A_b} &= \Big [ \hspace{0.25mm} -1.20 \cdot 10^{-4} \, c_{G} (\Lambda) - 0.0359 \, c_{uH}^{33} (\Lambda) + 5.91 \cdot 10^{-5} \, c_{HG} (\Lambda) \hspace{0.25mm} \Big ] \hspace{0.25mm} \cdot 10^{-4} \, \NLL \,, \\[1mm]
\frac{\Delta A_{\mathrm{FB}}^b}{A_{\mathrm{FB}}^b} &= \Big [ \hspace{0.25mm} -1.24 \cdot 10^{-4} \, c_{G} (\Lambda) - 3.20 \, c_{uH}^{33} (\Lambda) + 6.09 \cdot 10^{-5} \, c_{HG} (\Lambda) \hspace{0.25mm} \Big ] \hspace{0.25mm} \cdot 10^{-4} \, \NLL \,.
\end{align}
From the above results, it is clear that although the operators $Q_G$, $Q_{uH}^{33}$, and $Q_{HG}$ all start contributing at the $\NLL$ level, the~EWPO --- with the exception of $\Delta R_\ell$ --- are considerably more sensitive to $Q_{uH}^{33}$ than to $Q_G$ or $Q_{HG}$. Several effects are responsible for this hierarchy. First, only $Q_{uH}^{33}$ mixes into $Q_{HD}$ at the two-loop level, cf.~\eqref{eq:twolooptopyukawa}, while no comparable contribution appears for $Q_G$ or $Q_{HG}$ up to NLL order. Second, the top-quark Yukawa corrections to the left-handed $Z$-boson couplings to bottom quarks, entering through the combination $C_{Hq}^{(1)\hspace{0.25mm}33}(m_Z)+C_{Hq}^{(3)\hspace{0.25mm}33}(m_Z)$, cancel in the cases of $Q_G$ and $Q_{HG}$, as follows from~\eqref{eq:twolooptriplegluon} and~\eqref{eq:twohiggsgluon}. For $\Delta R_\ell$, by contrast, the combination of Wilson coefficients~$C_{HD}(m_Z)$, $C_{HW\!B}(m_Z)$, $C_{Hq}^{(1)\hspace{0.25mm}33}(m_Z)$, and $C_{Hq}^{(3)\hspace{0.25mm}33}(m_Z)$ entering the $Q_{uH}^{33}$ contribution largely cancels, resulting in a comparatively small coefficient. Notice also that, for $\Delta \Gamma_W$, $\Delta \Gamma_W^{\rm had}$, $\Delta \Gamma_Z$, and $\Delta \sigma_{\rm had}$, the coefficients associated with $Q_G$ and $Q_{HG}$ are of the same order as those of $Q_{uG}^{33}$. In the cases of $Q_G$ and $Q_{HG}$, this behavior is mainly driven by non-oblique contributions from $\psi^2 H^2 D$ operators involving first- and second-generation quark fields, whereas for $Q_{uG}^{33}$ the dominant $\NLL$ contribution originates from the oblique operator~$Q_{HD}$. We note that the $\NLL$ contributions to the $W$-boson mass and the weak mixing angle arising from the two-loop mixing of $Q_{uH}^{33}$ into $Q_{HD}$ and $Q_{HW\!B}$ have recently been investigated in~\cite{Born:2026tgm}. The results presented above extend this analysis to the complete set of relevant dimension-six operators as well as to the full set of~EWPO.

\section{NLL~EWPO fits} 
\label{sec:Zpolefits}

In this section, we perform dedicated~EWPO fits with particular emphasis on the classes of dimension-six operators that first contribute at the two-loop level, as discussed in~Section~\ref{sec:two}. To assess the sensitivity and complementarity of~EWPO, the resulting constraints are compared with bounds obtained from existing LHC measurements in the top-quark and Higgs sectors. We consider both the current experimental information from LEP and SLC as well as projected measurements at future high-luminosity $e^+e^-$ colliders. This~comparison allows us to quantify the extent to which~EWPO measurements can provide competitive or complementary sensitivity to BSM effects that are otherwise probed through direct collider~searches.

\subsection{EWPO and their SM predictions} 
\label{sec:SMexpEWPO}

{
\renewcommand{\arraystretch}{1.25}
\begin{table}[t!]
\centering
\begin{tabular}{|l|c|c|c|}
\hline
Observable & SM prediction & Experimental value & Future precision $\left (10^{-4} \right)$ \\
\hline \hline
$m_W \, [\mathrm{GeV}]$ & $80.356 \pm 0.004$ & $80.361 \pm 0.008$ & $\pm 0.049$ \\
$\Gamma_W \, [\mathrm{GeV}]$ & $2.090 \pm 0.001$ & $2.140 \pm 0.050$ & $\pm 1.61$ \\
$\Gamma_Z \, [\mathrm{GeV}]$ & $2.4942 \pm 0.0005$ & $2.4955 \pm 0.0023$ & $\pm 0.155$ \\
$\sigma_{\mathrm{had}} \, [\mathrm{nb}]$ & $41.492 \pm 0.008$ & $41.4802 \pm 0.0325$ & $\pm 0.433$ \\
$R_\ell$ & $20.749 \pm 0.006$ & $20.767 \pm 0.025$ & $\pm 0.220$ \\ 
$A_\ell$ & $0.1470 \pm 0.0004$ & $0.1500 \pm 0.0018$ & $\pm 8.79$ \\ 
$A_{\mathrm{FB}}^\ell$ & $0.01621 \pm 0.0001$ & $0.0171 \pm 0.0010$ & $\pm 12.5$ \\
$A_s$ & $0.9357 \pm 0.00004$ & $0.895 \pm 0.091$ & $\pm 1.42$ \\
$A_{\mathrm{FB}}^s$ & $0.1027 \pm 0.0002$ & $0.0976 \pm 0.0114$ & $\pm 8.82$ \\
$R_c$ & $0.17220 \pm 0.00003$ & $0.1721 \pm 0.0030$ & $\pm 10.0$ \\
$A_c$ & $0.6678 \pm 0.0002$ & $0.670 \pm 0.027$ & $\pm 1.69$ \\
$A_{\mathrm{FB}}^c$ & $0.0736 \pm 0.0002$ & $0.0707 \pm 0.0035$ & $\pm 8.91$ \\
$R_b$ & $0.21588 \pm 0.00003$ & $0.21629 \pm 0.00066$ & $\pm 2.01$ \\
$A_b$ & $0.93474 \pm 0.00003$ & $0.923 \pm 0.020$ & $\pm 1.17$ \\
$A_{\mathrm{FB}}^b$ & $0.1030 \pm 0.0003$ & $0.0996 \pm 0.0016$ & $\pm 8.78$ \\
\hline
\end{tabular}
\vspace{4mm} 
\caption{Overview of the EWPO in the LEP scheme. The first column lists the SM predictions, while the second gives the current experimental measurements. All values are taken from the 2025~PDG~update, except for $m_W$, which is determined as described in the text. The third column shows the projected relative uncertainties for the FCC-ee Tera-$Z$ stage in the S1 scenario of~\cite{Greljo:2025ggc}. Further details are provided in the main text.}
\label{tab:EWPOummary} 
\end{table}
}

Before presenting our EWPO fits, we briefly review the SM predictions and the corresponding experimental measurements of the observables entering our analysis. A summary is provided in~Table~\ref{tab:EWPOummary}. The values shown in the first two columns are taken from the 2025~PDG~update. Since the measurements of $R_\ell$ and $A_{\rm FB}^\ell$ for $\ell=e,\mu,\tau$ are consistent with lepton universality at the $68\%$~CL, the quoted experimental values correspond to the combined results. The value of $A_\ell$ is obtained from a weighted average of the~LEP and~SLD measurements, which are likewise compatible with lepton universality~\cite{ALEPH:2005ab}. Compared to the~EWPO~set introduced in~Section~\ref{sec:NLLEWPO}, the observables $\Gamma_W^{\rm had}$, $\Gamma_W^{\rm lep}$, and $R_s$ are not included in~Table~\ref{tab:EWPOummary}. We retain only the total $W$-boson width $\Gamma_W$, since the partial widths are not statistically independent observables, while $R_s$ is omitted because no sufficiently precise measurement is available from LEP or SLC. Since hadron colliders now provide the most precise determinations of $m_W$, we use the preliminary combination of the latest measurements from ATLAS, CMS, D\O, and LHCb, yielding $m_W=(80.361\pm0.008)\,{\rm GeV}$~\cite{Bozzi:2025xtr}. For the corresponding theoretical uncertainty, we follow~\cite{Awramik:2003rn}, which estimates a $4\,{\rm MeV}$ uncertainty dominated by the parametric dependence on the top-quark mass.

The third column in~Table~\ref{tab:EWPOummary} shows projected relative uncertainties for the FCC-ee Tera-$Z$ stage. These uncertainties have been obtained in~\cite{Greljo:2025ggc}, where they are called the~S1~scenario, by combining in quadrature three sources of error: the uncertainties associated with converting experimental measurements into~EWPO, the aggressive-variant uncertainties on SM theory predictions from missing higher orders, both taken from the EW Physics Preparatory Group~\cite{Freitas:2025ewppg}, and the experimental statistical and systematic uncertainties taken from the FCC-ee Feasibility Study Report~\cite{FCC:2025lpp}. This represents the most conservative and theory-limited of the three FCC-ee uncertainty scenarios considered in~\cite{Greljo:2025ggc}, corresponding to the regime where~EWPO conversion errors dominate. Whether these projected uncertainties are ultimately achievable will depend on substantial and sustained progress in SM~precision calculations over the coming decades.

\subsection{Third-generation four-quark operators} 
\label{sec:4HQfits}

The first class of dimension-six operators that we investigate are the third-generation four-quark operators listed in~\eqref{eq:2loop4top}. As discussed in~Section~\ref{sec:two} and explicitly illustrated in~Section~\ref{sec:semi4top} for the purely right-handed four-top operator $Q_{uu}^{3333}$, these operators contribute to~EWPO at $\LLtwo$, $\LLone$, and $\NLL$ order. Consequently, the constraints derived from~EWPO measurements depend sensitively on the scale~$\Lambda$ that characterizes the onset of BSM~physics and enters the logarithms defined in~\eqref{eq:defL}.

\begin{figure}[t!]
\centering
\includegraphics[width=\linewidth]{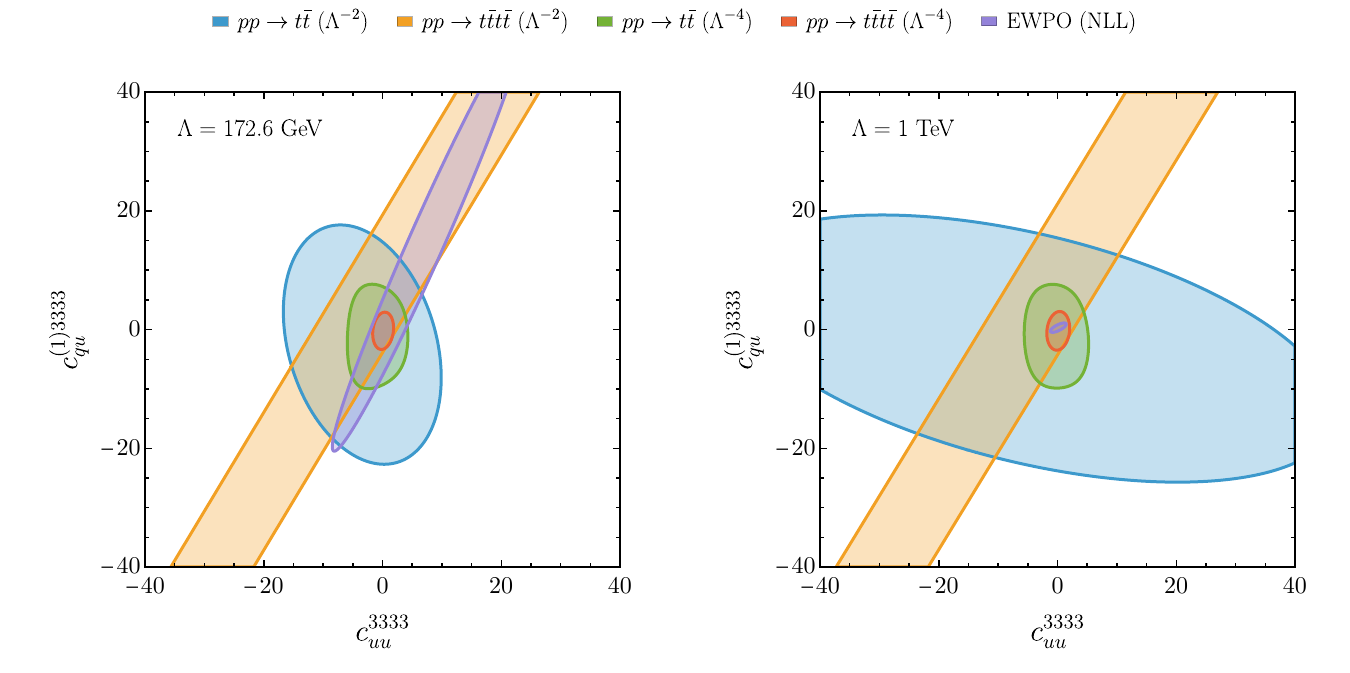}
\vspace{-6mm}
\caption{Impact of RG~evolution and EFT truncation effects on constraints on third-generation four-quark operators derived from~EWPO at NLL accuracy and from $pp \to t \bar t$ and $pp \to t \bar t t \bar t$ measurements at the LHC. The left (right) panel displays the $95\%$~CL constraints in the $c_{uu}^{3333} \hspace{0.25mm}$--$\hspace{0.5mm} c_{qu}^{(1)\hspace{0.25mm}3333}$ plane for $\Lambda = 172.6\,{\rm GeV}$ ($\Lambda = 1\,{\rm TeV}$). Further details are provided in the main text.}
\label{fig:RGEandEFTorderPlot}
\end{figure}

This behavior is illustrated in~Figure~\ref{fig:RGEandEFTorderPlot}. The left (right) panel shows the $95\%$~CL constraints in the $c_{uu}^{3333}\hspace{0.25mm}$--$\hspace{0.5mm}c_{qu}^{(1)\hspace{0.25mm}3333}$ plane for $\Lambda = 172.6\,{\rm GeV}$ ($\Lambda = 1\,{\rm TeV}$). The displayed Wilson coefficients are normalized according to~\eqref{eq:smallc} and evaluated at the high scale~$\Lambda$. The purple contours correspond to the~EWPO fits performed at NLL accuracy using the SM predictions together with the LEP and SLC measurements summarized in~Table~\ref{tab:EWPOummary}, with the full experimental correlation matrix consistently included. A comparison of the~EWPO constraints shows that the inclusion of RG~effects leads to a substantial strengthening of the bounds, reducing the size of the allowed parameter region and modifying the orientation of the confidence ellipse. We further note that the analysis of~\cite{DiNoi:2025uhu} includes certain NNLO fixed-order corrections to~EWPO induced by the third-generation four-quark operators in~\eqref{eq:4topoperators}, effectively evaluated at $\Lambda = 172.6\,{\rm GeV}$, and therefore does not incorporate the logarithmically enhanced RG~effects that are central to the present~work.

While the pronounced scale dependence of the~EWPO is an interesting feature, it is of limited significance on its own\footnote{In the spirit of Oscar Wilde’s dictum that “everything in life is relative.”} unless contrasted with the scale dependence of other observables sensitive to the operators introduced in~\eqref{eq:2loop4top}. For this purpose, we also include in~Figure~\ref{fig:RGEandEFTorderPlot} the $95\%$~CL constraints derived from existing $pp \to t\bar t$ and $pp \to t\bar t t\bar t$ measurements at the LHC. These bounds are based on the SMEFT predictions for top-quark processes presented in~Appendix~\ref{app:top} and the corresponding experimental data collected in~Appendix~\ref{app:LHC}. We refer the interested reader to these appendices for additional details. The~constraints are shown for an EFT truncation at linear ($\Lambda^{-2}$) and quadratic ($\Lambda^{-4}$) order in the Wilson coefficients, corresponding respectively to SM-BSM interference and squared BSM contributions. From the results it is apparent that the top-quark constraints are significantly less sensitive to the scale choice than the~EWPO bounds, with the only notable exception being the linear $pp \to t \bar t$ case, which weakens visibly for $\Lambda = 1\,{\rm TeV}$ compared to $\Lambda = 172.6\,{\rm GeV}$. More generally, the EFT truncation has a stronger impact than the scale variation, as flat directions in the $c_{uu}^{3333}\hspace{0.25mm}$--$\hspace{0.5mm}c_{qu}^{(1)\hspace{0.25mm}3333}$ plane are only lifted at ${\cal O}(\Lambda^{-4})$, a~well-known feature~\cite{Ethier:2021bye,Celada:2024mcf,Hartland:2019bjb,Degrande:2024mbg,DiNoi:2025uhu} that can affect the stability of EFT interpretations. By~contrast, in the NLL~EWPO analysis no quadratic terms in the Wilson coefficients are included, and such truncation ambiguities are therefore absent. This distinction should be kept in mind when comparing the~EWPO and LHC constraints.

\begin{figure}[t!]
\centering
\includegraphics[width=\linewidth]{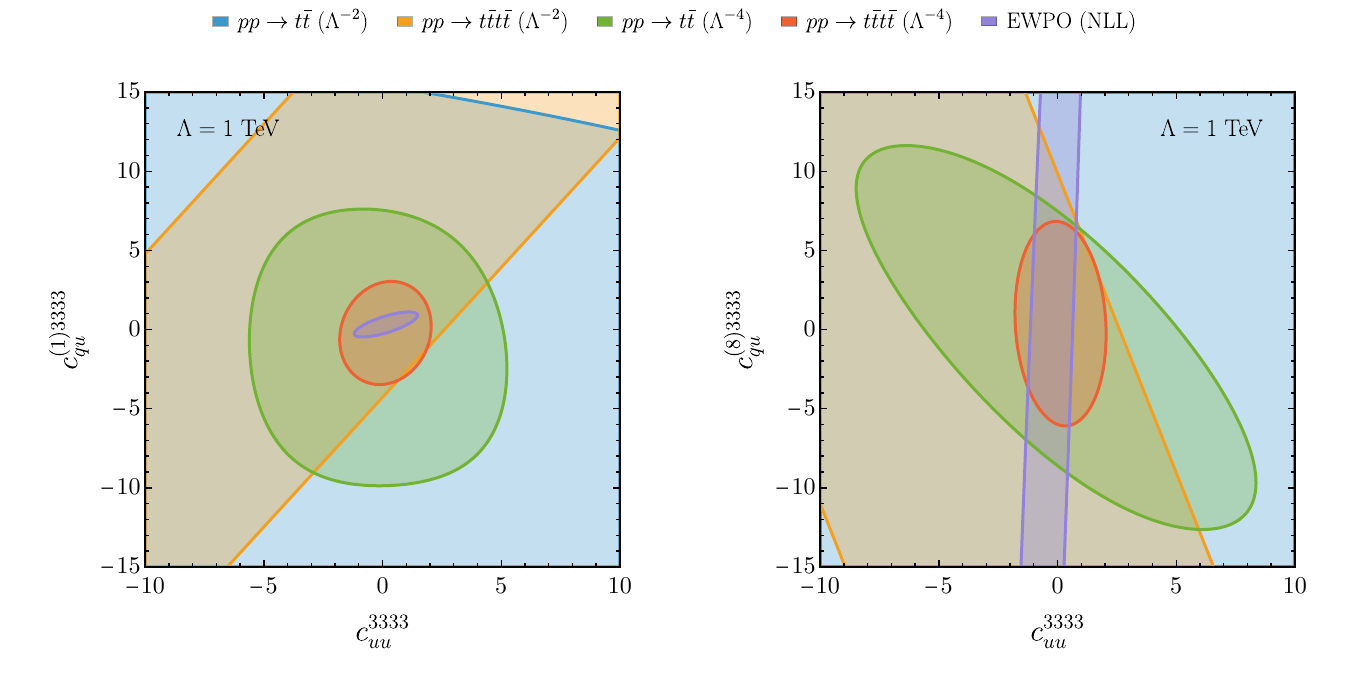}
\vspace{-6mm}
\caption{As in the right panel of~Figure~\ref{fig:RGEandEFTorderPlot}, but showing the various $95\%$~CL~constraints in the $c_{uu}^{3333}\hspace{0.25mm}$--$\hspace{0.5mm}c_{qu}^{(1)\hspace{0.25mm}3333}$ plane (left) and the $c_{uu}^{3333}\hspace{0.25mm}$--$\hspace{0.5mm}c_{qu}^{(8)\hspace{0.25mm}3333}$ plane (right).}
\label{fig:qu_vs_uu}
\end{figure}

Having established that RG~effects significantly impact the extraction of constraints on high-scale Wilson coefficients from~EWPO, while top-quark observables are comparatively less sensitive to such effects, we now present constraints on additional combinations of the dimensionless Wilson coefficients associated with the operators in~\eqref{eq:2loop4top}. Throughout, we adopt $\Lambda = 1\,{\rm TeV}$ as our default BSM~scale. For larger values of $\Lambda$, RG~effects in the NLL~EWPO fits become even more important. Our choice of scale is therefore conservative, since it avoids artificially enhancing the impact of logarithmically enhanced corrections. Our~fit results are shown in~Figures~\ref{fig:qu_vs_uu} and~\ref{fig:qqu1_vs_qq3} with additional plots provided in~Appendix~\ref{app:addplots}. The main message that emerges from the plots is that the LEP and SLC measurements of the~EWPO are already sufficiently precise for the resulting constraints on the Wilson coefficients of the third-generation four-quark operators~\eqref{eq:2loop4top} to be more stringent than those obtained from LHC top-quark measurements including terms up to order ${\cal O} (\Lambda^{-4})$ in the EFT expansion. The only exception is the operator $Q_{qu}^{(8)\hspace{0.25mm}3333}$, which is more strongly constrained by top-quark data than by~EWPO. This behavior can be understood from the fact that $Q_{qu}^{(8)\hspace{0.25mm}3333}$ contributes to $C_{Hq}^{(1)\hspace{0.25mm}33}(m_Z)$ and $C_{Hq}^{(3)\hspace{0.25mm}33}(m_Z)$ only at $\LLtwo$ order mixing via $Q_{qu}^{(1)\hspace{0.25mm}3333}$ with a small anomalous dimension, while the dominant oblique correction first arises at $\NLL$ through $C_{HW\!B}(m_Z)$. The corresponding two-loop beta function governing the mixing of $Q_{qu}^{(8)\hspace{0.25mm}3333}$ into $Q_{HW\!B}$ is, however, proportional to~$g_Y g_L \hspace{0.25mm} \alpha_t$ and is therefore numerically suppressed. As a result, the sensitivity of the~EWPO to the operator $Q_{qu}^{(8)\hspace{0.25mm}3333}$ remains limited even at NLL order, explaining the flat direction observed in the left panel of~Figure~\ref{fig:qu_vs_uu} as well as in several other plots shown in~Appendix~\ref{app:addplots}. Similar statements apply to the color-octet third-generation four-quark operators $Q_{qd}^{(8)\hspace{0.25mm}3333}$ and $Q_{ud}^{(8)\hspace{0.25mm}3333}$, whose mixing into operators relevant for the~$Z$-pole observables is numerically even smaller than in the case of $Q_{qu}^{(8)\hspace{0.25mm}3333}$. As a result, these operators are even more difficult to probe through~EWPO, and we therefore do not present two-dimensional constraints involving their Wilson coefficients.

\begin{figure}[t!]
\centering
\includegraphics[width=\linewidth]{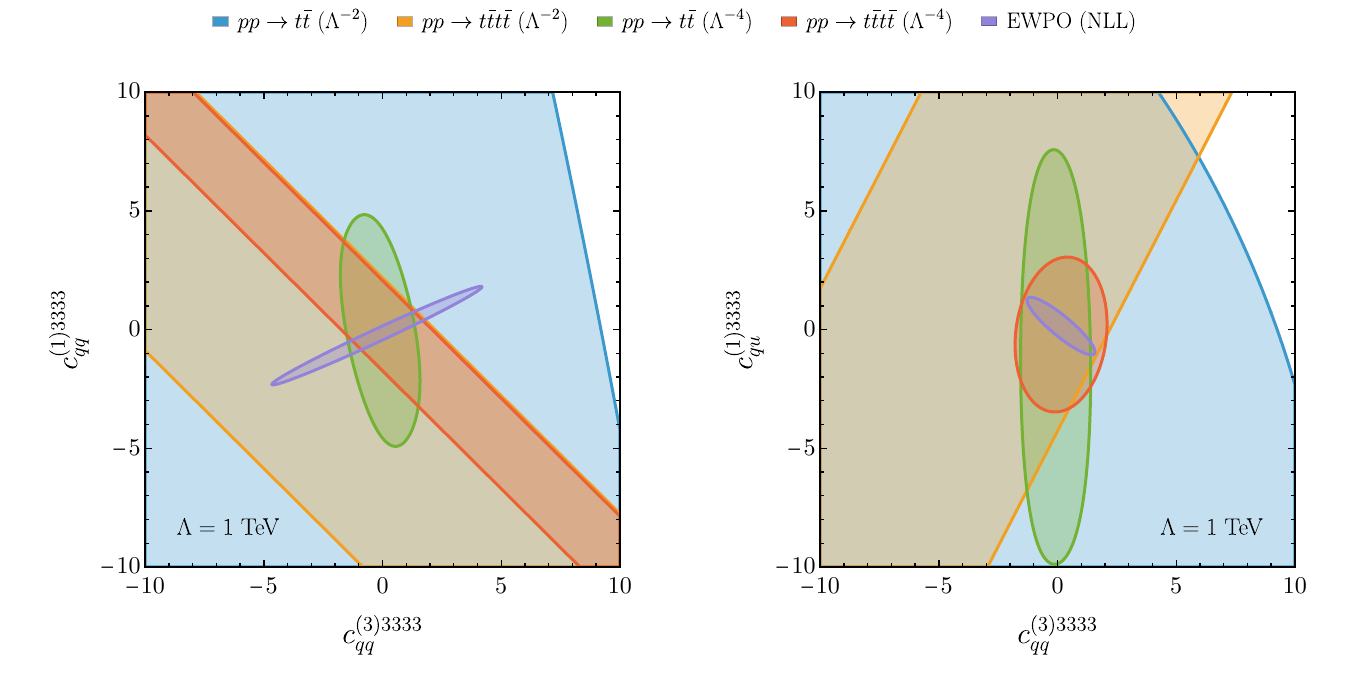}
\vspace{-6mm}
\caption{As in~Figure~\ref{fig:qu_vs_uu}, but showing $95\%$~CL constraints in the $c_{qq}^{(3)\hspace{0.25mm}3333}\hspace{0.25mm}$--$\hspace{0.5mm}c_{qq}^{(1)\hspace{0.25mm}3333}$ plane~(left) and the $c_{qq}^{(3)\hspace{0.25mm}3333}\hspace{0.25mm}$--$\hspace{0.5mm}c_{qu}^{(1)\hspace{0.25mm}3333}$ plane~(right).}
\label{fig:qqu1_vs_qq3}
\end{figure}

\subsection{Bosonic and two-top-quark operators} 
\label{sec:bosonic_2tfits}

The second class of dimension-six operators we consider consists of those highlighted in green in~Tables~\ref{tab:no4ferm}, \ref{tab:4ferm}, and \ref{tab:uXoperators}, which contribute to~EWPO for the first time at NLL order. Unlike the third-generation four-quark operators studied previously, RG~evolution generates neither $\LLtwo$ nor $\LLone$ terms, but only $\NLL$ contributions. The absence of $\LLtwo$ terms implies that current~EWPO measurements are in general not precise enough to yield constraints on the relevant Wilson coefficients competitive with those derived from other collider observables, such as Higgs and top-quark measurements. For this reason, we present constraints based on both current data and future collider projections in all cases. For the~EWPO, we use the projected FCC-ee Tera-$Z$ relative uncertainties summarized in~Table~\ref{tab:EWPOummary} and briefly discussed in~Section~\ref{sec:SMexpEWPO}. Details~on~the high-luminosity LHC~(HL-LHC) projections are given instead~in~Appendix~\ref{app:LHC}.

\begin{figure}[t!]
\centering
\includegraphics[width=\linewidth]{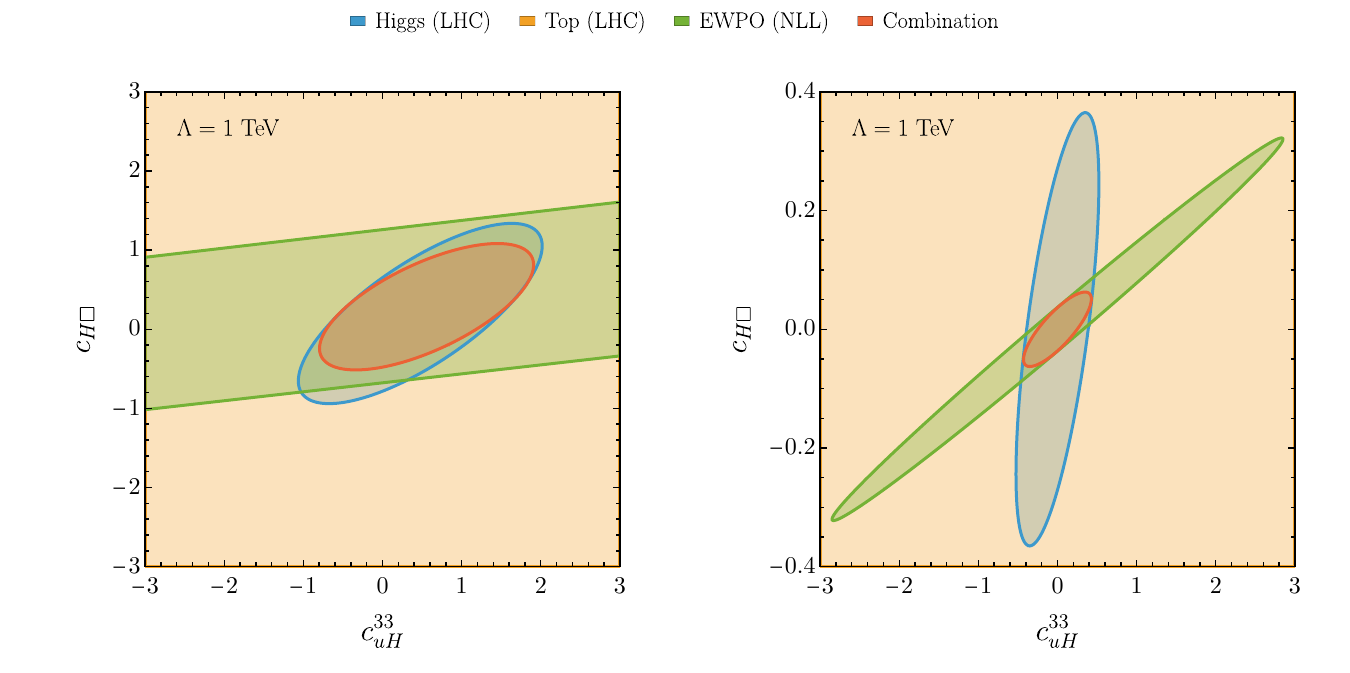}
\vspace{-6mm}
\caption{Comparison of $95\%$~CL constraints in the $c_{uH}^{33}\hspace{0.25mm}$--$\hspace{0.5mm}c_{H\Box}$ plane from current data~(left) and from projected FCC-ee Tera-$Z$ and HL-LHC data~(right), assuming a high scale of $\Lambda=1\,{\rm TeV}$. See main text for details.}
\label{fig:cHbox_vs_cuH33}
\end{figure}

As a first example, we consider the operators $Q_{uH}^{33}$ and $Q_{H\Box}$. The operator $Q_{uH}^{33}$ modifies the relation between the top-quark Yukawa coupling and the top-quark mass, whereas $Q_{H\Box}$ induces a universal rescaling of all on-shell Higgs amplitudes proportional to the number of external Higgs fields. Figure~\ref{fig:cHbox_vs_cuH33} compares the $95\%$~CL constraints in the $c_{uH}^{33}\hspace{0.25mm}$--$\hspace{0.5mm}c_{H\Box}$ plane obtained from current data (left panel) and from projected  FCC-ee Tera-$Z$ and HL-LHC measurements (right panel). In both cases, the high scale is set to $\Lambda = 1\,{\rm TeV}$. The left panel shows that, at present, Higgs measurements provide by far the strongest constraints on the two Wilson coefficients. This is not surprising, since both $Q_{uH}^{33}$ and $Q_{H\Box}$ affect the Higgs observables at the same perturbative order as the SM, whereas their contributions to~EWPO arise only at the one- and two-loop level. As illustrated by the right panel, this situation may change in the future, as the extremely precise~EWPO measurements expected at the FCC-ee can at least partially overcome the loop suppression of the SMEFT effects. As a result,~EWPO will provide a powerful complementary probe that helps disentangle the effects of $Q_{uH}^{33}$ and $Q_{H\Box}$. We add that, as discussed below~(\ref{eq:higgskinetic}) and explicitly shown in~(\ref{eq:twolooptopyukawa}), both $Q_{H\Box}$ and $Q_{uH}^{33}$ induce important contributions to $Q_{HD}$ through RG~evolution. Consequently, the sensitivity of future~EWPO measurements to these operators is dominated by the $W$-boson mass, whose projected FCC-ee Tera-$Z$ precision in the S1 scenario corresponds to an improvement of approximately a factor of $30$ relative to the current experimental uncertainty. The contribution of the remaining~EWPO to the resulting constraints is comparatively small.

\begin{figure}[t!]
\centering
\includegraphics[width=\linewidth]{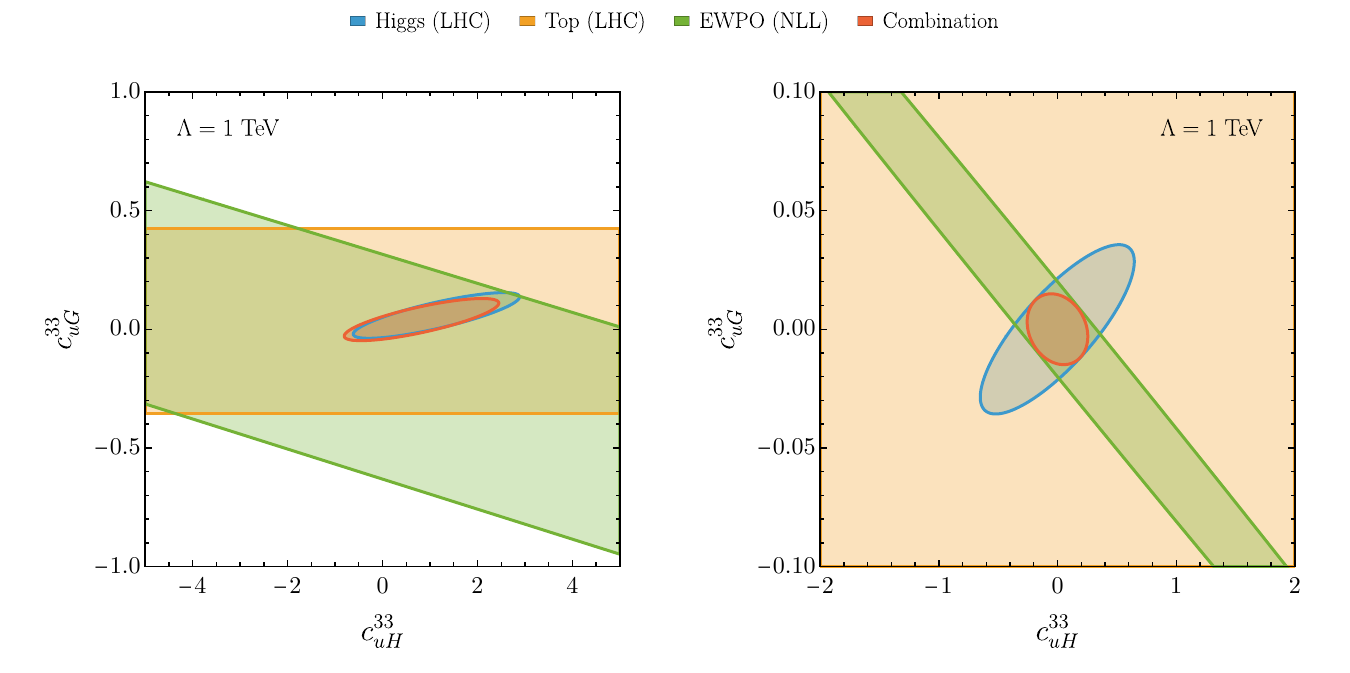}
\vspace{-6mm}
\caption{As~in~Figure~\ref{fig:cHbox_vs_cuH33}, but showing $95\%$~CL constraints in the $c_{uH}^{33}\hspace{0.25mm}$--$\hspace{0.5mm}c_{uG}^{33}$ plane.}
\label{fig:cuG33_vs_cuH33}
\end{figure}

Figure~\ref{fig:cuG33_vs_cuH33} shows, as a second example, the $95\%$~CL constraints in the $c_{uH}^{33}\hspace{0.25mm}$--$\hspace{0.5mm}c_{uG}^{33}$ plane obtained from current data (left panel) and from projected FCC-ee Tera-$Z$ and HL-LHC measurements (right panel), assuming $\Lambda = 1\,{\rm TeV}$. The left panel demonstrates that, at present, Higgs measurements provide the dominant constraints on both Wilson coefficients. In the case of the top-quark chromomagnetic dipole operator $Q_{uG}^{33}$, the sensitivity is driven by its sizable $\LLone$-enhanced one-loop RG~mixing into $Q_{HG}$, which modifies Higgs production via gluon fusion. As illustrated by the right panel,~EWPO are expected to play a much more important role at the FCC-ee. We recall that, as discussed in~Section~\ref{sec:semitopdipole}, the NLL contributions are numerically larger than the corresponding LL effects in this case. This~behavior originates from a large two-loop beta function proportional to $y_t^3$ that drives the mixing of $Q_{uG}^{33}$ into $Q_{HD}$, thereby enhancing the NLL corrections despite their higher-loop origin. Since the $W$-boson mass provides the strongest individual constraint on both $c_{uH}^{33}$ and $c_{uG}^{33}$, and~EWPO predictions depend only linearly on the Wilson coefficients through ${\cal O}(\Lambda^{-2})$ terms, the FCC-ee~EWPO fit exhibits an approximate flat direction in the two-dimensional parameter space. Combining the~EWPO information with the projected HL-LHC Higgs constraints breaks this degeneracy, yielding a single closed elliptical region in the $c_{uH}^{33}\hspace{0.25mm}$--$\hspace{0.5mm}c_{uG}^{33}$~plane. The remaining~EWPO play only a subleading role in the resulting~constraints.

\begin{figure}[t!]
\centering
\includegraphics[width=\linewidth]{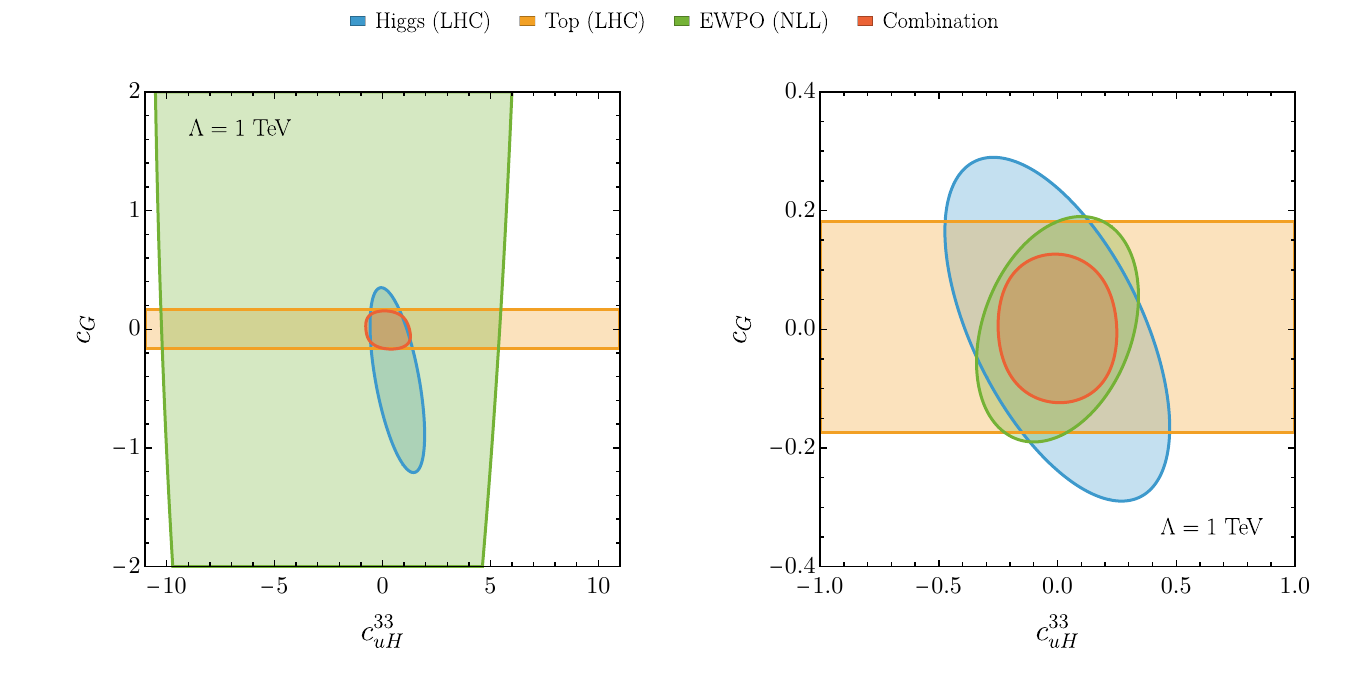}
\vspace{-6mm}
\caption{As~in~Figure~\ref{fig:cHbox_vs_cuH33}, but showing $95\%$~CL constraints in the $c_{uH}^{33}\hspace{0.25mm}$--$\hspace{0.5mm}c_{G}$ plane.}
\label{fig:cG_vs_cuH33}
\end{figure}

\begin{figure}[t!]
\centering
\includegraphics[width=\linewidth]{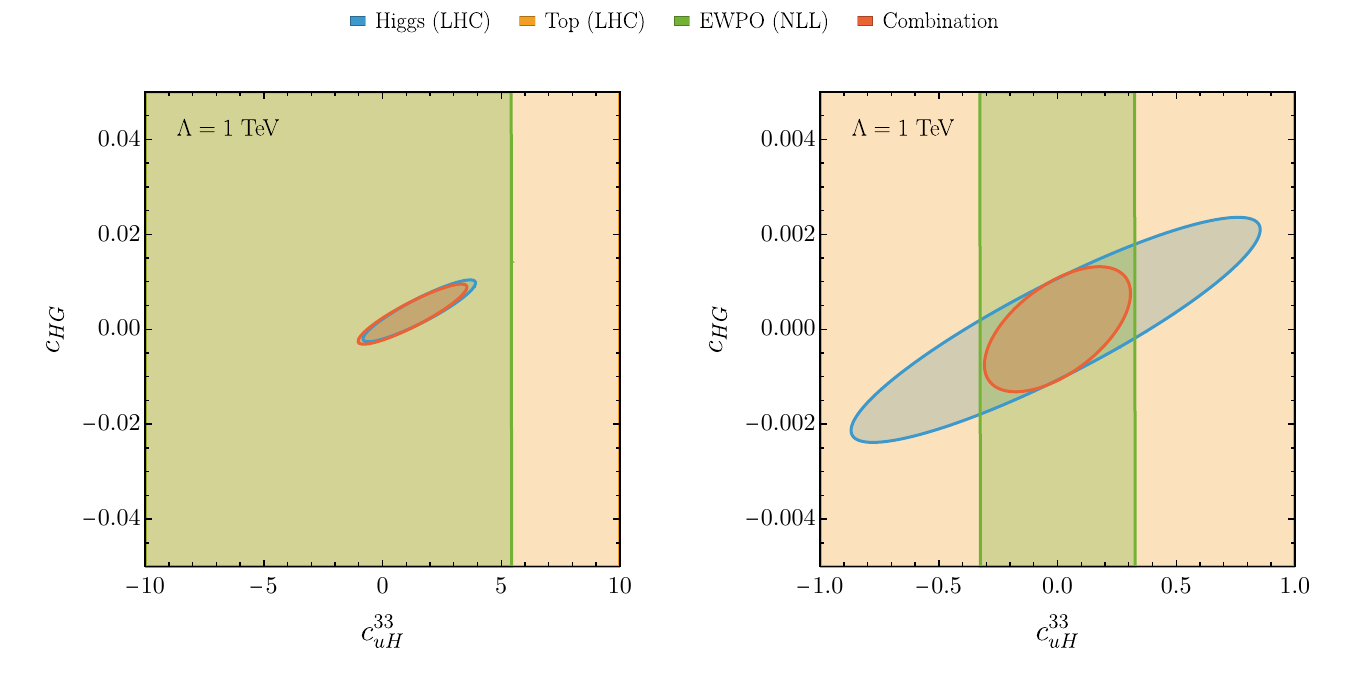}
\vspace{-6mm}
\caption{As~in~Figure~\ref{fig:cHbox_vs_cuH33}, but showing $95\%$~CL constraints in the $c_{uH}^{33}\hspace{0.25mm}$--$\hspace{0.5mm}c_{HG}$ plane.}
\label{fig:cHG_vs_cuH33}
\end{figure}

In all examples discussed so far, the~EWPO fits have been dominated by the constraint from the $W$-boson mass. As explained in~Section~\ref{sec:semiNLL}, this is no longer the case for the operators $Q_G$ and $Q_{HG}$, since they do not induce contributions to $Q_{HD}$ up to NLL order. Meaningful constraints on these operators therefore require sensitivity from other~EWPO. In~Figures~\ref{fig:cG_vs_cuH33} and~\ref{fig:cHG_vs_cuH33}, we present the $95\%$~CL constraints in the $c_{uH}^{33}\hspace{0.25mm}$--$\hspace{0.5mm}c_G$ and $c_{uH}^{33}\hspace{0.25mm}$--$\hspace{0.5mm}c_{HG}$ planes, respectively, comparing current data (left panels) with projected FCC-ee Tera-$Z$ and HL-LHC measurements (right panels) for $\Lambda = 1\,{\rm TeV}$. For $c_G$, the strongest current bounds arise from top-quark observables, where the triple-gluon operator contributes already at tree level. In contrast, $c_{HG}$ is most tightly constrained by Higgs measurements, in particular by $gg \to h$, which also receives a tree-level contribution from $Q_{HG}$. The greatly improved precision of FCC-ee~EWPO measurements can nevertheless alter this picture. As shown in the right panel of~Figure~\ref{fig:cG_vs_cuH33},~EWPO provide complementary sensitivity to $c_G$ which, when combined with future top-quark and Higgs data, leads to significantly stronger constraints in the $c_{uH}^{33}\hspace{0.25mm}$--$\hspace{0.5mm}c_G$ plane. For $c_{HG}$, the impact of~EWPO remains more modest, as this operator is already tightly constrained by existing Higgs data. In both cases, the~EWPO sensitivity is driven primarily by non-oblique corrections affecting $\Gamma_Z$ and $R_\ell$, while the contribution from the remaining observables is comparatively small.

\subsection{Single-operator limits} 
\label{sec:singleOp}

Up to this point, we have focused on constraints in two-dimensional Wilson-coefficient planes. To facilitate comparison with global SMEFT fits and analyses of specific classes of dimension-six operators, we now also present single-operator limits. Table~\ref{tab:single_op_bounds} summarizes the corresponding $95\%$~CL bounds on selected bosonic, two-top-quark, and third-generation four-quark operators derived from~EWPO measurements at LEP and SLC, as well as from top-quark and Higgs data at the LHC. All Wilson coefficients are renormalized at the high scale $\Lambda = 1\,{\rm TeV}$. For each coefficient, we additionally identify the observable that provides the strongest constraint within the respective data set. The limits are quoted in terms of the effective scale $|C_i|^{-1/2}$, choosing the sign of $C_i$ that leads to the weakest bound. For the top-quark analysis, the results obtained when quadratic ${\cal O}(\Lambda^{-4})$ contributions are omitted are shown in parentheses. Wilson coefficients highlighted in orange are most strongly constrained by~EWPO measurements. Details on the LHC data sets used in our analysis can be found in~Appendix~\ref{app:LHC}.

An interesting observation from~Table~\ref{tab:single_op_bounds} is that the existing LEP and SLC measurements of the~EWPO already provide the strongest current constraints on the third-generation four-quark operators $Q_{qq}^{(1)\hspace{0.25mm}3333}$, $Q_{qq}^{(3)\hspace{0.25mm}3333}$, $Q_{qu}^{(1)\hspace{0.25mm}3333}$, and $Q_{uu}^{3333}$. In all cases, these indirect constraints probe effective scales above $1\,{\rm TeV}$. The resulting limits are significantly stronger than those obtained in~\cite{Dawson:2022bxd,DiNoi:2025uhu}, further underscoring the important role of logarithmically enhanced RG effects in our NLL~EWPO analysis. The inclusion of RG~evolution also explains why we are able to derive a constraint on the operator $Q_{qu}^{(8)\hspace{0.25mm}3333}$. In contrast, the fixed-order~EWPO analyses of~\cite{Dawson:2022bxd,DiNoi:2025uhu}, performed directly at the EW scale, are insensitive to this operator because its effects enter only through RG-induced mixing. Table~\ref{tab:single_op_bounds} also shows that, for the bosonic operators $Q_{HG}$, $Q_G$, and $Q_{H\Box}$ as well as the two-top-quark operators $Q_{uG}^{33}$ and $Q_{uH}^{33}$, the most stringent current constraints arise from top-quark and, in particular, Higgs measurements at the LHC. This is expected, since these operators contribute directly to the corresponding collider observables, whereas their effects on~EWPO enter only through loop-induced corrections. As a result, the precision achieved by the LEP and SLC measurements is not yet sufficient for~EWPO to provide the dominant constraints in these cases.

\begin{table}[t!]
\centering
\scriptsize
\renewcommand{\arraystretch}{1.2}
\setlength{\tabcolsep}{4pt}
\resizebox{\textwidth}{!}{%
\begin{tabular}{|l|c c|l c|l c|c|}
\hline
& LEP and SLC & $\Lambda_{\rm bound}$ [TeV]
& Top (LHC) & $\Lambda_{\rm bound}$ [TeV]
& Higgs (LHC) & $\Lambda_{\rm bound}$ [TeV]
& $\Lambda_{\rm best}$ [TeV] \\
\hline
$C_{HG}$ & $R_\ell$ & 0.17
& $d\sigma_{t\bar t}/d\cos\theta_\ast$ (CMS) & 0.37 (0.38)
& $\mu_{\mathrm{ggF}}^{ZZ}$ (ATLAS) & 21.0
& 21.0 \\

$C_{uG}^{33}$ & $m_W$ & 1.35
& $d\sigma_{t\bar t}/d\cos\theta_\ast$ (CMS) & 1.72 (1.76)
& $\mu_{\mathrm{ggF}}^{ZZ}$ (ATLAS) & 4.36
& 4.36 \\

$C_G$ & $R_\ell$ & 0.24
& $d\sigma_{t\bar t}/dm_{t\bar t}$ (CMS) & 2.60 (0.71)
& $\mu_{\mathrm{ggF}}^{ZZ}$ (ATLAS) & 1.21
& 2.60 \\

\cellcolor{orange!20}$C_{qq}^{(1)\hspace{0.25mm}3333}$ & $R_\ell$ & 1.66
& $\sigma_{t\bar t t\bar t}$ (ATLAS; CMS) & 0.72 (0.32)
& $\mu_{\mathrm{VBF}}^{WW}$ (ATLAS) & 0.15
& 1.66 \\

\cellcolor{orange!20}$C_{qu}^{(1)\hspace{0.25mm}3333}$ & $m_W$ & 1.26
& $\sigma_{t\bar t t\bar t}$ (ATLAS; CMS) & 0.55 (0.27)
& $\mu_{\mathrm{ggF}}^{WW}$ (ATLAS) & 0.31
& 1.26 \\

$C_{H\Box}$ & $m_W$ & 0.96
& \hspace{1cm} --- & ---
& $\mu_{\mathrm{ggF}}^{\gamma\gamma}$ (ATLAS) & 1.22
& 1.22 \\

\cellcolor{orange!20}$C_{qq}^{(3)\hspace{0.25mm}3333}$ & $R_\ell$ & 1.14
& $\sigma_{t\bar t t\bar t}$ (ATLAS) & 0.85 (0.34)
& $\mu_{WH}^{WW}$ (ATLAS) & 0.08
& 1.14 \\

\cellcolor{orange!20}$C_{uu}^{3333}$ & $m_W$ & 1.09
& $\sigma_{t\bar t t\bar t}$ (CMS) & 0.72 (0.30)
& $\mu_{\mathrm{VBF}}^{WW}$ (ATLAS) & 0.16
& 1.09 \\

$C_{uH}^{33}$ & $m_W$ & 0.34
& \hspace{1cm} --- & ---
& $\mu_{\mathrm{ggF}}^{WW}$ (ATLAS) & 0.95
& 0.95 \\

$C_{qu}^{(8)\hspace{0.25mm}3333}$ & $A_\ell$ & 0.18
& $\sigma_{t\bar t t\bar t}$ (ATLAS) & 0.42 (0.16)
& $\mu_{\mathrm{ggF}}^{WW}$ (ATLAS) & 0.29
& 0.42 \\
\hline
\end{tabular}%
}
\vspace{2mm}
\caption{Single-operator $95\%$~CL bounds from current~EWPO, top-quark, and Higgs~data. All Wilson coefficients are renormalized at $\Lambda=1\,{\rm TeV}$. For each coefficient, the most constraining observable in a given data set is indicated. Limits are quoted as effective scales $|C_i|^{-1/2}$, choosing the sign of $C_i$ that gives the weakest bound. For the top-quark data set, results obtained without quadratic ${\cal O}(\Lambda^{-4})$ terms are shown in parentheses. Entries denoted by ``---'' indicate that no constraint can be derived at the perturbative order considered. Coefficients highlighted in orange are most strongly constrained by~EWPO. Details of the notation and definitions of the observables used in our analysis are provided in Appendix~\ref{app:LHC}.}
\label{tab:single_op_bounds}
\end{table}

Table~\ref{tab:single_op_bounds_future} summarizes the projected $95\%$~CL bounds on selected bosonic, two-top-quark, and third-generation four-quark operators from FCC-ee~EWPO and HL-LHC top-quark and Higgs measurements. The FCC-ee Tera-$Z$ projections assume the S1 scenario defined in~Section~\ref{sec:SMexpEWPO}, while the HL-LHC projections follow the methodology described in~Appendix~\ref{app:LHC}. All other notation is as in~Table~\ref{tab:single_op_bounds}, except that the Wilson coefficients are renormalized at the high scale $\Lambda=5\,\mathrm{TeV}$. The projected sensitivities demonstrate the exceptional discovery potential of future precision measurements. In particular, FCC-ee~EWPO are expected to provide the strongest constraints on all third-generation four-quark operators, probing effective scales of up to about $20\,\mathrm{TeV}$. They also surpass the projected HL-LHC sensitivity to the bosonic operators $Q_G$ and $Q_{H\Box}$ as well as the two-top-quark operators $Q_{uG}^{33}$ and $Q_{uH}^{33}$. This improvement is driven by the unprecedented precision anticipated for FCC-ee~EWPO, which, when combined with the logarithmically enhanced RG effects included in our NLL~analysis, substantially extends the indirect sensitivity to high-scale new physics. The notable exception is $Q_{HG}$, for which the projected HL-LHC Higgs measurements remain by far the most sensitive probe due to its direct contribution to the effective Higgs-gluon coupling.

\begin{table}[t!]
\centering
\scriptsize
\renewcommand{\arraystretch}{1.2}
\setlength{\tabcolsep}{4pt}
\resizebox{\textwidth}{!}{%
\begin{tabular}{|l|c c|l c|l c|c|}
\hline
&
FCC-ee
& $\Lambda_{\rm bound}$ [TeV]
& Top (HL-LHC)
& $\Lambda_{\rm bound}$ [TeV]
& Higgs (HL-LHC)
& $\Lambda_{\rm bound}$ [TeV]
& $\Lambda_{\rm best}$ [TeV]
\\
\hline

$C_{HG}$
& $R_\ell$
& 2.37
& $d\sigma_{t\bar t}/dm_{t\bar t}$ (CMS)
& 0.65 (0.67)
& $\mu_{\mathrm{ggF}}^{\gamma\gamma}$ (ATLAS)
& 36.2
& 36.2
\\

\cellcolor{orange!25}$C_{qq}^{(1)\hspace{0.25mm}3333}$
& $\Gamma_Z$
& 20.9
& $\sigma_{t\bar t t\bar t}$ (CMS)
& 0.95 (0.65)
& $\mu_{\mathrm{VBF}}^{WW}$ (ATLAS)
& 0.48
& 20.9
\\

\cellcolor{orange!25}$C_{qu}^{(1)\hspace{0.25mm}3333}$
& $\Gamma_Z$
& 15.8
& $\sigma_{t\bar t t\bar t}$ (CMS)
& 0.75 (0.48)
& $\mu_{\mathrm{ggF}}^{WW}$ (CMS)
& 0.66
& 15.8
\\

\cellcolor{orange!25}$C_{qq}^{(3)\hspace{0.25mm}3333}$
& $R_\ell$
& 14.6
& $\sigma_{t\bar t t\bar t}$ (CMS)
& 0.99 (0.65)
& $\mu_{\mathrm{ggF}}^{ZZ}$ (ATLAS)
& 0.30
& 14.6
\\

\cellcolor{orange!25}$C_{uu}^{3333}$
& $m_W$
& 13.1
& $\sigma_{t\bar t t\bar t}$ (CMS)
& 0.95 (0.59)
& $\mu_{\mathrm{VBF}}^{WW}$ (ATLAS)
& 0.50
& 13.1
\\

\cellcolor{orange!25}$C_{uG}^{33}$
& $m_W$
& 10.8
& $d\sigma_{t\bar t}/dm_{t\bar t}$ (CMS)
& 2.22 (2.27)
& $\mu_{\mathrm{ggF}}^{\gamma\gamma}$ (ATLAS)
& 10.1
& 10.8
\\

\cellcolor{orange!25}$C_{H\Box}$
& $m_W$
& 6.25
& \hspace{1cm} --- & ---
& $\mu_{\mathrm{ggF}}^{\gamma\gamma}$ (ATLAS)
& 2.05
& 6.25
\\

\cellcolor{orange!25}$C_G$
& $R_\ell$
& 3.38
& $d\sigma_{t\bar t}/dm_{t\bar t}$ (CMS)
& 2.30 (0.81)
& $\mu_{\mathrm{ggF}}^{\gamma\gamma}$ (ATLAS)
& 3.19
& 3.38
\\

\cellcolor{orange!25}$C_{uH}^{33}$
& $m_W$
& 2.53
& \hspace{1cm} --- & ---
& $\mu_{\mathrm{ggF}}^{ZZ}$ (ATLAS)
& 1.73
& 2.53
\\

\cellcolor{orange!25}$C_{qu}^{(8)3333}$
& $m_W$
& 1.41
& $\sigma_{t\bar t t\bar t}$ (CMS)
& 0.56 (0.34)
& $\mu_{\mathrm{ggF}}^{ZZ}$ (ATLAS)
& 0.69
& 1.41
\\

\hline
\end{tabular}%
}
\vspace{2mm}
\caption{Same as Table~\ref{tab:single_op_bounds}, but showing the projected future constraints from~EWPO, top-quark, and Higgs measurements. All Wilson coefficients are renormalized at $\Lambda=5\,\mathrm{TeV}$. The FCC-ee Tera-$Z$ projections assume the S1 scenario defined in~Section~\ref{sec:SMexpEWPO}, while the methodology used for the HL-LHC projections is described in~Appendix~\ref{app:LHC}. Further details are provided in the main text.}
\label{tab:single_op_bounds_future}
\end{table}

\section{BSM model examples} 
\label{sec:BSM}

Having established model-independent SMEFT constraints in the previous section, we now turn to illustrative examples within concrete BSM~scenarios. In particular, we consider models that generate sizable tree-level contributions to the third-generation four-quark operators introduced in~\eqref{eq:4topoperators}. As already briefly mentioned in~Section~\ref{sec:two}, two representative realizations of this kind are RS and CH models. 
\subsection{Custodial~RS model} 
\label{sec:CRS}

As a first example of a framework capable of generating sizable tree-level contributions to four-quark operators involving third-generation fields, we consider the custodially protected version of the RS model proposed in~\cite{Agashe:2003zs}. Compared to the original RS model~\cite{Randall:1999ee}, the extended bulk gauge symmetry ensures custodial protection of the $W$- and $Z$-boson mass ratio --- equivalently, the Peskin-Takeuchi $T$ parameter --- as well as of the $Z$-boson coupling to left-handed bottom quarks~\cite{Agashe:2006at}, which we refer to in the following as the $Z \bar b_L b_L$ coupling or vertex. We derive the tree-level matching conditions for all dimension-six SMEFT operators relevant in the context of~EWPO, building on the results of~\cite{Agashe:2006wa,Casagrande:2008hr,Agashe:2003zs,Blanke:2008zb,Bauer:2009cf,Casagrande:2010si}. Although our presentation is self-contained, it relies on several analytical results and intermediate steps established in these works, to which we refer the reader for further technical details.

In order to explain the hierarchy between the EW scale $m_W$ and the Planck scale $m_{\mathrm{Pl}}$, the volume of the warped extra dimension in the RS model must satisfy
\beq \label{eq:volume}
{\mathcal{V}} \simeq \ln \left( \frac{m_{\mathrm{Pl}}}{m_W} \right) \simeq \ln \left( 10^{16} \right) \simeq 37 \,,
\eeq
so that the corresponding warp factor $\exp(-{\mathcal{V}})$ naturally generates the exponential suppression of the EW scale relative to the Planck scale. This warp factor also determines the mass scale of the low-lying Kaluza-Klein~(KK) excitations. In particular, the mass of the first KK gluon is given by
\beq \label{eq:mG}
m_G \simeq j_{0,1} \hspace{0.25mm} m_{\mathrm{KK}} \,, \qquad j_{0,1} \simeq 2.40 \,.
\eeq
Here, $j_{0,1}$ denotes the first zero of the Bessel function $J_0(x)$, and $m_{\mathrm{KK}}$ represents the KK~scale, typically of the order of a few TeV.

The first tree-level matching conditions we consider involve $Q_{HD}$ and $Q_{HW\!B}$, which respectively modify the Peskin-Takeuchi parameters $T$ and $S$. We find the corresponding expressions for the relevant Wilson coefficients 
\beq \label{eq:CRSmatchingoblique}
C_{HD} (m_G) \simeq \frac{g_Y^2}{8 \hspace{0.125mm} m_{\mathrm{KK}}^2} \hspace{0.25mm} \frac{1}{\mathcal{V}} \simeq \frac{2.7 \cdot 10^{-3}}{m_G^2} \,, \qquad C_{HW\!B} (m_G) \simeq \frac{g_Y g_L}{8 \hspace{0.25mm} m_{\mathrm{KK}}^2} \left ( 1 - \frac{1}{\mathcal{V}} \right ) \simeq \frac{0.17}{m_G^2} \,,
\eeq
where we have used~\eqref{eq:numerics2}, \eqref{eq:volume}, and~\eqref{eq:mG} to obtain the final numerical results. From~\eqref{eq:CRSmatchingoblique}, it is evident that $C_{HD}(m_G)$ is volume suppressed, whereas $C_{HW\!B}(m_G)$ is not. This demonstrates the custodial protection of the $T$ parameter in the custodial~RS model, in contrast to the original RS setup, where it would scale as ${\mathcal{V}}$. By contrast, the result for $C_{HW\!B}(m_G)$, or equivalently for $S$, is the same in both scenarios.

The tree-level matching conditions for operators involving quarks depend, in addition to the volume factor ${\mathcal{V}}$ and the KK-gluon mass $m_G$, on the entries of the five-dimensional~(5D) Yukawa matrices ${\mathcal{Y}}_q$ and the bulk-mass parameters ${\mathcal{c}}_q$. For the third-generation quarks, however, the warped-space Froggatt-Nielsen mechanism~\cite{Casagrande:2008hr,Blanke:2008zb} implies that the bulk-mass parameters ${\mathcal{c}}_{q_3}$ and ${\mathcal{c}}_{d_3}$ of the left-handed third-generation quark doublet and the right-handed bottom quark, respectively, can be expressed in terms of the bulk-mass parameter~${\mathcal{c}}_{u_3}$ of the right-handed top quark, the relevant entries of the 5D Yukawa matrices~${\mathcal{Y}}_u$ and ${\mathcal{Y}}_d$, and the observed top- and bottom-quark Yukawa couplings:
\beq \label{eq:RSFN}
y_t \simeq \big | {\mathcal{Y}}_u^{33} \big | \hspace{0.5mm} {\mathcal{F}}({\mathcal{c}}_{u_3}) \hspace{0.25mm} {\mathcal{F}}({\mathcal{c}}_{q_3}) \,, \qquad 
y_b \simeq \big | {\mathcal{Y}}_d^{33} \big | \hspace{0.5mm} {\mathcal{F}}({\mathcal{c}}_{d_3}) \hspace{0.25mm} {\mathcal{F}}({\mathcal{c}}_{q_3}) \,.
\eeq
Here, ${\mathcal{F}}({\mathcal{c}})$ denotes the fermionic zero-mode profile~\cite{Grossman:1999ra,Gherghetta:2000qt}, which for ${\mathcal{c}} > -1/2$, corresponding to fermions localized towards the infrared (IR) brane, is given by ${\mathcal{F}}({\mathcal{c}}) \simeq \sqrt{1 + 2\hspace{0.25mm}{\mathcal{c}}}$.

Using the relations~(\ref{eq:RSFN}) to eliminate ${\mathcal{c}}_{q_3}$ and ${\mathcal{c}}_{d_3}$, we obtain the following expressions for the Wilson coefficients that modify the couplings of the $Z$ boson to third-generation quarks:
\begin{align}
C_{Hq}^{(1) \hspace{0.25mm} 33} (m_G) & \simeq \frac{g_L^2 \hspace{0.125mm} y_t^2}{16 \hspace{0.25mm} m_{\mathrm{KK}}^2} \hspace{0.25mm} \frac{1}{ \big | {\mathcal{Y}}_u^{33} \big |^2 \left ( 1 + 2 \hspace{0.25mm} {\mathcal{c}}_{u_3} \right )} \left ( {\mathcal{V}} + \frac{s_w^2}{3 \hspace{0.125mm} c_w^2} \right ) \nonumber \\[1mm]
& \phantom{xx} + \frac{y_t^2}{m_{\mathrm{KK}}^2} \frac{1}{2 \left ( 3 - 4 \hspace{0.25mm} {\mathcal{c}}_{u_3} \left ( 1 + {\mathcal{c}}_{u_3} \right ) \right )} \simeq -\frac{0.25}{m_G^2} \,, \label{eq:CRSmatchingZcouplings1} \\[1mm]
C_{Hq}^{(3) \hspace{0.25mm} 33} (m_G) & \simeq -\frac{g_L^2 \hspace{0.125mm} y_t^2}{16 \hspace{0.25mm} m_{\mathrm{KK}}^2} \hspace{0.25mm} \frac{1}{ \big | {\mathcal{Y}}_u^{33} \big |^2 \left ( 1 + 2 \hspace{0.25mm} {\mathcal{c}}_{u_3} \right )} \left ( {\mathcal{V}} - 1 \right ) \nonumber \\[1mm]
& \phantom{xx} - \frac{y_t^2}{m_{\mathrm{KK}}^2} \frac{1}{2 \left ( 3 - 4 \hspace{0.25mm} {\mathcal{c}}_{u_3} \left ( 1 + {\mathcal{c}}_{u_3} \right ) \right )} \simeq \frac{0.26}{m_G^2} \,, \label{eq:CRSmatchingZcouplings2} \\[1mm]
C_{Hu}^{33} (m_G) & \simeq \frac{g_L^2 \hspace{0.125mm} s_w^2}{12 \hspace{0.125mm} c_w^2 \hspace{0.25mm} m_{\mathrm{KK}}^2} \hspace{0.25mm} \frac{\left ( 1 + 2 \hspace{0.25mm} {\mathcal{c}}_{u_3} \right ) \left ( 5 + 2 \hspace{0.25mm} {\mathcal{c}}_{u_3} \right )}{\left ( 3 + 2 \hspace{0.25mm} {\mathcal{c}}_{u_3} \right )^2} \nonumber \\[1mm]
& \phantom{xx} -\frac{y_t^2}{4 \hspace{0.25mm} m_{\mathrm{KK}}^2} \left ( 1 - \frac{y_t^2}{2 \hspace{0.25mm} \big | {\mathcal{Y}}_u^{33} \big |^2 \left ( 1 + 2 \hspace{0.25mm} {\mathcal{c}}_{u_3} \right )} \right ) \simeq -\frac{0.93}{m_G^2} \,, \label{eq:CRSmatchingZcouplings3} \\[1mm]
C_{Hd}^{33} (m_G) & \simeq \frac{g_L^2 \hspace{0.125mm} y_b^2}{8 \hspace{0.125mm}  y_t^2 \hspace{0.25mm} m_{\mathrm{KK}}^2} \hspace{0.25mm} \frac{\big | {\mathcal{Y}}_u^{33} \big |^2 \left ( 1 + 2 \hspace{0.25mm} {\mathcal{c}}_{u_3} \right )}{ \big | {\mathcal{Y}}_d^{33} \big |^2 } \left (  {\mathcal{V}} - \frac{s_w^2}{3 c_w^2} \right ) \nonumber \\[1mm]
& \phantom{xx} + \frac{y_b^2}{4 \hspace{0.125mm} y_t^2 \hspace{0.25mm} m_{\mathrm{KK}}^2} \hspace{0.25mm} \sum_{r=1,2,3} \frac{\big | {\mathcal{Y}}_d^{r3} \big |^2}{\big | {\mathcal{Y}}_d^{33} \big |^2} \, \big | {\mathcal{Y}}_u^{33} \big |^2 \left ( 1 + 2 \hspace{0.25mm} {\mathcal{c}}_{u_3} \right ) \simeq \frac{0.11}{m_G^2} \,, \label{eq:CRSmatchingZcouplings4}
\end{align}
Here, we have employed~\eqref{eq:volume} and~\eqref{eq:mG}, together with ${\mathcal{c}}_{u_3} \simeq 1$, $ \left | {\mathcal{Y}}_u^{33} \right | \simeq 3$, $\left | {\mathcal{Y}}_d^{r3} \right | \simeq 1$, $g_Y \simeq 0.37$, $g_L \simeq 0.65$, $y_t \simeq 0.83$, and $y_b \simeq 0.014$, to obtain the final numerical results. Note~that ${\mathcal{c}}_{u_3} > 1$ signifies that the bulk mass exceeds the curvature scale, implying that the right-handed top quark is effectively a brane-localized rather than a bulk field. Accordingly, our choice of ${\mathcal{c}}_{u_3}$ corresponds to a fully composite bulk field. The 5D~Yukawa couplings are taken under the flavor-anarchy assumption, while the SM Yukawa couplings are evaluated in the $\overline{\rm MS}$ scheme at a scale of roughly~$1 \, \mathrm{TeV}$. From the expressions~\eqref{eq:CRSmatchingZcouplings1} to \eqref{eq:CRSmatchingZcouplings4}, it is clear that all Wilson coefficients except $C_{Hu}^{33}(m_G)$ receive volume-enhanced contributions from the gauge-boson terms in the first lines. The~$Z \bar b_L b_L$~coupling, however, is proportional to the sum $C_{Hq}^{(1) \hspace{0.25mm} 33}(m_G) + C_{Hq}^{(3) \hspace{0.25mm} 33}(m_G)$, in which both the volume-enhanced gauge-boson and fermionic contributions cancel. This~reflects the custodial protection of the $Z \bar b_L b_L$ vertex in the custodial~RS~model. The same symmetry also shields the $Z \bar t_R t_R$ vertex, proportional to~$C_{Hu}^{33}(m_G)$, from corrections scaling with~${\mathcal{V}}$. In contrast, the~$Z \bar b_R b_R$~coupling, proportional to~$C_{Hd}^{33}(m_G)$, and the $Z \bar t_L t_L$~vertex, proportional to $C_{Hq}^{(1) \hspace{0.25mm} 33}(m_G) - C_{Hq}^{(3) \hspace{0.25mm} 33}(m_G)$, are not custodially protected and thus receive volume-enhanced contributions. Nevertheless, the~$Z \bar b_R b_R$~vertex is strongly suppressed by the bottom-quark Yukawa coupling due to the warped-space Froggatt-Nielsen mechanism.

We finally turn to the tree-level matching contributions to the complete set of third-generation four-quark operators listed in~\eqref{eq:4topoperators}. Restricting the discussion to the numerically dominant effects from KK-gluon exchange and retaining only the volume-enhanced terms, we obtain the following non-vanishing Wilson coefficients:
\begin{align} 
C_{qq}^{(1)\hspace{0.25mm}3333} (m_G) & \simeq -\frac{g_s^2 \hspace{0.125mm} y_t^4}{48 \hspace{0.25mm} m_{\mathrm{KK}}^2} \frac{{\mathcal{V}}}{ \big | {\mathcal{Y}}_u^{33} \big |^4 \left ( 1 + 2 \hspace{0.25mm} {\mathcal{c}}_{u_3} \right )^2} \simeq -\frac{2.5 \cdot 10^{-3}}{m_G^2} \,,\label{eq:CRSmatching4top1} \\[1mm]
\hspace{0.25mm} C_{qq}^{(3)\hspace{0.25mm}3333} (m_G) & \simeq -\frac{g_s^2 \hspace{0.125mm} y_t^4}{16 \hspace{0.25mm} m_{\mathrm{KK}}^2} \frac{{\mathcal{V}}}{ \big | {\mathcal{Y}}_u^{33} \big |^4 \left ( 1 + 2 \hspace{0.25mm} {\mathcal{c}}_{u_3} \right )^2} \simeq -\frac{7.6 \cdot 10^{-3}}{m_G^2} \,, \label{eq:CRSmatching4top2} \\[1mm]
C_{qu}^{(8) \hspace{0.25mm} 3333} (m_G) & \simeq -\frac{g_s^2 \hspace{0.125mm} y_t^2}{2 \hspace{0.25mm} m_{\mathrm{KK}}^2} \frac{\left ( 5 + 2 \hspace{0.25mm} {\mathcal{c}}_{u_3} \right )}{\big | {\mathcal{Y}}_u^{33} \big |^2 \left ( 3 + 2 \hspace{0.25mm} {\mathcal{c}}_{u_3} \right )^2} \, {\mathcal{V}} \simeq -\frac{2.0}{m_G^2} \,, \label{eq:CRSmatching4top3} \\[1mm]
C_{qd}^{(8) \hspace{0.25mm} 3333} (m_G) & \simeq -\frac{g_s^2 \hspace{0.125mm} y_b^2}{2 \hspace{0.25mm} m_{\mathrm{KK}}^2} \frac{{\mathcal{V}}}{\big | {\mathcal{Y}}_d^{33} \big |^2} \simeq -\frac{1.8 \cdot 10^{-2}}{m_G^2} \,, \label{eq:CRSmatching4top4} \\[1mm]
C_{ud}^{(8) \hspace{0.25mm} 3333} (m_G) & \simeq -\frac{g_s^2 \hspace{0.125mm} y_b^2}{4 \hspace{0.125mm} y_t^2 \hspace{0.25mm} m_{\mathrm{KK}}^2} \frac{\big | {\mathcal{Y}}_u^{33} \big |^2 \left ( 1 + 2 \hspace{0.25mm} {\mathcal{c}}_{u_3} \right )^2 \left ( 11 + 6 \hspace{0.125mm} {\mathcal{c}}_{u_3} \right )}{\big | {\mathcal{Y}}_d^{33} \big |^2 \left ( 3 + 2 \hspace{0.25mm} {\mathcal{c}}_{u_3} \right )^2} \, {\mathcal{V}} \simeq -\frac{0.72}{m_G^2} \,, \label{eq:CRSmatching4top5} \\[1mm]
C_{uu}^{3333} (m_G) & \simeq -\frac{g_s^2}{6 \hspace{0.25mm} m_{\mathrm{KK}}^2} \frac{\left ( 1 + 2 \hspace{0.25mm} {\mathcal{c}}_{u_3} \right )^2 \left ( 5 + 2 \hspace{0.25mm} {\mathcal{c}}_{u_3} \right )}{\left ( 3 + 2 \hspace{0.25mm} {\mathcal{c}}_{u_3} \right )^3} \, {\mathcal{V}} \simeq -\frac{16}{m_G^2} \,, \label{eq:CRSmatching4top6} \\[1mm]
C_{dd}^{3333} (m_G) & \simeq -\frac{g_s^2 \hspace{0.125mm} y_b^4}{12 \hspace{0.125mm} y_t^4 \hspace{0.25mm} m_{\mathrm{KK}}^2} \frac{ \big | {\mathcal{Y}}_u^{33} \big |^4 \left ( 1 + 2 \hspace{0.25mm} {\mathcal{c}}_{u_3} \right )^2}{\big | {\mathcal{Y}}_d^{33} \big |^4} \, {\mathcal{V}}  \simeq -\frac{8.9 \cdot 10^{-4}}{m_G^2} \,. \label{eq:CRSmatching4top7}
\end{align}
Since the KK gluon transforms as a color-octet state, one furthermore has $C_{\psi_A \psi_B}^{(1)\hspace{0.25mm}3333}(m_G)=0$ for $\psi_A \psi_B = qu, qd, ud$. In obtaining the numerical values quoted above, we have employed the input parameters given below~\eqref{eq:CRSmatchingZcouplings4}, together with $g_s \simeq 0.94$, corresponding to the strong coupling evaluated at a scale of approximately $1 \, {\rm TeV}$. One observes that the Wilson coefficients involving right-handed bottom-quark fields are suppressed by a factor of $y_b$ for each such field. This suppression is a direct consequence of the warped-space Froggatt–Nielsen mechanism and renders these contributions phenomenologically negligible. By contrast, the remaining Wilson coefficients exhibit a pronounced hierarchy, with $C_{uu}^{3333}(m_G) > C_{qu}^{(8) \hspace{0.25mm} 3333}(m_G) \gg C_{qq}^{(1)\hspace{0.25mm}3333}(m_G)$ for the custodial~RS benchmark considered here. More quantitatively, the hierarchy among these Wilson coefficients is approximately characterized by
\beq \label{eq:4topcompositeness}
\frac{C_{uu}^{3333} (m_G)}{C_{qu}^{(8) \hspace{0.25mm} 3333} (m_G)} \simeq \frac{2}{3 \hspace{0.125mm} y_t^2} \hspace{0.5mm} \big | {\mathcal{Y}}_u^{33} \big |^2 \hspace{0.5mm} {\mathcal{c}}_{u_3} \,, \qquad
\frac{C_{qu}^{(8) \hspace{0.25mm} 3333} (m_G)}{C_{qq}^{(1)\hspace{0.25mm}3333} (m_G)} \simeq \frac{48}{y_t^2} \hspace{0.5mm} \big | {\mathcal{Y}}_u^{33} \big |^2 \hspace{0.5mm} {\mathcal{c}}_{u_3} \,,
\eeq
showing that the relative enhancement of $C_{uu}^{3333}(m_G)$ with respect to $C_{qu}^{(8)\hspace{0.25mm}3333}(m_G)$ grows as the right-handed top quark becomes more composite, corresponding to increasing values of the bulk-mass parameter ${\mathcal{c}}_{u_3} > 0$. In particular, for ${\mathcal{c}}_{u_3} \gtrsim 1/\left | {\mathcal{Y}}_u^{33} \right |^2$, the operator~$Q_{uu}^{3333}$~receives the dominant matching contribution. This feature provides an a posteriori justification for our emphasis in~Sections~\ref{sec:two} and~\ref{sec:semi4top} on the purely right-handed four-top operator.

In addition, RS models predict potentially sizable deviations in Higgs observables from the SM~\cite{Casagrande:2010si,Azatov:2010pf,Goertz:2011hj,Carena:2012fk,Frank:2013un,Malm:2013jia,Hahn:2013nza,Malm:2014gha,Archer:2014jca}. The tree-level modifications of the Higgs couplings to EW~gauge bosons and fermions are largely model independent, since they arise from universal KK-mixing effects and are therefore largely insensitive to the details of the Higgs sector. By~contrast, the loop-induced Higgs couplings to gluon and photon pairs depend on the full~KK~spectrum and are sensitive to whether the Higgs field is localized on the IR brane or propagates in the bulk, as well as to the structure of the 5D Yukawa sector and the embedding of the fermion fields. To maintain the model independence of the numerical analysis presented in Section~\ref{sec:BSMnumerics}, we therefore restrict our analysis to the universal tree-level matching corrections and neglect the model-dependent loop-induced contributions. This leads to the following Wilson coefficients:
\begin{align} 
C_{H\Box} (m_G) & \simeq -\frac{g_L^2}{2 \hspace{0.25mm} m_{\mathrm{KK}}^2} \hspace{0.25mm} {\mathcal{V}} \simeq -\frac{45}{m_G^2} \,, \label{eq:RSCHbox} \\[1mm]
C_{uH}^{33} (m_G) & \simeq \frac{2}{3 \hspace{0.25mm} m_{\mathrm{KK}}^2} \frac{({\mathcal{Y}}_u \hspace{0.125mm} {\mathcal{Y}}_u^\dagger \hspace{0.25mm} {\mathcal{Y}}_u)_{33}}{{\mathcal{Y}}_u^{33}} \simeq \frac{87}{m_G^2} \,, \label{eq:RSCuH} \\[1mm]
C_{dH}^{33} (m_G) & \simeq \frac{2}{3 \hspace{0.25mm} m_{\mathrm{KK}}^2} \frac{({\mathcal{Y}}_d {\mathcal{Y}}_d^\dagger {\mathcal{Y}}_d)_{33}}{{\mathcal{Y}}_d^{33}} \simeq \frac{10}{m_G^2} \label{eq:RSCdH} \,.
\end{align}
Here, the numerical estimate on the right-hand side of~\eqref{eq:RSCHbox} has been obtained using the relations in~\eqref{eq:numerics2}, \eqref{eq:volume}, and~\eqref{eq:mG}. To determine the size of the Yukawa structures entering~\eqref{eq:RSCuH} and~\eqref{eq:RSCdH}, we follow the standard anarchic RS assumption. Namely, the entries of the 5D Yukawa matrices are taken to be random complex numbers with magnitudes bounded by $|{\mathcal{Y}}_q^{\hspace{0.25mm}r\hspace{-0.125mm}s}| \leq Y_q^\ast$ where $q = u, d$ and $r,s=1,2,3$. For sufficiently large ensembles of random Yukawa matrices generated in this way, the average value of the relevant Yukawa combination is found to obey~\cite{Malm:2013jia,Hahn:2013nza,Malm:2014gha}
\beq \label{eq:yukav}
\left \langle \frac{({\mathcal{Y}}_q {\mathcal{Y}}_q^\dagger {\mathcal{Y}}_q)_{33}}{{\mathcal{Y}}_q^{33}} \right \rangle = \frac{5}{2} \left(Y_q^\ast\right)^2 \,,
\eeq
where imposing the observed quark masses and flavor-mixing angles has only a minor effect on the average. The numerical estimates in~\eqref{eq:RSCuH} and~\eqref{eq:RSCdH} correspond to the choices $Y_u^\ast \simeq 3$ and $Y_d^\ast \simeq 1$, respectively. Notably, the correction in~\eqref{eq:RSCHbox} is strictly negative, whereas those in~\eqref{eq:RSCuH} and~\eqref{eq:RSCdH} are strictly positive. As a result, the Higgs couplings to EW gauge bosons are universally suppressed relative to their SM values, while the Higgs couplings to top and bottom quarks are enhanced. The suppression of the Higgs couplings to EW gauge bosons is a rather generic consequence of Higgs-sector modifications as has been emphasized, for example, in~\cite{Low:2009di}.

\subsection{Custodial CH model} 
\label{sec:MCH}

CH models~\cite{Agashe:2004rs,Agashe:2005dk,Contino:2006nn,Panico:2015jxa} provide a purely four-dimensional (4D) realization of the dynamics underlying RS constructions. In these theories, the Higgs boson emerges as a pseudo-Nambu-Goldstone boson (PNGB) of a strongly interacting sector with a spontaneously broken global symmetry, most commonly based on the minimal coset $SO(5)/SO(4)$. The masses of the SM fermions are generated through the mechanism of partial compositeness~\cite{Kaplan:1991dc}, in which the elementary fermions mix linearly with composite operators. Through the AdS/CFT correspondence~\cite{Maldacena:1997re,Gubser:1998bc,Witten:1998qj}, the heavy spin-1 and spin-$1/2$ resonances of the composite sector are identified with the KK excitations of the RS framework: bulk gauge symmetries are mapped onto global symmetries of the strong sector, while the localization of the Higgs field towards the IR brane reflects its composite nature. In custodially protected realizations, the enlarged symmetry suppresses corrections to the Peskin-Takeuchi~$T$~parameter~\cite{Agashe:2003zs} as well as to the $Z\bar b_Lb_L$ coupling~\cite{Agashe:2006at}. At the same time, reproducing the large top-quark Yukawa coupling requires a sizable degree of compositeness for the third generation, leading to enhanced interactions with the heavy resonances and, after integrating out these states, to sizable tree-level four-quark operators.

We assume the composite sector is characterized by a single resonance mass scale $m_\ast$ and a typical strong coupling $g_\ast$, related through
\beq \label{eq:mgf}
m_\ast = g_\ast f \, ,
\eeq
where $f$ denotes the PNGB decay constant associated with the spontaneous breaking of the global symmetry and therefore sets the characteristic symmetry-breaking scale of the strong sector.

The degree of compositeness of the top quark is parameterized by the mixing parameters~$\epsilon_{q_3}$ and~$\epsilon_{u_3}$, which describe the linear mixings of the left-handed top-quark $SU(2)_L$ doublet and the right-handed top-quark singlet with operators of the composite sector. Within~the paradigm of partial compositeness, these mixings generate the top-quark Yukawa coupling
\beq \label{eq:partialcomposite}
y_t \simeq g_\ast \hspace{0.25mm} \epsilon_{q_3} \hspace{0.125mm} \epsilon_{u_3} \,.
\eeq
Consequently, the third-generation sector of a broad class of CH models can be parameterized in terms of the three quantities $m_\ast$, $g_\ast$, and $\epsilon_{u_3}$. By contrast, the much smaller bottom-quark Yukawa coupling implies a correspondingly weaker mixing of the right-handed bottom quark with the composite sector. Operators involving the field $d_3$ are therefore parametrically suppressed and, unlike in the custodial~RS matching calculation of~Section~\ref{sec:CRS}, will be neglected in the following.

The dominant contributions to the Wilson coefficients of the third-generation four-quark operators in~\eqref{eq:4topoperators} arise from integrating out the heavy color-octet vector resonance~$\rho_\mu^A$.\footnote{In our numerical analysis in~Section~\ref{sec:BSMnumerics}, we include the contributions from the tree-level exchange of all $SU(3)_C \times SO(4) \times U(1)_X$ vector resonances to the third-generation four-quark operators.} Its interactions with the third-generation quark currents are described by
\beq \label{eq:Lrho}
{\cal L}_{\rho} \supset \sqrt{r_s}\, g_\ast
\sum_{\psi=q_3,u_3,d_3}
\epsilon_\psi^2 \,
\bar{\psi}\gamma^\mu T^A \psi \,
\rho_\mu^A \,,
\eeq
where $T^A$ denote the generators of $SU(3)_C$ and $r_s=g_s^2/g_L^2$ is the ratio of the QCD and $SU(2)_L$ gauge couplings squared. The factor $\sqrt{r_s}$ originates from our choice to express all vector-resonance interactions in terms of a universal strong-sector coupling $g_\ast$. This coupling is identified with the physical interaction strength of the vector resonances associated with the custodial symmetry group $SO(4)\supset SU(2)_L\times SU(2)_R$.

In the normalization of the dimension-six SMEFT Lagrangian introduced in~\eqref{eq:L6}, the non-vanishing tree-level Wilson coefficients of the third-generation four-quark operators are
\begin{align} 
C_{qq}^{(1)\hspace{0.25mm}3333} (m_\ast) & \simeq -\frac{r_s \hspace{0.125mm} y_t^4}{24 \hspace{0.25mm} g_\ast^2 \hspace{0.25mm} \epsilon_{u_3}^4 m_{\ast}^2} \simeq -\frac{8.9 \cdot 10^{-4}}{m_\ast^2} \,, \label{eq:CCHmatching4top1} \\[1mm]
C_{qq}^{(3)\hspace{0.25mm}3333} (m_\ast) & \simeq -\frac{r_s \hspace{0.125mm} y_t^4}{8 \hspace{0.25mm} g_\ast^2 \hspace{0.25mm} \epsilon_{u_3}^4 m_{\ast}^2} \simeq -\frac{2.7 \cdot 10^{-3}}{m_\ast^2} \,, \label{eq:CCHmatching4top2} \\[1mm]
C_{qu}^{(8) \hspace{0.25mm} 3333} (m_\ast) & \simeq -\frac{r_s \hspace{0.125mm}y_t^2}{m_{\ast}^2} \simeq -\frac{1.43}{m_\ast^2} \,, \label{eq:CCHmatching4top3} \\[1mm]
C_{uu}^{3333} (m_\ast) & \simeq -\frac{r_s \hspace{0.125mm} g_\ast^2 \hspace{0.25mm} \epsilon_{u_3}^4}{6 \hspace{0.25mm} m_{\ast}^2} \simeq -\frac{16}{m_\ast^2} \,. \label{eq:CCHmatching4top4} 
\end{align}

The numerical values quoted above are obtained by fixing the strong-sector coupling~$g_\ast$ through matching the CH and RS predictions for the Wilson coefficient of the operator~$Q_{uu}^{3333}$, i.e.~by equating~\eqref{eq:CRSmatching4top6} and~\eqref{eq:CCHmatching4top4}. Using $g_s\simeq0.94$, $g_L\simeq0.65$, ${\cal c}_{u_3}=\epsilon_{u_3}=1$, and identifying the characteristic resonance masses, $m_\ast=m_G$, we obtain
\beq \label{eq:gstarRS}
{\cal V} \simeq 37 \qquad \Longrightarrow \qquad g_\ast \simeq 6.8 \,.
\eeq
This result implies that an RS model addressing the full hierarchy between the EW and Planck scales is dual to a CH theory with an almost maximally strongly coupled composite sector. By contrast
\beq \label{eq:gstarlittleRS}
g_\ast \simeq 3 \qquad \Longrightarrow \qquad {\cal V} \simeq 7.3 \,,
\eeq
which is characteristic of a little RS scenario~\cite{Davoudiasl:2008hx,Bauer:2008xb}. In this case, the warped volume is significantly reduced, implying a lower ultraviolet (UV) cutoff of order ${\rm PeV}$, corresponding to approximately $400\,{\rm TeV}$ for the benchmark considered above. More generally, the matching relation $g_\ast \propto g_s\sqrt{\cal V}$ follows directly from the holographic relation between the 5D gauge coupling and the coupling of the dual 4D composite sector.

Although the numerical values in~\eqref{eq:CCHmatching4top1} to \eqref{eq:CCHmatching4top4} do not exactly coincide with the corresponding RS predictions in~\eqref{eq:CRSmatching4top1} to \eqref{eq:CRSmatching4top7}, the differences are readily understood. They~arise from the distinct parameterizations of the two frameworks. While the CH~framework is characterized by a single strong-sector coupling $g_\ast$, the RS setup contains two independent 5D Yukawa matrices, ${\cal Y}_u$ and ${\cal Y}_d$, which determine the quark mass hierarchies. As a result, the matching coefficients agree only at the parametric level. Nevertheless, both frameworks predict the same characteristic hierarchy among the dominant Wilson coefficients. In particular, one~finds
\beq \label{eq:CHhierarchy}
\frac{C_{uu}^{3333} (m_\ast)}{C_{qu}^{(8) \hspace{0.25mm} 3333} (m_\ast)} \simeq \frac{g_\ast^2 \hspace{0.25mm} \epsilon_{u_3}^4}{6 \hspace{0.25mm} y_t^2} \,, \qquad
\frac{C_{qu}^{(8) \hspace{0.25mm} 3333} (m_\ast)}{C_{qq}^{(1)\hspace{0.25mm}3333} (m_\ast)} \simeq \frac{24 \hspace{0.125mm} g_\ast^2 \hspace{0.25mm} \epsilon_{u_3}^4}{y_t^2} \,.
\eeq
The first relation demonstrates that the relative importance of $Q_{uu}^{3333}$ grows with both the strong-sector coupling and the degree of compositeness of the right-handed top quark. In~particular, for $g_\ast \hspace{0.125mm}\epsilon_{u_3}^2 \gtrsim 2$, the Wilson coefficient $C_{uu}^{3333}(m_\ast)$ exceeds $C_{qu}^{(8)\hspace{0.25mm}3333}(m_\ast)$, implying that $Q_{uu}^{3333}$ provides the dominant tree-level matching contribution. CH models that address either the hierarchy or little hierarchy problem therefore generically predict the ordering $C_{uu}^{3333}(m_\ast) > C_{qu}^{(8)\hspace{0.25mm}3333}(m_\ast) \gg C_{qq}^{(1)\hspace{0.25mm}3333}(m_\ast)$. This hierarchy further motivates the emphasis placed on the purely right-handed four-top operator $Q_{uu}^{3333}$ in~Sections~\ref{sec:two} and~\ref{sec:semi4top}. Over a broad and phenomenologically relevant region of parameter space, it constitutes the dominant low-energy imprint of top-quark compositeness in third-generation four-quark interactions.

In addition to the third-generation four-quark operators given in~\eqref{eq:CCHmatching4top1} to~\eqref{eq:CCHmatching4top4}, the tree-level exchange of spin-$1$ resonances associated with the strong sector generates further dimension-six operators in CH models. Among these, the operators that modify the couplings of the $Z$ boson to third-generation quarks receive the following contributions:
\beq \label{eq:ZcouplingsCH}
C_{Hq}^{(1) \hspace{0.25mm} 33} (m_\ast) \simeq \frac{y_t^2}{4 \hspace{0.25mm} \epsilon_{u_3}^2 m_{\ast}^2} \simeq \frac{0.17}{m_\ast^2} \,, \qquad C_{Hq}^{(3) \hspace{0.25mm} 33} (m_\ast) \simeq -\frac{y_t^2}{4 \hspace{0.25mm} \epsilon_{u_3}^2 m_{\ast}^2} \simeq -\frac{0.17}{m_\ast^2} \,.
\eeq
Here, we have used $y_t \simeq 0.83$ and $\epsilon_{u_3}=1$. The relation $C_{Hq}^{(1)\hspace{0.25mm}33}(m_\ast)=-C_{Hq}^{(3)\hspace{0.25mm}33}(m_\ast)$ is a consequence of the custodial symmetry of the composite sector and ensures the cancellation of the tree-level corrections to the $Z\bar b_Lb_L$ coupling arising from the exchange of $SU(2)_L$ and $SU(2)_R$ vector resonances. The same custodial protection mechanism also protects the $Z\bar t_Rt_R$ coupling, leading to $C_{Hu}^{33}(m_\ast)=0$ at tree level. In contrast, the $Zt_L\bar t_L$ coupling, which depends on the combination $C_{Hq}^{(1)\hspace{0.25mm}33}(m_\ast)-C_{Hq}^{(3)\hspace{0.25mm}33}(m_\ast)$, is not protected by custodial symmetry and therefore receives non-zero tree-level corrections. The $Z\bar b_Rb_R$ coupling, governed by $C_{Hd}^{33}(m_\ast)$, is likewise not protected in general. However, its contribution is strongly suppressed by the small bottom-quark Yukawa coupling resulting from partial compositeness and can therefore be neglected in practice.

To derive the tree-level matching conditions for the operators $Q_{HD}$ and $Q_{HW\!B}$, or equivalently for the Peskin-Takeuchi parameters $T$ and $S$, it is necessary to go beyond the fermionic sector and include the non-linear dynamics of the PNGB Higgs fields arising from the spontaneous breaking of the global symmetry. These interactions are systematically described within the Callan-Coleman-Wess-Zumino~(CCWZ) formalism~\cite{Coleman:1969sm,Callan:1969sn}. The spin-$1$ resonances that generate the leading tree-level contributions can be efficiently incorporated using the hidden local symmetry approach~\cite{Bando:1984ej,Bando:1987br}, which is equivalent to the two-site formulation of CH models and provides the 4D dual description of the KK gauge sector in RS models. The resulting matching conditions are
\beq \label{eq:STCH}
C_{HD} (m_\ast) = 0 \,, \qquad C_{HW\!B} (m_\ast) \simeq \frac{g_Y g_L}{2 m_\ast^2} \simeq \frac{0.12}{m_\ast^2} \,,
\eeq
where the numerical estimate uses $g_Y\simeq0.37$ and $g_L\simeq0.65$. The absence of a tree-level contribution to $C_{HD}$ follows from the custodial $SO(4)$ symmetry of the strong sector, which protects the $T$ parameter. In contrast, tree-level exchange of the heavy vector resonances generates the custodially symmetric operator $Q_{HW\!B}$, leading to a non-zero contribution to the $S$ parameter.

Three additional dimension-six operators are generated at tree level exclusively through the CCWZ construction and therefore represent universal predictions of CH models, independent of the details of the resonance spectrum. The corresponding matching conditions~are
\beq \label{eq:CHCHboxCuH}
C_{H\Box} (m_\ast) \simeq -\frac{g_\ast^2}{2 \hspace{0.25mm} m_\ast^2} \simeq -\frac{23}{m_\ast^2} \,, \qquad C_{uH}^{33} (m_\ast) \simeq \frac{g_\ast^2 \hspace{0.25mm} y_t}{m_\ast^2} \simeq \frac{38}{m_\ast^2} \,,
\eeq
while $C_H(m_\ast)=0$ at tree level. The absence of a tree-level contribution to this Wilson coefficient is a direct consequence of the PNGB nature of the Higgs: the Higgs potential is not generated at tree level and arises only radiatively from interactions that explicitly break the global symmetry. In obtaining the numerical estimates above, we have used the relation in~(\ref{eq:gstarRS}). Notice that the matching conditions in~\eqref{eq:CHCHboxCuH} exhibit the same pattern of suppressed and enhanced Higgs couplings as those in~\eqref{eq:RSCHbox} to~\eqref{eq:RSCdH}. This~provides further evidence of the duality between CH and RS models, extending also to the dimension-six~SMEFT operators $Q_{H\Box}$ and $Q_{qH}^{33}$.

We emphasize that the tree-level matching conditions derived in this section capture the universal contributions arising from the CCWZ realization of the PNGB Higgs dynamics and the exchange of spin-$1$ resonances in CH models. They are therefore largely insensitive to the details of the composite spectrum and the specific implementation of partial compositeness, making them applicable to a broad class of CH scenarios. In particular, they reproduce the universal contribution in the minimal CH model with composite fermions embedded in the~$\bm{5}$~representation of $SO(5)$~aka the MCHM$_5$~\cite{Agashe:2004rs}. Additional contributions from the fermion sector are model dependent and can be incorporated separately. In order to maintain the model-independent character of the numerical analysis presented below, we nevertheless neglect such effects.

\subsection{Numerical analysis} 
\label{sec:BSMnumerics}

With the tree-level SMEFT matching conditions for the custodial RS and CH models established, we now investigate the resulting constraints on their parameter spaces from Higgs measurements, top-quark observables, and~EWPO. Following the recent article~\cite{Stefanek:2024kds}, we~consider three benchmark scenarios corresponding to right-handed, left-handed, and mixed top-quark compositeness. In the first two cases, one top-quark chirality is assumed to be fully composite, while the other is taken to be as elementary as possible subject to reproducing the observed top-quark Yukawa coupling. In the RS model, these benchmarks are obtained by solving the first relation in~\eqref{eq:RSFN} for ${\mathcal{c}}_{q_3}$ with ${\mathcal{c}}_{u_3}=1$ and ${\mathcal{Y}}_u^{33}=3$ in the right-handed compositeness scenario~(RHCS), and vice versa for the left-handed compositeness scenario~(LHCS). In the CH model, they follow from solving~\eqref{eq:partialcomposite} for $\epsilon_{q_3}$ with~$\epsilon_{u_3}=1$ for each value of $g_\ast$ in the RHCS, with $\epsilon_{q_3}$ and $\epsilon_{u_3}$ interchanged in the LHCS. The~mixed compositeness scenario (MCS) is defined by ${\mathcal{c}}_{q_3}={\mathcal{c}}_{u_3}$ in the RS model and~$\epsilon_{q_3}=\epsilon_{u_3}$ in the~CH~model.

\begin{figure}[t!]
\centering
\begin{subfigure}[t]{0.315\textwidth}
\centering
\includegraphics[width=\linewidth]{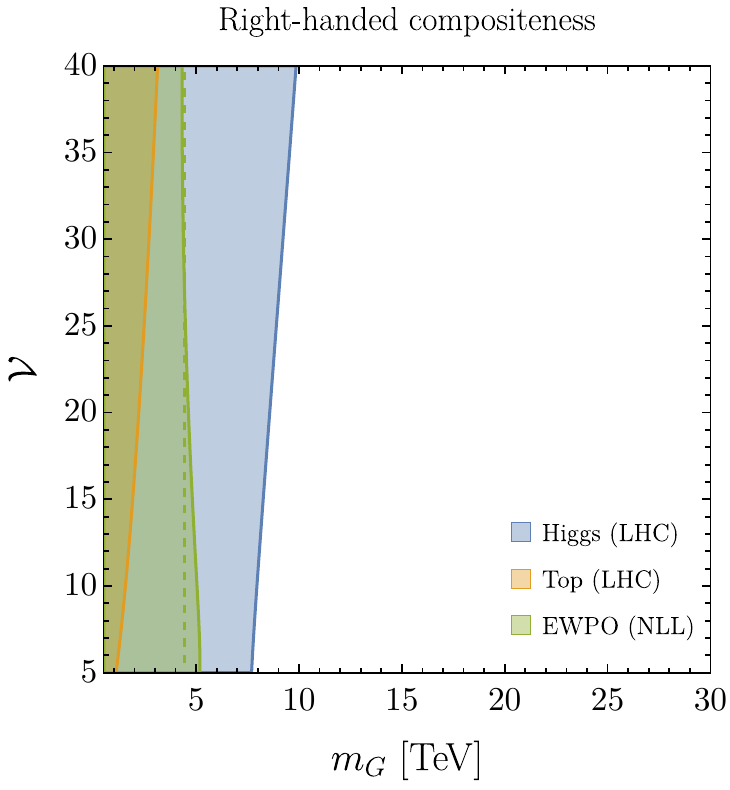}
\end{subfigure}
\hfill
\begin{subfigure}[t]{0.315\textwidth}
\centering
\includegraphics[width=\linewidth]{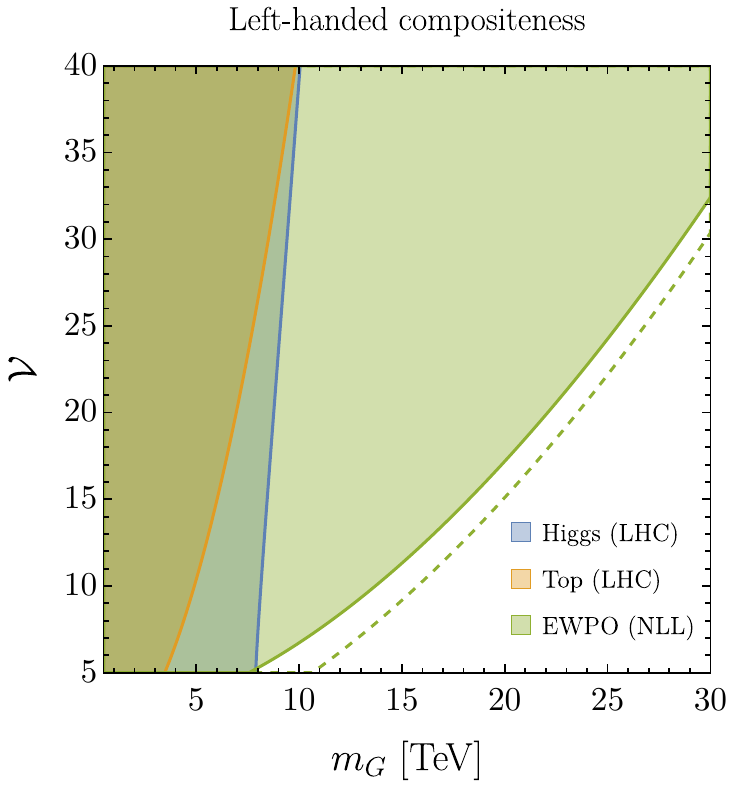}
\end{subfigure}
\hfill
\begin{subfigure}[t]{0.315\textwidth}
\centering
\includegraphics[width=\linewidth]{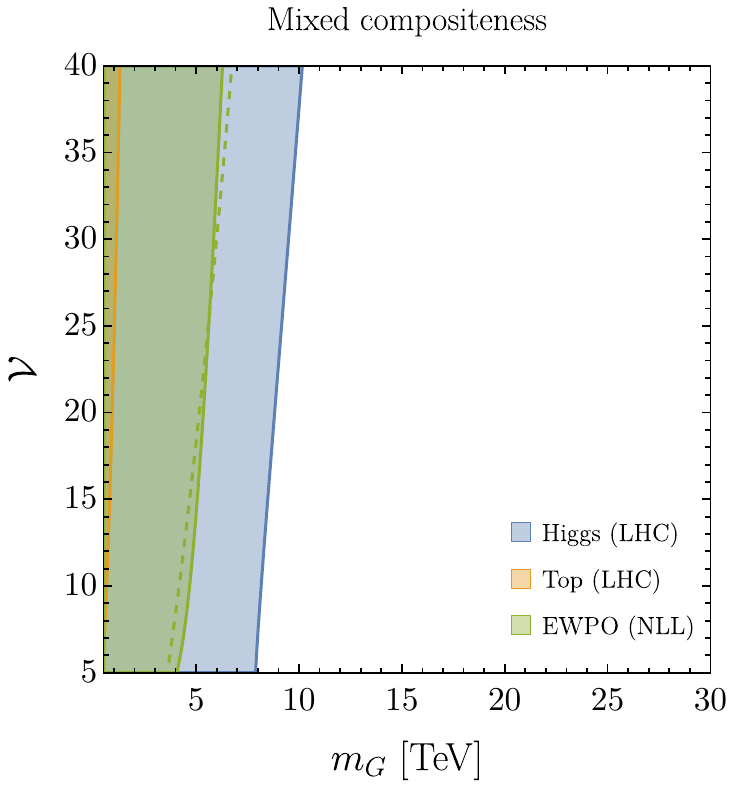}
\end{subfigure}
\vspace{0mm}
\caption{Constraints on the custodial RS model from LHC Higgs measurements, top-quark observables, and an NLL analysis of~EWPO based on LEP and SLC data. The left, middle, and right panels show the resulting $95\%$~CL constraints for benchmark scenarios with a fully composite right-handed top quark, a fully composite left-handed top quark, and comparable degrees of left- and right-handed compositeness, respectively. Further details are provided in the main text.}
\label{fig:RSplots}
\end{figure}

\begin{figure}[t!]
\centering
\begin{subfigure}[t]{0.315\textwidth}
\centering
\includegraphics[width=\linewidth]{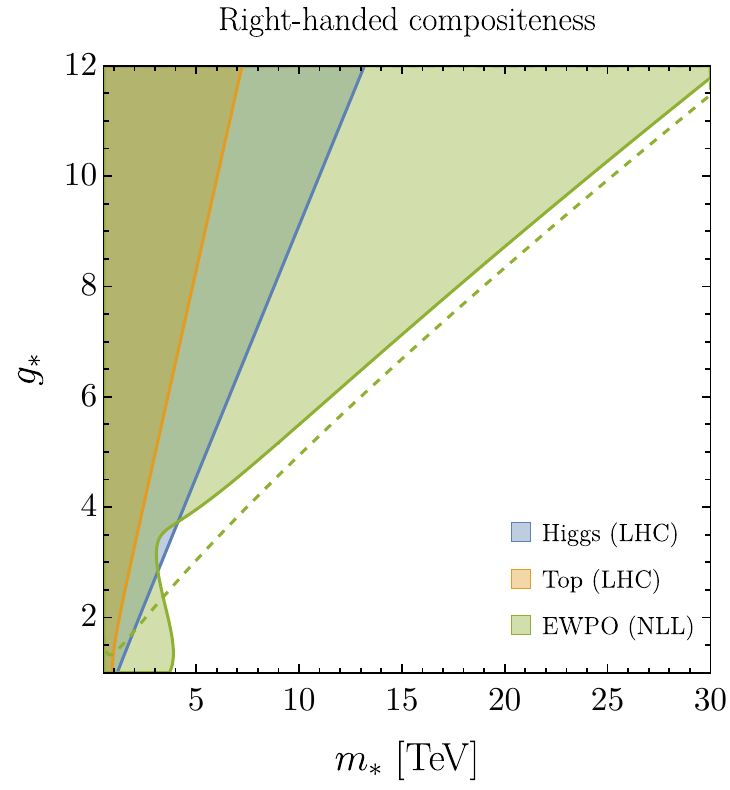}
\end{subfigure}
\hfill
\begin{subfigure}[t]{0.315\textwidth}
\centering
\includegraphics[width=\linewidth]{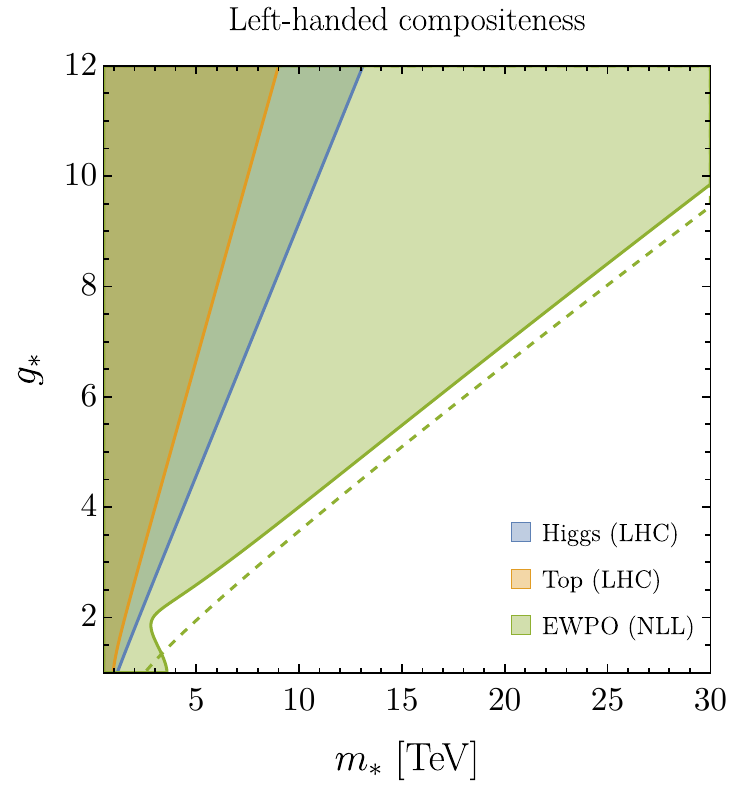}
\end{subfigure}
\hfill
\begin{subfigure}[t]{0.315\textwidth}
\centering
\includegraphics[width=\linewidth]{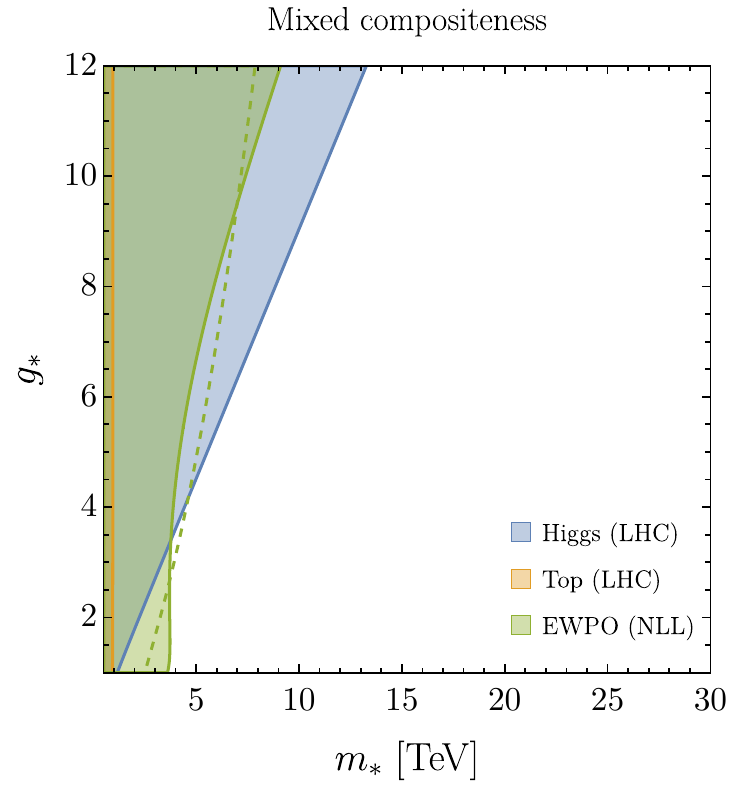}
\end{subfigure}
\vspace{0mm}
\caption{Same as Figure~\ref{fig:RSplots}, but for the custodial CH model. For additional explanations consult the main text.}
\label{fig:CHplots}
\end{figure}

\begin{figure}[t!]
\centering
\begin{subfigure}[t]{0.315\textwidth}
\centering
\includegraphics[width=\linewidth]{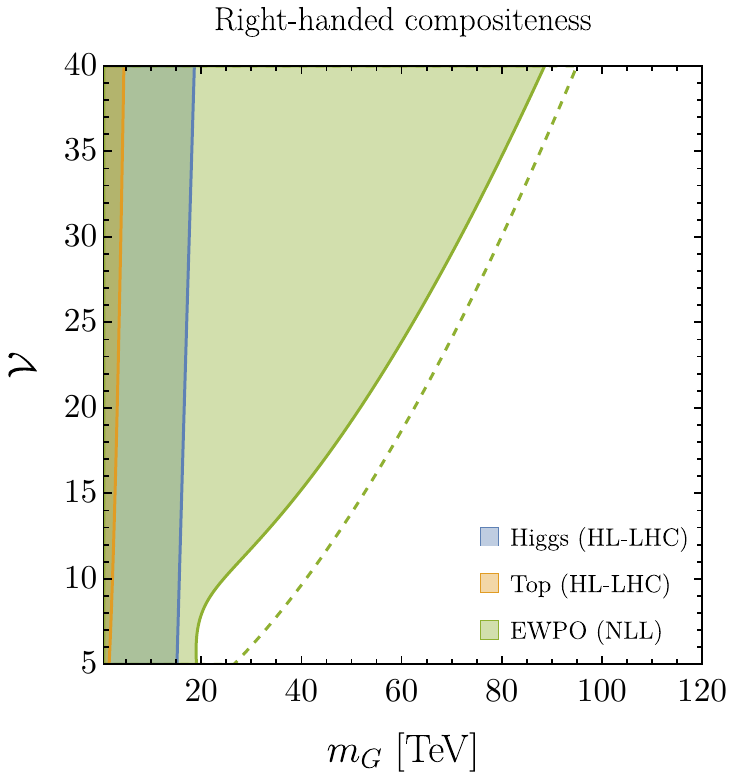}
\end{subfigure}
\hfill
\begin{subfigure}[t]{0.315\textwidth}
\centering
\includegraphics[width=\linewidth]{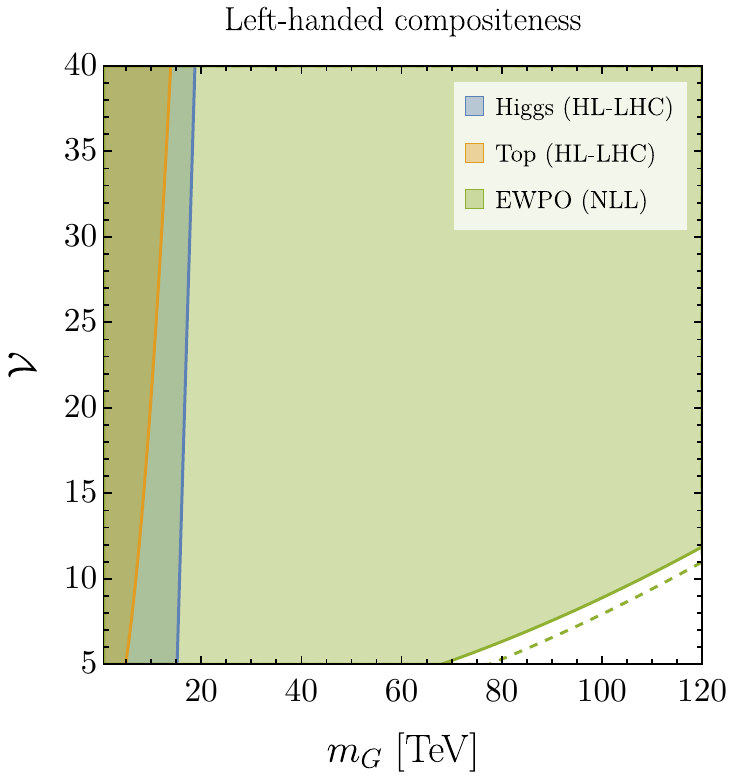}
\end{subfigure}
\hfill
\begin{subfigure}[t]{0.315\textwidth}
\centering
\includegraphics[width=\linewidth]{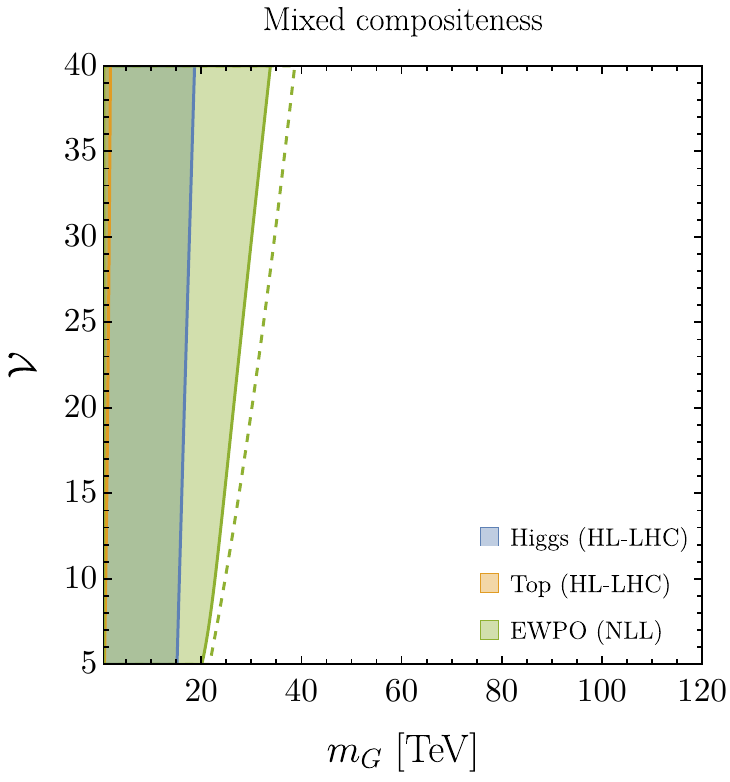}
\end{subfigure}
\vspace{0mm}
\caption{Same as Figure~\ref{fig:RSplots}, but for the projected constraints on the custodial RS model from HL-LHC Higgs measurements, top-quark observables, and an NLL analysis of~EWPO at the FCC-ee Tera-$Z$. Further details can be found in the main text.}
\label{fig:RSplots_future}
\end{figure}

\begin{figure}[t!]
\centering
\begin{subfigure}[t]{0.315\textwidth}
\centering
\includegraphics[width=\linewidth]{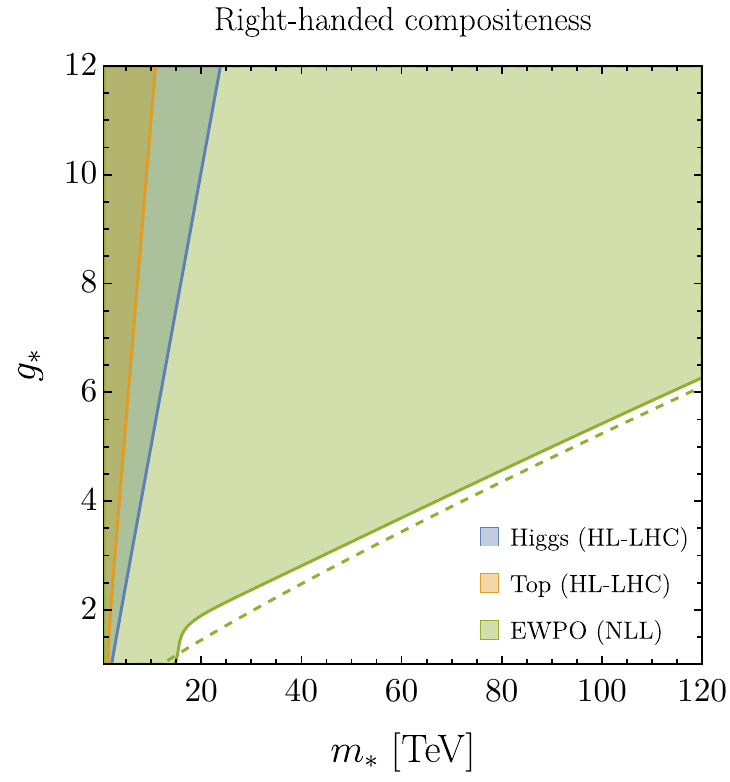}
\end{subfigure}
\hfill
\begin{subfigure}[t]{0.315\textwidth}
\centering
\includegraphics[width=\linewidth]{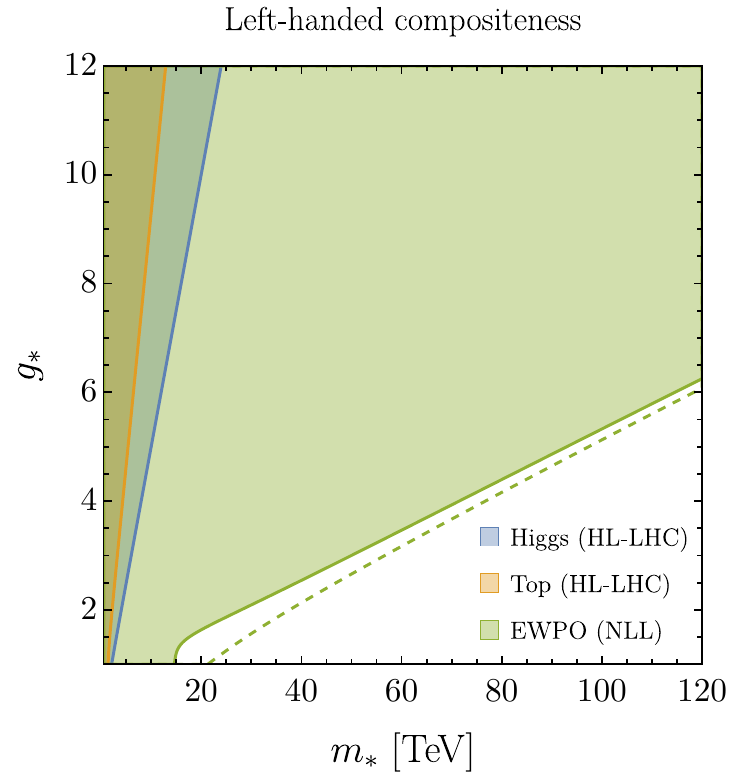}
\end{subfigure}
\hfill
\begin{subfigure}[t]{0.315\textwidth}
\centering
\includegraphics[width=\linewidth]{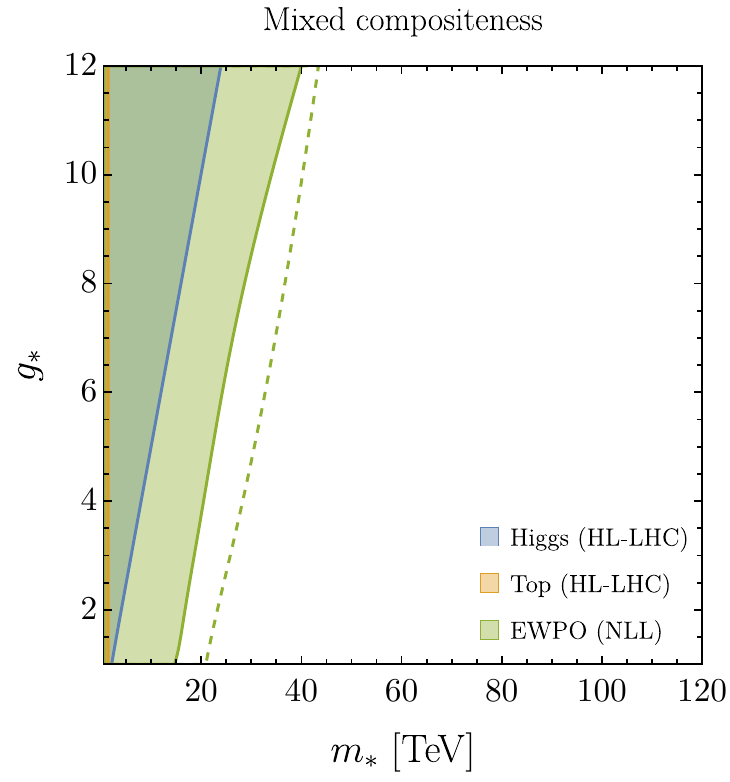}
\end{subfigure}
\vspace{0mm}
\caption{Same as Figure~\ref{fig:CHplots}, but showing the projected constraints on the custodial CH model from HL-LHC Higgs measurements, top-quark observables, and an NLL analysis of~EWPO at the FCC-ee Tera-$Z$. Additional details are given in the main text.}
\label{fig:CHplots_future}
\end{figure}

The constraints on the custodial RS and CH models derived from LHC Higgs measurements, top-quark observables, and an NLL analysis of~EWPO based on LEP and SLC data are shown in Figures~\ref{fig:RSplots} and~\ref{fig:CHplots}, respectively. The corresponding projections for the~HL-LHC and the~Tera-$Z$ run at FCC-ee are presented in Figures~\ref{fig:RSplots_future} and~\ref{fig:CHplots_future}. In all figures, the dashed green curve shows the~EWPO constraints obtained in simplified benchmark scenarios in which only the dominant class of Wilson coefficients is retained at the high scale $\Lambda$, identified with $m_G$ in the RS model and $m_\ast$ in the CH model. Specifically, in the RHCS we retain only $C_{uu}^{3333}(\Lambda)$ or $C_{Hu}^{33}(\Lambda)$, while in the LHCS and MCS we keep only $C_{Hq}^{(1)\hspace{0.25mm}33}(\Lambda)=-C_{Hq}^{(3)\hspace{0.25mm}33}(\Lambda)$ together with $C_{qq}^{(1)\hspace{0.25mm}3333}(\Lambda)$ and $C_{qq}^{(3)\hspace{0.25mm}3333}(\Lambda)$. All remaining Wilson coefficients are set to zero. These benchmark scenarios are motivated by the fact that they capture the dominant contributions to the Peskin-Takeuchi parameter $T$ and the~$Z\bar b_L b_L$~coupling. At one loop, the relevant corrections are given by 
\begin{align}
T^{(1)} & \simeq -\frac{v^2}{2\alpha} \hspace{0.25mm} \frac{6\alpha_t}{\pi} \hspace{0.5mm}  \left[ -C_{Hq}^{(1)\hspace{0.25mm}33}(\Lambda) + C_{Hu}^{33}(\Lambda)\right] \LLone\,, \label{eq:T1loop}  \\[1mm]
\delta g_{L}^{b \hspace{0.25mm} (1)} & \simeq -\frac{v^2}{2} \hspace{0.25mm} \frac{3\alpha_t}{\pi } \left[  -C_{qq}^{(1)\hspace{0.25mm}3333}(\Lambda)+\frac{1}{3} \hspace{0.25mm} C_{qq}^{(3)\hspace{0.25mm}3333}(\Lambda) \right] \LLone \,. \label{eq:deltaZbb1}
\end{align} 
Here, $\delta g_{L}^{b\hspace{0.25mm}(1)}$ denotes the relative additive one-loop shift in the tree-level SM coupling of the $Z$ boson to left-handed bottom quarks, $g_L^b = -1/2 + s_w^2/3 \simeq -0.42$. At the two-loop level, the only relevant correction arises in $T$, which is given by
\beq \label{eq:T2loop}
\begin{split}
T^{(2)} & \simeq -\frac{v^2}{2 \alpha} \, \Bigg \{ \hspace{0.25mm} \frac{21 \alpha_t^2}{2\pi^2} \hspace{0.5mm} \left [ \LLtwo - \frac{1}{7} \hspace{0.25mm} \NLL \right ] C_{qq}^{(1)\hspace{0.25mm}3333}(\Lambda) \\[1mm]
& \hspace{1.5cm} + \frac{9 \alpha_t^2}{2\pi^2} \hspace{0.5mm} \left [ \LLtwo -  \hspace{0.25mm} \NLL \right ] C_{qq}^{(3)\hspace{0.25mm}3333}(\Lambda) + \frac{12 \alpha_t^2}{\pi^2} \hspace{0.5mm} \left [ \LLtwo - \frac{1}{4} \hspace{0.25mm} \NLL \right ] C_{uu}^{3333}(\Lambda) \hspace{0.25mm} \Bigg \} \,.
\end{split}
\eeq
Below, we investigate how accurately these simplified benchmark scenarios and formulas reproduce the impact of the complete set of tree-level matching corrections entering the~NLL~EWPO analysis of the custodial~RS and~CH models.

\paragraph{Current bounds on the custodial RS model:}

Focusing first on the custodial~RS model, we retain only the contributions to the third-generation four-quark operators induced by KK-gluon exchange and neglect tree-level matching corrections proportional to the bottom-quark Yukawa coupling. Both approximations are numerically well justified. The situation is different for the operators modifying the $Z$-boson couplings, which receive additional contributions from fermionic KK resonances. While custodial symmetry protects the $Z\bar b_Lb_L$ coupling at tree level, it cannot simultaneously protect the $Z\bar t_Lt_L$ coupling, corresponding to the combination $C_{Hq}^{(1)\hspace{0.25mm}33}(m_G)-C_{Hq}^{(3)\hspace{0.25mm}33}(m_G)$, as follows from~\eqref{eq:CRSmatchingZcouplings1} and~\eqref{eq:CRSmatchingZcouplings2}. This is a consequence of the fact that, although the bulk theory preserves custodial symmetry, the elementary-composite fermion mixing required to generate the top-quark Yukawa coupling breaks this symmetry. Consequently, the tree-level matching generates a non-vanishing coefficient $C_{Hu}^{33}(m_G)$ proportional to $y_t^2$, as is evident from~\eqref{eq:CRSmatchingZcouplings3}.

Starting with the LHCS shown in the central panel of~Figure~\ref{fig:RSplots}, we find that the approximation obtained by retaining only $C_{Hq}^{(1)\hspace{0.25mm}33}(m_G)=-C_{Hq}^{(3)\hspace{0.25mm}33}(m_G)$, $C_{qq}^{(1)\hspace{0.25mm}3333}(m_G)$, and $C_{qq}^{(3)\hspace{0.25mm}3333}(m_G)$ in~\eqref{eq:T1loop} to \eqref{eq:T2loop} reproduces the full~NLL~EWPO constraint very well. The~small residual difference arises from $\mathcal{V}$-independent contributions to the $S$ and $T$ parameters induced by~$C_{HWB}(m_G)$ and $C_{uH}^{33}(m_G)$, respectively. The~EWPO bound in the~LHCS is also consistently stronger than the constraint from current LHC top-quark measurements, even when quadratic ${\cal O}(\Lambda^{-4})$ terms are included. In contrast, in the RHCS and MCS, shown in the left and right panels of Figure~\ref{fig:RSplots}, respectively, the dashed green curves are obtained by retaining only $C_{Hu}^{33}(m_G)$ in the RHCS, whereas in the~MCS the same set of operators as in the LHCS is switched on. In both cases, the~$\mathcal{V}$-enhanced contributions entering the full~NLL~EWPO analysis partially cancel, yielding approximately $\mathcal{V}$-independent bounds that are weaker than those from LHC Higgs measurements, which are themselves largely insensitive to the choice of compositeness scenario. This behavior can be traced to the tree-level matching conditions: while the custodial RS contribution to~$Q_{H\Box}$ is enhanced by the volume factor, as shown in~\eqref{eq:RSCHbox}, the matching contributions to~$Q_{uH}^{33}$ and $Q_{dH}^{33}$ scale with the squares of the corresponding 5D Yukawa couplings, see~\eqref{eq:RSCuH} and~\eqref{eq:RSCdH}, and are independent of the bulk-mass parameters ${\mathcal{c}}_q$. Finally, the top-quark constraint is significantly stronger in the LHCS than in the RHCS, reflecting the enhanced tree-level matching contribution to $Q_{qq}^{(3)\hspace{0.25mm}3333}$, which directly affects top-quark pair production through the $b\bar b\to t\bar t$ channel. By contrast, the sensitivity of top-quark measurements is substantially reduced in the~MCS. This can be understood from the pattern of the third-generation four-quark operator coefficients, where only $Q_{qu}^{(8)\hspace{0.25mm}3333}$ receives a numerically relevant matching contribution in the mixed top-quark compositeness scenario.

\paragraph{Current bounds on the custodial CH model:}

The corresponding constraints for the custodial CH model are shown in~Figure~\ref{fig:CHplots}. In the numerical analysis, we include all tree-level SMEFT matching contributions derived in~Section~\ref{sec:MCH}, including those to the third-generation four-quark operators induced by the exchange of the complete set of $SU(3)_C \times SO(4) \times U(1)_X$ vector resonances. The dashed green approximation in the RHCS, shown in the left panel of Figure~\ref{fig:CHplots}, is obtained by retaining only $C_{uu}^{3333}(m_\ast)$ in~\eqref{eq:T2loop}. It reproduces the full~NLL~EWPO constraint well, with the small residual deviations originating from contributions to the $S$ and $T$ parameters induced by $C_{HW\!B}(m_\ast)$ and $C_{H\Box}(m_\ast)$, respectively, which are omitted in the approximation. The largest deviations occur around $g_\ast \simeq 3.5$, where the full NLL result exhibits a partial cancellation, driven primarily by the interplay between the $g_\ast$-independent contribution to the $S$ parameter induced by $C_{HW\!B}(m_\ast)$ and the $g_\ast$-dependent contribution to the $T$ parameter arising from $C_{uu}^{3333}(m_\ast)$. In contrast to the RS model, the~EWPO constraint in the CH model is generally the most stringent in the RHCS, surpassing both the LHC top-quark and Higgs constraints. The weaker Higgs constraints in the CH model can be traced to the different parametric scaling of the relevant tree-level matching corrections. In particular, the PNGB nature of the Higgs boson implies that all shift-symmetry-breaking couplings are proportional to the corresponding~SM~couplings, thereby suppressing the induced deviations in Higgs observables. For example, in the custodial CH model both~$C_{uH}^{33}(m_\ast) \simeq g_\ast^2 \hspace{0.25mm} y_t/m_\ast^2$ and the contributions to the third-generation four-quark operators~\eqref{eq:CCHmatching4top1} to \eqref{eq:CCHmatching4top4} scale as~$g_\ast^2$. By contrast, in the custodial~RS model, while the four-quark operators are enhanced by ${\mathcal{V}}$, the Wilson coefficient $C_{uH}^{33}(m_G)$ is instead proportional to the square of the 5D up-type Yukawa couplings and remains ${\mathcal{V}}$-independent, as shown in~\eqref{eq:RSCuH}. Consequently, in the CH model the Higgs-sector and four-quark contributions decouple in the same way for decreasing $g_\ast$, whereas in the RS model only the four-quark contributions are suppressed in the small-${\mathcal{V}}$ limit, while the Higgs-sector corrections remain unsuppressed. In this context, it is worth recalling that, in the MCHM$_5$, single-Higgs production via gluon fusion is insensitive to the spectrum of fermionic resonances at leading order in the heavy-mass expansion~\cite{Low:2010mr,Gillioz:2012se}. In the SMEFT description, this follows from an exact cancellation between the explicit loop contributions of the heavy fermions and the contribution encoded in the Wilson coefficient~$C_{uH}^{33}(m_\ast)$, leaving only the universal correction associated with the CCWZ sector. Consequently, the result~\eqref{eq:CHCHboxCuH} correctly reproduces the universal contribution to the effective~Higgs-gluon~coupling.

The central panel of~Figure~\ref{fig:CHplots} shows that, as in the custodial~RS model, the~EWPO constraint in the LHCS is stronger than in the RHCS and exceeds the bounds from both Higgs and top-quark measurements. The dominant effect arises from the $g_\ast^2$-enhanced~$\LLone$~corrections to the $T$ parameter induced by the modified $Z\bar t_Lt_L$ coupling. In addition, the operators $C_{qq}^{(1)\hspace{0.25mm}3333}(m_\ast)$ and $C_{qq}^{(3)\hspace{0.25mm}3333}(m_\ast)$ generate $g_\ast^2$-enhanced non-oblique $\LLone$ corrections to the $Z\bar b_Lb_L$ vertex through~\eqref{eq:deltaZbb1}, as well as $\LLtwo$ and $\NLL$ contributions to $T$ through~\eqref{eq:T2loop}. Including these effects leads to the dashed green curve, which reproduces the full~NLL~EWPO constraint very well. The remaining small deviations are due to contributions from the matching conditions $C_{HW\!B}(m_\ast)$ and $C_{H\Box}(m_\ast)$, which are not included in the approximation but have the same origin as in the RHCS. As in the custodial~RS model, the top-quark constraints are stronger in the LHCS than in the RHCS and MCS. In the LHCS, this enhanced sensitivity originates from the larger tree-level matching contribution to the left-handed third-generation four-quark operator $Q_{qq}^{(3)\hspace{0.25mm}3333}$. By contrast, the reduced sensitivity in the~MCS can be traced to the fact that only the mixed-chirality operator $Q_{qu}^{(8)\hspace{0.25mm}3333}$ receives a numerically relevant contribution, which in the CH model is relatively small and independent of $g_\ast$. Finally, we note that the Higgs constraint is essentially independent of the compositeness scenario in the custodial~CH model, since the relevant Wilson coefficients in~\eqref{eq:CHCHboxCuH} do not depend on the compositeness parameters $\epsilon_q$. As a result, the same Higgs bound is obtained in all scenarios. In fact, Higgs observables provide the strongest constraints for $g_\ast \gtrsim 3.5$ in the MCS shown in the right panel of~Figure~\ref{fig:CHplots}, where the full~NLL~EWPO result again exhibits an approximate cancellation of $g_\ast$-enhanced contributions.

\paragraph{Interpretation of current bounds on the custodial RS and CH models:}

Taken together, the results in~Figures~\ref{fig:RSplots} and~\ref{fig:CHplots} show that the RHCS and MCS are subject to the weakest overall constraints from Higgs, top-quark, and~EWPO. We note, however, that additional strong constraints, not considered here, can arise from flavor-changing neutral currents in the down-quark sector, particularly in CH models with left-handed or mixed partial compositeness~\cite{Glioti:2024hye,Stefanek:2024kds}. The RHCS therefore permits the lowest masses of the heavy BSM resonances in both the custodial~RS and~CH models once all relevant constraints are taken into account. It consequently offers the largest allowed parameter space for the volume factor $\mathcal{V}$ and the strong-sector coupling $g_\ast$, which control the interaction strengths of the KK and composite sectors, respectively. Interestingly, in the custodial~CH model the weakest constraints in the RHCS are obtained for $g_\ast \in [2,4]$, a range that coincides with the values typically preferred for generating a realistic Higgs potential in CH models~\cite{Giudice:2007fh,Bellazzini:2014yua,Fuentes-Martin:2022xnb}.

\paragraph{Projected constraints on the custodial RS and CH models:}

In~Figures~\ref{fig:RSplots_future} and~\ref{fig:CHplots_future}, we present the projected constraints on the custodial~RS and~CH models, respectively, obtained from HL-LHC Higgs measurements, top-quark observables, and an NLL analysis of~EWPO at the FCC-ee Tera-$Z$ stage. The HL-LHC projections follow the methodology described in~Appendix~\ref{app:LHC}, while the FCC-ee projections assume the S1 scenario defined in~Section~\ref{sec:SMexpEWPO}. The simplified estimates based only on the leading contributions in~\eqref{eq:T1loop},~\eqref{eq:deltaZbb1}, and~\eqref{eq:T2loop}, indicated by the dashed green curves, have the same interpretation as for the current constraints discussed above. The six panels demonstrate the exceptional sensitivity expected from future precision measurements. For all compositeness scenarios considered, FCC-ee~EWPO are projected to provide by far the strongest constraints on the parameter spaces of both the custodial~RS and~CH models. This substantial improvement is a direct consequence of the unprecedented precision anticipated at FCC-ee and highlights the importance of consistently including the full logarithmically enhanced~RG~effects in the NLL analysis. These results emphasize the unique potential of FCC-ee $Z$-pole physics as an indirect probe of BSM scenarios and demonstrate the crucial role of precision measurements in extending the reach for new physics well beyond the direct energy frontier.

\section{Conclusions} 
\label{sec:conclusions}

In this article, we have demonstrated how two-loop RG~evolution within the SMEFT can be exploited to systematically improve predictions for~EWPO to NLL accuracy. Combining the recently computed two-loop beta functions for the complete set of dimension-six operators~\cite{Born:2026xkr} with the NLO fixed-order matching corrections derived in~\cite{Biekotter:2025nln}, we have obtained NLL expressions for all EWPO, including their dependence on the complete set of 210 independent Wilson coefficients identified in~Section~\ref{sec:one}. These results are provided as ancillary electronic files accompanying the arXiv submission of this work.

A central outcome of our analysis is a systematic classification of dimension-six operators according to the perturbative order at which they first contribute to~EWPO. Operators entering already at tree level are tightly constrained by the precision measurements performed at LEP and SLC, while one-loop effects have been investigated extensively in previous fixed-order analyses~\cite{Dawson:2019clf,Dawson:2022bxd,Bellafronte:2023amz,Biekotter:2025nln,Bellafronte:2026jic}. The main conceptual advance of the present work is the complete treatment of operators that become accessible to~EWPO constraints only through genuine two-loop effects. These include the top-quark chromomagnetic dipole operator $Q_{uG}^{33}$, the triple-gluon operator~$Q_G$, the top-quark Yukawa operator $Q_{uH}^{33}$~\cite{Born:2026tgm}, and the Higgs-gluon operator~$Q_{HG}$, all of which generate logarithmically enhanced contributions to~EWPO for the first time at NLL order. By contrast, the Higgs self-interaction operator~$Q_H$ remains inaccessible at this level of precision, since its leading-logarithmic contributions first arise at NNLL order, for instance through four-loop RG mixing.

We have described a systematic methodology for obtaining NLL-accurate predictions for~EWPO within the SMEFT. While logarithmically enhanced effects arise both from RG~evolution and EW-scale matching, we have shown that two-loop matching corrections are not required for NLL accuracy. Using the contribution of the purely right-handed four-top operator $Q_{uu}^{3333}$ to the Peskin-Takeuchi parameter $T$ as an explicit example, we demonstrated that two-loop matching corrections do not generate the large logarithms associated with the mass gap between the BSM and EW  scales and therefore contribute only at NNLL order. Consequently, NLL predictions can be obtained by combining tree-level and one-loop matching conditions with the SMEFT RG~evolution, which is now known to two-loop order~\cite{Born:2026xkr}. This observation is particularly relevant since complete one-loop matching results are available for the full set of~EWPO~\cite{Dawson:2019clf,Dawson:2022bxd,Bellafronte:2023amz,Biekotter:2025nln,Bellafronte:2026jic}, whereas two-loop matching calculations remain restricted to a small subset of operators~\cite{Haisch:2024wnw,Degrassi:2017ucl,Kribs:2017znd}.

On the phenomenological side, we have performed a comprehensive study of~EWPO at NLL accuracy, focusing on operators whose effects first arise beyond the one-loop level and comparing the resulting constraints with those obtained from Higgs and top-quark observables at the LHC. Using both current LEP and SLC data and projected Tera-$Z$ measurements at the FCC-ee, we have demonstrated that RG-enhanced effects can substantially improve the sensitivity to BSM physics. A particularly striking result concerns third-generation four-quark operators, for which logarithmically enhanced RG effects increase the sensitivity of~EWPO such that even existing LEP and SLC measurements often yield constraints that surpass current LHC bounds. Future FCC-ee data further improve this sensitivity, probing effective scales of at least $20\,{\rm TeV}$. These results highlight the strong complementarity of~EWPO measurements and high-energy collider observables.

We also find that operators such as $Q_{uH}^{33}$ and $Q_{uG}^{33}$, which are currently constrained predominantly through Higgs and top-quark observables, can receive competitive or even stronger indirect bounds from future~EWPO through two-loop RG-induced mixing. In~particular, the projected FCC-ee Tera-$Z$ determination of the $W$-boson mass provides a powerful probe of these operators. A similar improvement is observed for $Q_G$, for which future~EWPO measurements become competitive with direct collider probes through non-oblique corrections, most notably via the observables $\Gamma_Z$ and $R_\ell$. In contrast, the sensitivity to $Q_{HG}$ remains dominated by Higgs measurements due to its tree-level contribution to the effective Higgs-gluon coupling. Overall, these results demonstrate the strong complementarity of future~EWPO measurements and high-energy collider observables in probing BSM physics within the SMEFT framework.

The model-independent~EWPO constraints derived in this work were subsequently applied to the custodial~RS and~CH models, for which we obtained the corresponding tree-level SMEFT matching conditions. Both frameworks exhibit a characteristic hierarchy among the third-generation four-quark operators, with the purely right-handed four-top operator~$Q_{uu}^{3333}$ emerging as the dominant low-energy signature of right-handed top-quark compositeness over a wide range of parameter space. In addition, we identified the leading tree-level contributions to the EW and Higgs sectors, thereby providing a unified EFT description of the dominant low-energy effects in custodial~RS and~CH scenarios. Combining our NLL~SMEFT analysis of~EWPO with current LHC Higgs and top-quark measurements, we~derived the resulting constraints on the custodial~RS and~CH models. We find that~EWPO provide the most stringent bounds in both frameworks, in particular for scenarios with left-handed top-quark compositeness. The main features of the complete NLL~analysis are well reproduced by simple analytical approximations based on the dominant tree-level matching contributions, although additional RG-induced effects and cancellations can lead to visible~deviations.

Once flavor constraints are taken into account, right-handed compositeness emerges as the least constrained scenario in both BSM models, allowing for the lowest resonance masses and the largest viable parameter space. In the custodial~CH model, this scenario remains weakly constrained for $g_\ast \in [2,4]$, a range compatible with realistic Higgs-potential considerations~\cite{Giudice:2007fh,Bellazzini:2014yua,Fuentes-Martin:2022xnb}, while avoiding the severe flavor constraints in the down-quark sector typically associated with left-handed or mixed partial compositeness~\cite{Glioti:2024hye,Stefanek:2024kds}. Our~projections for the custodial~RS and~CH models show that future FCC-ee~EWPO measurements can provide the dominant constraints across all compositeness scenarios, substantially surpassing the sensitivity expected from HL-LHC Higgs and top-quark measurements. The~exceptional precision of the FCC-ee Tera-$Z$ stage, combined with large double-logarithmic RG effects, makes the inclusion of full NLL RG~evolution essential for reliably assessing the reach of future precision experiments.

Several important directions for future work emerge from our analysis. The NLL results obtained here provide a systematic framework for incorporating two-loop RG effects into global SMEFT studies. Such precision will become essential for future high-luminosity $e^+e^-$ collider programs, including FCC-ee and CEPC, where~EWPO are expected to be measured at the subpermille level. At this accuracy, NLL corrections are no longer a subleading theoretical refinement but an essential component of reliable SMEFT predictions. A natural next step is the combination of the~EWPO constraints derived here with complementary information from Higgs, top-quark, and flavor observables in global SMEFT fits, thereby exploiting the interplay between precision measurements and high-energy collider probes.

More generally, our results demonstrate that higher-order RG effects are becoming a central ingredient of precision SMEFT phenomenology. Logarithmically enhanced contributions that first arise at two-loop order can compete with, and in some cases dominate over, formally lower-order effects. Whenever one-loop SMEFT matching conditions are available, the complete two-loop anomalous dimensions~\cite{Born:2026xkr} allow existing NLO analyses to be systematically extended to NLL accuracy. The results presented here, together with the accompanying ancillary files, provide the necessary ingredients for this program and establish a general framework for future precision studies of BSM physics within the SMEFT.

\acknowledgments{UAH gratefully acknowledges Tom Green for the entertainment. BAS thanks Admir Greljo and Victor Maura for helpful discussions.}

\begin{appendix}

\section{Operator building blocks} 
\label{app:definitions}

This appendix collects the field-theoretic building blocks entering the dimension-six SMEFT operators in the Warsaw basis relevant for the present analysis. The dimension-six contribution to the SMEFT Lagrangian reads
\beq \label{eq:L6}
{\mathcal{L}}_6 = \sum_{X \in \mathrm{real}} C_X \hspace{0.125mm} O_X + \sum_{X \in \mathrm{complex}} \big( C_X \hspace{0.125mm} O_X + \mathrm{h.c.} \big) \,,
\eeq
where $O_X$ are the operators and $C_X$ the corresponding dimensionful Wilson coefficients. For complex operators, Hermitian conjugates are included to ensure the reality of the Lagrangian.

The operators $O_X$ relevant for this work are listed in~Tables~\ref{tab:no4ferm},~\ref{tab:4ferm}, and~\ref{tab:uXoperators}. Here, $H$ denotes the SM Higgs doublet, $\widetilde{H}_i = \epsilon_{ij}(H^j)^\ast$ with $\epsilon_{ij}$ the antisymmetric Levi-Civita tensor of the fundamental $SU(2)_L$ indices normalized by $\epsilon_{12} = 1$. The fermion fields $q$ and $\ell$ are left-handed, while $u$, $d$, and $e$ are right-handed. The covariant derivative acts as
\beq \label{eq:covariant}
D_\mu q^{ai} = \partial_\mu q^{ai} + i g_Y Y B_\mu \hspace{0.25mm} q^{ai} + i g_L t^I_{ij} \hspace{0.25mm} W^I_\mu \hspace{0.25mm} q^{aj} + i g_s T^A_{ab} \hspace{0.25mm} G^A_\mu \hspace{0.25mm} q^{bi} \,,
\eeq
where $B_\mu$, $W^I_\mu$, and $G^A_\mu$ are the $U(1)_Y$, $SU(2)_L$, and $SU(3)_C$ gauge fields, $Y$ is the hypercharge, and $t^I = \tau^I/2$, $T^A = \lambda^A/2$ are the fundamental-representation generators of $SU(2)_L$ and $SU(3)_C$, with $\tau^I$ and $\lambda^A$ the Pauli and Gell-Mann matrices, normalized as $\mathrm{tr}\left(t^I t^J \right) = \delta^{IJ}/2$ and $\mathrm{tr}\left(T^A T^B\right) = \delta^{AB}/2$.

The corresponding field-strength tensors are
\beq \label{eq:BWG}
\begin{split}
B_{\mu\nu} &= \partial_\mu B_\nu - \partial_\nu B_\mu \,, \\[1mm]
W^I_{\mu\nu} &= \partial_\mu W^I_\nu - \partial_\nu W^I_\mu - g_L \hspace{0.25mm} \epsilon^{IJK} \hspace{0.25mm} W^J_\mu \hspace{0.25mm} W^K_\nu \,, \\[1mm]
G^A_{\mu\nu} &= \partial_\mu G^A_\nu - \partial_\nu G^A_\mu - g_s \hspace{0.25mm} f^{ABC} \hspace{0.25mm} G^B_\mu \hspace{0.25mm} G^C_\nu \,,
\end{split}
\eeq
with $\epsilon^{IJK}$ and $f^{ABC}$ the fully antisymmetric structure constants of $SU(2)_L$ and $SU(3)_C$, respectively. We further define $\sigma_{\mu\nu} = i/2\,[\gamma_\mu,\gamma_\nu]$, where $[a,b] = ab - ba$ denotes the commutator and $\gamma_\mu$ are the Dirac matrices, and the Hermitian covariant derivatives are defined as $a^\ast \raisebox{2mm}{\boldmath ${}^\leftrightarrow$}\hspace{-4mm} D_\mu b = a^\ast(D_\mu - \raisebox{2mm}{\boldmath${}^\leftarrow$}\hspace{-4mm} D_\mu) \hspace{0.25mm} b$ and $a^\ast \raisebox{2mm}{\boldmath ${}^\leftrightarrow$}\hspace{-4mm} D_\mu^{\,I} b = a^\ast( \tau^I D_\mu - \raisebox{2mm}{\boldmath${}^\leftarrow$}\hspace{-4mm} D_\mu \tau^I) \hspace{0.25mm} b$.

\section{Constraints from individual~EWPO}
\label{app:4topEWPO}

In~Section~\ref{sec:semi4top}, we presented NLL-accurate formulas for the~EWPO describing their dependence on the Wilson coefficient of the purely right-handed four-top operator $Q_{uu}^{3333}$. These~formulas were then used in~Section~\ref{sec:4HQfits} to derive two-dimensional constraints on the Wilson coefficients of the third-generation four-quark operators from a global~EWPO analysis. In this appendix, we illustrate the constraints arising from individual~EWPO, using the examples of $Q_{uu}^{3333}$ together with $Q_{qu}^{(1)\hspace{0.25mm}3333}$ and $Q_{qq}^{(1)\hspace{0.25mm}3333}$.

\begin{figure}[t!]
\centering
\includegraphics[width=\linewidth]{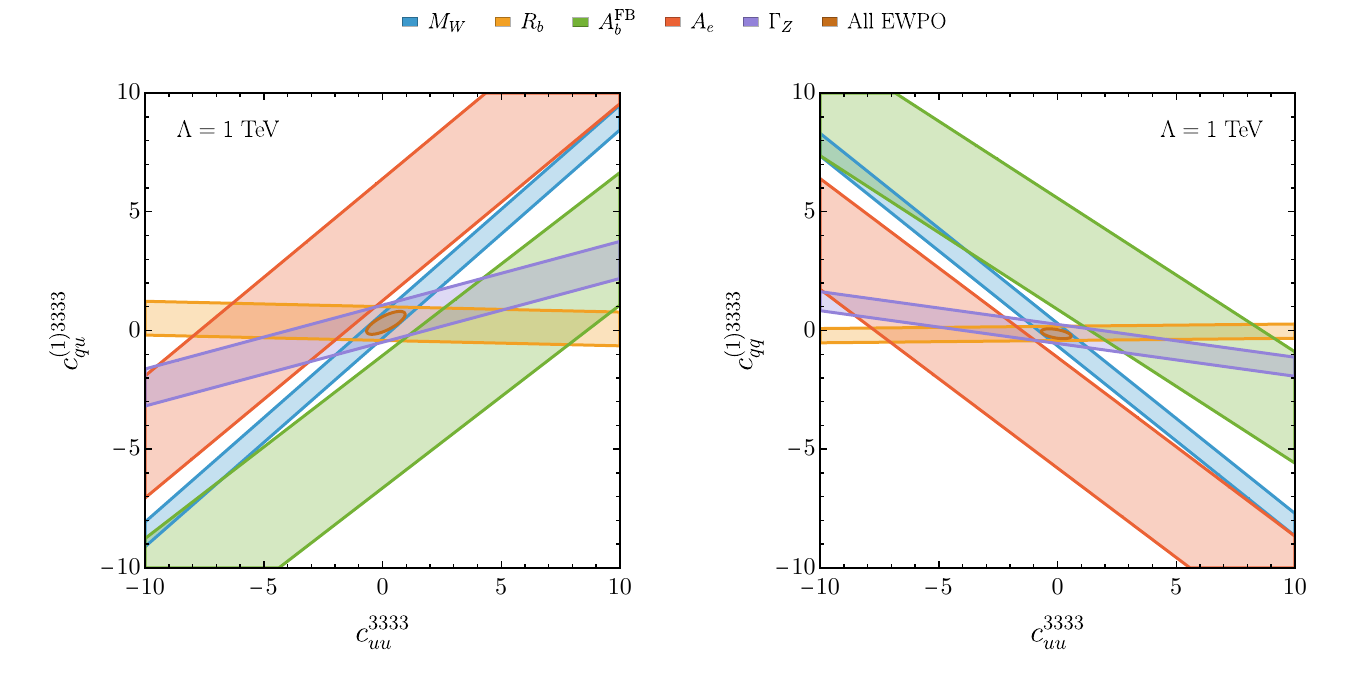}
\caption{$68\%$~CL constraints from individual~EWPO in the $c_{uu}^{3333}\hspace{0.25mm}$--$\hspace{0.5mm}c_{qu}^{(1)\hspace{0.25mm}3333}$ plane (left) and the $c_{uu}^{3333}\hspace{0.25mm}$--$\hspace{0.5mm}c_{qq}^{(1)\hspace{0.25mm}3333}$ plane (right). The plots are obtained assuming a common high scale of $\Lambda = 1\,{\rm TeV}$. Consult the main text for more details.}
\label{fig:topOpsEWsensitivity}
\end{figure}

Figure~\ref{fig:topOpsEWsensitivity} shows the $68\%$~CL constraints in the $c_{uu}^{3333}\hspace{0.25mm}$--$\hspace{0.5mm}c_{qu}^{(1)\hspace{0.25mm}3333}$ plane (left) and the $c_{uu}^{3333}\hspace{0.25mm}$--$\hspace{0.5mm}c_{qq}^{(1)\hspace{0.25mm}3333}$ plane (right), obtained from the observables $m_W$, $R_b$, $A_{\rm FB}^b$, $A_e$, $\Gamma_Z$, and from the combined fit of all~EWPO. The corresponding SM predictions and experimental measurements are listed in the first and second columns of Table~\ref{tab:EWPOummary}, respectively. The~dimensionless Wilson coefficients are defined in~\eqref{eq:smallc} and the plots are obtained assuming a common high scale of $\Lambda = 1\,{\rm TeV}$. From~both panels it is apparent that the strongest individual constraints arise from measurements of the $W$-boson mass, followed by the ratio $R_b$ and the total $Z$-boson decay width. The~forward-backward asymmetry of bottom quarks and the left-right asymmetry of electrons are instead significantly less constraining. One also observes that all individual constraints exhibit flat directions, and only their combination lifts this degeneracy, with the global fit being essentially driven by $m_W$, $R_b$, and $\Gamma_Z$. This~implies that, in order to fully exploit the constraining power of NLL~EWPO in SMEFT fits, both oblique and non-oblique corrections need to be consistently included.

\section{SMEFT effects in top-quark physics}
\label{app:top}

The third-generation four-quark operators introduced in~\eqref{eq:4topoperators} can be constrained at the~LHC through measurements of $pp \to t \bar t$ and $pp \to t \bar t t \bar t$ production~\cite{Ethier:2021bye,Aoude:2022deh,ATLAS:2023ajo,Degrande:2024mbg,DiNoi:2025uhu,ATLAS:2018fwl,CMS:2019rvj,CMS:2019eih,ATLAS:2020hpj,Banelli:2020iau,ATLAS:2021kqb,CMS:2023zdh,CMS-PAS-TOP-24-008,CMS:2025ugn,Subba:2026opu,Dawson:2022bxd,Degrande:2020evl,Brivio:2019ius}. In~Section~\ref{sec:Zpolefits}, we employed the available experimental information on these production channels to constrain the relevant Wilson coefficients and compared the resulting bounds with those derived from~EWPO. To make our discussion self-contained, this appendix provides additional details on the calculations underlying our SMEFT predictions for $pp \to t \bar t$ and $pp \to t \bar t t \bar t$ production.

\subsection{Top-quark pair production} 
\label{eq:2top}

The third-generation four-quark operators in~\eqref{eq:4topoperators} contribute to $b \bar b \to t \bar t$ scattering already at tree level, whereas their effects in $q \bar q \to t \bar t$ and $g g \to t \bar t$, with $q=u,d,s,c$, arise only at the one-loop level. After renormalizing the top-quark mass in the on-shell~(OS) scheme, the loop-induced $g g \to t \bar t$ amplitudes are UV finite. By contrast, the corresponding $q \bar q \to t \bar t$ amplitudes develop UV divergences in dimensional regularization with~$d=4-2\hspace{0.125mm}\epsilon$. These~singularities are proportional to QCD penguin-type amplitudes and are removed through operator renormalization in the $\overline{\rm MS}$ scheme, introducing the renormalization scale~$\mu_R$. In~our calculation, $\gamma_5$ is treated using a naive anti-commuting prescription~(NDR). We have explicitly verified that the divergence structure of our results agrees with the direct calculations of the relevant one-loop beta functions first presented in~\cite{Alonso:2013hga}, and that the use of the NDR~scheme yields consistent results for the observables considered here. 
In our calculation, the Cabibbo-Kobayashi-Maskawa matrix is approximated by the identity, and only the top-quark mass is retained as non-zero, while the bottom-quark and all remaining quark masses are set to zero. The generation and evaluation of the relevant amplitudes were performed using the {\tt Mathematica} packages {\tt FeynRules}~\cite{Alloul:2013bka}, {\tt FeynArts}~\cite{Hahn:2000kx}, {\tt FormCalc}~\cite{Hahn:2016ebn}, and~{\tt Package-X}~\cite{Patel:2015tea}. In the results presented below and employed in~Section~\ref{sec:Zpolefits}, we include both the interference of the tree-level and one-loop~SMEFT amplitudes with the SM contribution and the squared contributions of all~SMEFT amplitudes.

\paragraph{Analytic results:}

In $t\bar t$ production, the partonic cross section is commonly decomposed into components that are even or odd under the exchange of the $t$ and $\bar t$, equivalently under $\cos\theta \to -\cos\theta$ in the partonic centre-of-mass~(COM) frame, where $\theta$ denotes the scattering angle. The corresponding inclusive cross sections are obtained by integrating over the invariant mass range~$M \in [2m_t,\sqrt{S}]$, where $M^2$ is the invariant mass squared of the~$t\bar t$~system and $S$~denotes the hadronic COM energy squared. Explicitly,
\beq \label{eq:sigmaF}
\sigma_f = \frac{8 \pi \alpha_s^2 \beta}{3 S} \sum_{i,j} \, \int_{2 m_t}^{\sqrt{S}} \frac{dM}{M} \hspace{0.25mm} f\hspace{-1.3mm}f_{ij}\!\left(\frac{M^2}{S}, \mu_F \right) F_{ij} \,,
\eeq
where $f=s,a$ and $F=S,A$ correspond to the symmetric and asymmetric contributions, respectively. Here $f\hspace{-1.3mm}f_{ij}(x,\mu_F)$ denotes the parton luminosity functions defined by
\beq \label{eq:lumis}
f\hspace{-1.3mm}f_{ij}(x,\mu_F) = \int_{x}^{1} \frac{dy}{y}\, f_{i/p}(y,\mu_F)\, f_{j/p}\!\left(\frac{x}{y},\mu_F\right) \,.
\eeq
The functions $f_{i/p}(x,\mu_F)$ are the universal non-perturbative proton parton distribution functions (PDFs), describing the probability of finding a parton $i$ inside the proton carrying a longitudinal momentum fraction $x$. The parameter $\mu_F$ denotes the factorization scale. In~\eqref{eq:sigmaF} we have further introduced
\beq \label{eq:betarho}
\beta = \sqrt{1 - \rho}\,, \qquad \rho = \frac{4m_t^2}{M^2}\,.
\eeq

Interestingly, the coefficients $S_{ij}$ and $A_{ij}$ in~\eqref{eq:sigmaF} can be derived analytically up to one-loop order for all channels and operators appearing in~\eqref{eq:4topoperators}. As a representative example, we present below the results for the operator $Q_{qu}^{(8)\hspace{0.25mm} 3333}$. The non-vanishing symmetric coefficients associated with the $b \bar b \to t \bar t$ channel, linear or quadratic in the Wilson coefficient~$C_{qu}^{(8) \hspace{0.25mm} 3333}$, are given by
\beq \label{eq:Sbb}
S_{b \bar b}^{(1)} = \frac{m_t^2}{18 \hspace{0.125mm} \pi \alpha_s} \frac{3 - \beta^2}{1 - \beta^2} \, C_{qu}^{(8) \hspace{0.25mm} 3333} \,, \qquad 
S_{b \bar b}^{(2)} = \frac{m_t^4}{72 \hspace{0.125mm} \pi \alpha_s} \frac{3 + \beta^2}{\left( 1 - \beta^2 \right)^2} \, \big( C_{qu}^{(8) \hspace{0.25mm} 3333} \big)^2 \,,
\eeq

In the case of $q \bar q \to t \bar t$ with $q = u, d, s, c$, we instead obtain
\beq \label{eq:Sqq}
\begin{split}
S_{q \bar q}^{(1)} & = -\frac{m_t^2}{18 \hspace{0.125mm} \pi^2} \frac{3 - \beta^2}{1 - \beta^2} \hspace{0.5mm} \Bigg [ \hspace{0.25mm} \frac{9 - 5 \beta^2 + \beta^4}{9 - 3 \beta^2} + \frac{1}{6} \hspace{0.25mm} L_\beta + \frac{3 - \beta^2}{6} \, {\rm Re} \hspace{0.25mm} B + L_\mu \hspace{0.25mm} \Bigg ] \hspace{0.5mm} C_{qu}^{(8) \hspace{0.25mm} 3333} \,, \\[2mm]
S_{q \bar q}^{(2)} & = \frac{m_t^4}{1296 \hspace{0.125mm} \pi^4 \left ( 1 - \beta^2 \right )^2} \hspace{0.5mm} \Bigg \{ \hspace{0.25mm} \frac{486 - 353 \beta^2 + 135 \beta^4 - 18 \beta^6}{18} + \frac{27 - 10 \beta^2 + 3 \beta^4}{3} \hspace{0.5mm} L_\beta \\[1mm] 
& \phantom{xx} + \left ( 27 - 24 \beta^2 + 8 \beta^4 - \beta^6 \right ) {\rm Re} \hspace{0.25mm} B + \frac{1}{4} \left ( 3 + \beta^2 \right ) L_\beta^2 + \frac{1}{2} \left ( 3 - \beta^2 \right )^2 L_\beta \hspace{0.5mm} {\rm Re} \hspace{0.25mm} B \\[1mm]
& \phantom{xx} + \frac{1}{4} \left ( 3 - \beta^2 \right )^3 \left ( {\rm Re} \hspace{0.25mm} B \right )^2 + \frac{\pi^2}{4} \left ( 48 - 38 \beta^2 + 11 \beta^4 - \beta^6 \right ) \\[1mm]
& \phantom{xx} + \left [ \frac{162 - 80 \beta^2 + 18 \beta^4}{3} + \left ( 9 - \beta^2 \right ) L_\beta + 3 \left ( 3 - \beta^2 \right )^2 {\rm Re} \hspace{0.25mm} B \right ] L_\mu \\[1mm]
& \phantom{xx} + \left ( 27 - 7 \beta^2 \right ) L_\mu^2 \hspace{0.25mm} \Bigg \} \hspace{0.5mm} \big( C_{qu}^{(8) \hspace{0.25mm} 3333} \big)^2 \,.
\end{split}
\eeq
The abbreviations appearing in~\eqref{eq:Sqq} are defined as follows:
\beq \label{eq:funcqq}
L_\beta = \ln \left ( \frac{1 - \beta^2}{4} \right ) \,, \qquad L_\mu = \ln \frac{\mu_R}{m_t} \,, \qquad {\rm Re} \hspace{0.25mm} B = - \beta \ln \left( \frac{1 + \beta}{1 - \beta} \right) \,.
\eeq

For the $gg \to t \bar t$ channel the relevant expressions are
\beq \label{eq:Sgg}
\begin{split}
S_{g g}^{(1)} & = -\frac{9 \hspace{0.125mm} m_t^2}{128 \hspace{0.125mm} \pi^2} \hspace{0.5mm} \Bigg[ \hspace{0.25mm} \frac{19}{12} - \beta^2 - \frac{61 - 38 \beta^2}{27 \beta} \hspace{0.5mm} T_\beta + \frac{1 - \beta^2}{2 \beta} \left( \beta - T_\beta \right) {\rm Re} \hspace{0.25mm} B \\[1mm]
& \phantom{xx} + \left( 1 - \frac{64 - 42 \beta^2 + 22 \beta^4}{27 \beta (1 - \beta^2)} \hspace{0.5mm} T_\beta \right) {\rm Re} \hspace{0.5mm} C \hspace{0.25mm} \Bigg] \hspace{0.5mm} C_{qu}^{(8)\hspace{0.25mm}3333} \,, \\[2mm]
S_{g g}^{(2)} & = \frac{3 \hspace{0.125mm} m_t^4}{1024 \hspace{0.125mm} \pi^4} \hspace{0.5mm} \Bigg \{ \hspace{0.25mm} \frac{15345 - 16137 \beta^2 + 3006 \beta^4 - 1944 \beta^6}{512 \left ( 1 - \beta^2 \right )^2} \\[1mm]
& \phantom{xx} - \frac{153 - 8415 \beta^2 + 9234 \beta^4 - 1944 \beta^6}{512 \hspace{0.25mm} \beta^2 \left ( 1 - \beta^2 \right)} \hspace{0.5mm} {\rm Re} \hspace{0.25mm} B + \frac{2601 - 3249 \beta^2 - 4302 \beta^4}{64 \left ( 1 - \beta^2 \right)^2 } \hspace{0.5mm} {\rm Re} \hspace{0.5mm} C \\[1mm]
& \phantom{xx} + \frac{243}{64} \frac{3 - \beta^2}{1 - \beta^2} \hspace{0.5mm} \pi \hspace{0.25mm} {\rm Im} \hspace{0.5mm} C - \frac{333 - 1395 \beta^2 + 486 \beta^4}{128 \hspace{0.25mm} \beta^2 \left ( 1 - \beta^2 \right )} \left ( {\rm Re} \hspace{0.25mm} B \right ) \left ( {\rm Re} \hspace{0.5mm} C \right ) \\[1mm]
& \phantom{xx} + \frac{243}{256} \left ( 3 - \beta^2 \right ) \left [ \left ( {\rm Re} \hspace{0.25mm} B \right )^2 + \pi^2 \right ] \\[1mm]
& \phantom{xx} + \frac{2691 - 621 \beta^2 + 324 \beta^4 + 2394 \beta^6}{32 \left ( 1 - \beta^2 \right )^3} \left [ \left ( {\rm Re} \hspace{0.5mm} C \right )^2 - \left ( {\rm Im} \hspace{0.5mm} C \right )^2 \right ] \hspace{0.25mm} \Bigg \} \hspace{0.5mm} \big( C_{qu}^{(8) \hspace{0.25mm} 3333} \big)^2 \,. 
\end{split}
\eeq
The abbreviations appearing in~\eqref{eq:Sgg} are defined as
\beq \label{eq:funcgg}
\begin{split}
T_\beta & = \tanh^{-1} \beta \,, \\[1mm]
\quad {\rm Re} \hspace{0.5mm} C & = \frac{1}{8} \left ( 1 - \beta^2 \right ) \left [ \ln^2 \left( \frac{1 + \beta}{1 - \beta} \right) - \pi^2 \right ] \,, \\[1mm]
{\rm Im} \hspace{0.5mm} C & = -\frac{\pi}{4} \left ( 1 - \beta^2 \right ) \ln \left( \frac{1 + \beta}{1 - \beta} \right) \,,
\end{split}
\eeq

In the case of the asymmetric coefficients, only the contributions associated with the $b \bar b \to t \bar t$ channel are non-vanishing
\beq \label{eq:Abb}
A_{b \bar b}^{(1)} = -\frac{m_t^2}{12 \hspace{0.125mm} \pi \alpha_s} \frac{\beta}{1 - \beta^2} \, C_{qu}^{(8) \hspace{0.25mm} 3333} \,, \qquad 
A_{b \bar b}^{(2)} = -\frac{m_t^4}{24 \hspace{0.125mm} \pi \alpha_s} \frac{\beta}{\left ( 1 - \beta^2 \right )^2} \, \big ( C_{qu}^{(8) \hspace{0.25mm} 3333} \big )^2 \,,
\eeq
while all remaining coefficients vanish identically.

{
\def\arraystretch{1.25}
\begin{table}[t!]
\centering
\begin{tabular}{|c|cc|cc|}
\hline
\multirow{2}{*}{$\sigma \, [{\rm pb}]$} & \multicolumn{2}{c|}{symmetric} & \multicolumn{2}{c|}{asymmetric} \\
\cline{2-5}
& $\Lambda^{-2}$ & $\Lambda^{-4}$ & $\Lambda^{-2}$ & $\Lambda^{-4}$ \\
\hline
$b \bar b$ & $0.59$ & $0.60$ & $-0.25$ & $-0.40$ \\
$q \bar q$ & $0.65$ & $0.24$ & 0 & 0 \\
$gg$ & $-2.29$ & $1.05$ & 0 & 0 \\
\hline
\end{tabular}
\vspace{2mm}
\caption{SMEFT predictions for the total inclusive symmetric and asymmetric contributions to the $t\bar t$ production cross section are presented. Results are shown for the EFT expansion truncated at $\Lambda^{-2}$ and $\Lambda^{-4}$ order, separately for the $b\bar b \to t\bar t$, $q\bar q \to t\bar t$, and $gg \to t\bar t$ channels. All predictions are obtained assuming $c_{qu}^{(8)\hspace{0.25mm}3333} = 10$. Further details are provided in the main text.}
\label{tab:sigmasa}
\end{table}
}

A few comments concerning the results in~\eqref{eq:Sbb}, \eqref{eq:Sqq}, \eqref{eq:Sgg}, and~\eqref{eq:Abb} are in order. First, as already noted, only the coefficients in~\eqref{eq:Sqq}, which encode the one-loop contributions to $q \bar q \to t \bar t$, depend on the renormalization scale~$\mu_R$. This scale dependence originates from the UV divergences generated by the QCD penguin-type amplitudes, which require operator renormalization in the $\overline{\rm MS}$ scheme. By contrast, the loop-induced contributions to $gg \to t \bar t$ in~\eqref{eq:Sgg} are UV finite after OS renormalization of the top-quark mass and therefore do not introduce any residual dependence on~$\mu_R$ at this order. Second, in the case of the asymmetric contributions, the one-loop-induced $q \bar q \to t \bar t$ and $gg \to t \bar t$ channels vanish identically. For the $q \bar q$ channel, this can be understood from the fact that the generated QCD penguin-type amplitudes couple vectorially to the light quarks and therefore do not generate terms that are odd under $\cos\theta \to -\cos\theta$. In the $gg$ channel, the absence of an asymmetric contribution follows instead from the symmetry properties of the initial state, which is charge symmetric and hence cannot induce a asymmetry. Finally, we note that our results for $S_{q \bar q}^{(1)}$ and $S_{gg}^{(1)}$ agree with the corresponding expressions presented in~\cite{Degrande:2024mbg}.\footnote{We have independently cross-checked all analytic expressions in~Sections~3.1.1 and~3.1.2 of~\cite{Degrande:2024mbg} and find agreement in all cases except~(3.20) and~(3.21), which appear to be missing an overall factor of $1/2$.} The~contributions $S_{q \bar q}^{(2)}$ and $S_{gg}^{(2)}$ instead constitute, to the best of our knowledge, new results and cannot be computed using the existing {\tt SMEFT@NLO} implementation~\cite{Degrande:2020evl}.

\paragraph{Numerical results:}

\begin{figure}[t!]
\centering
\begin{subfigure}{0.55\textwidth}
\centering
\includegraphics[width=\linewidth]{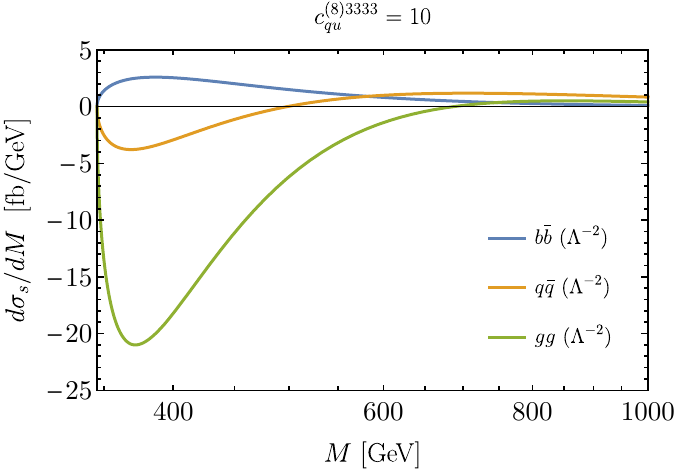}
\end{subfigure}

\vspace{2mm}

\begin{subfigure}{0.55\textwidth}
\centering
\includegraphics[width=\linewidth]{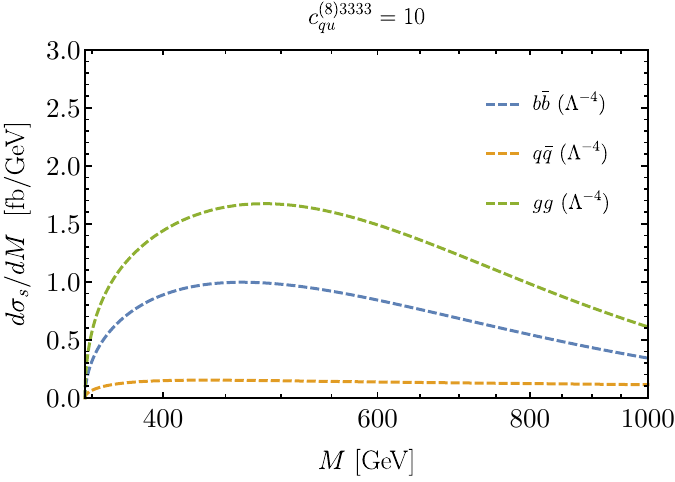}
\end{subfigure}

\vspace{2mm}

\begin{subfigure}{0.55\textwidth}
\centering
\includegraphics[width=\linewidth]{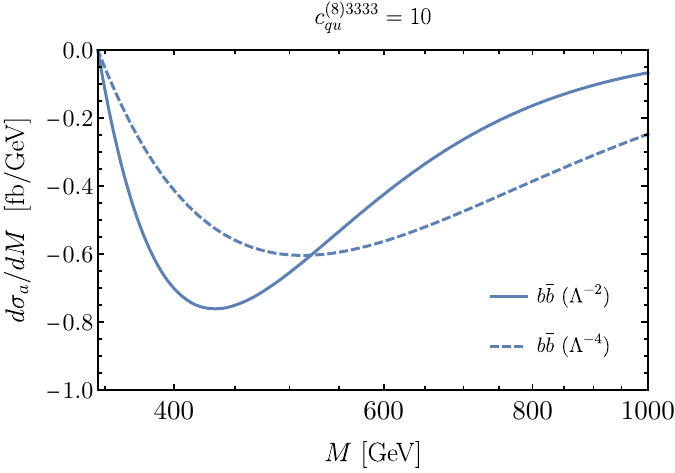}
\end{subfigure}
\vspace{2mm}
\caption{As in~Table~\ref{tab:sigmasa}, but now differential in the invariant mass~$M$ of the~$t\bar t$~system for the $t\bar t$ production cross section. Additional explanations can be found in the main text.}
\label{fig:ttplots}
\end{figure}

In~Table~\ref{tab:sigmasa}, we present SMEFT predictions for the symmetric and asymmetric components of the total inclusive $t\bar t$ production cross section. The results correspond to LHC collisions at a COM energy of $\sqrt{S} = 13\,{\rm TeV}$ and are obtained using the {\tt NNPDF31\_nlo\_as\_0118} PDF set~\cite{NNPDF:2017mvq}, with both renormalization and factorization scales set to the top-quark mass as given in~\eqref{eq:numerics1}. The PDFs are accessed in our computations via~{\tt ManeParse}~\cite{Clark:2016jgm}. Results are shown for an EFT expansion truncated at both~$\Lambda^{-2}$ and~$\Lambda^{-4}$ order, separately for the $b\bar b \to t\bar t$, $q\bar q \to t\bar t$, and $gg \to t\bar t$ production channels. All~predictions are obtained assuming $c_{qu}^{(8)\hspace{0.25mm}3333} = 10$, where the dimensionless Wilson coefficient is defined in~\eqref{eq:smallc}.

From the numerical results several features emerge. First, despite being loop-suppressed, the $q\bar q \to t\bar t$ and $gg \to t\bar t$ channels yield contributions to the symmetric cross section that are comparable to, and in the latter case even larger than, those from the $b\bar b \to t\bar t$ channel, which already occurs at tree level. This behavior is driven by the strong suppression of the bottom-quark PDF, which compensates the loop suppression. Second, for the chosen benchmark value of the Wilson coefficient, the linear and quadratic EFT contributions are of comparable size. Third, for the inclusive asymmetric cross section only the $b\bar b \to t\bar t$ channel contributes, as expected, and the linear and quadratic terms are again of similar magnitude, with the latter even exceeding the former in absolute value.

To gain further insight into the origin of the relatively large $\Lambda^{-4}$ contributions, we display in~Figure~\ref{fig:ttplots} the corresponding SMEFT predictions for the symmetric and asymmetric components of the $t\bar t$ production cross section as functions of the invariant mass~$M$~of the~$t\bar t$~system. The same parameter choices as in~Table~\ref{tab:sigmasa} are used to produce the plots. From all three panels it is evident that the $\Lambda^{-4}$ distributions exhibit significantly enhanced high-mass tails compared to the corresponding $\Lambda^{-2}$ spectra, in both the symmetric and asymmetric cases and across all production channels. This is a well-known feature~\cite{Ethier:2021bye,Celada:2024mcf,Hartland:2019bjb,Degrande:2024mbg,DiNoi:2025uhu}, which can be understood via naive dimensional analysis: the~squared dimension-six contributions scale effectively as dimension-eight operators, leading to a stronger growth with energy and thus a harder spectrum than the interference terms at order $\Lambda^{-2}$. In view of the robustness required for EFT interpretations, this behavior is clearly an undesirable~aspect.

We note that the results presented in~Table~\ref{tab:sigmasa} and Figure~\ref{fig:ttplots} are obtained using the scale choice $\mu_R=\mu_F=m_t$ for $t\bar t$ production, such that the logarithms $L_\mu$ appearing in~\eqref{eq:Sqq} vanish. The same choice is adopted in the numerical analysis of Section~\ref{sec:NLLEWPO}. To~derive bounds on the Wilson coefficients, we subsequently evolve these parameters from $m_t$ to the high scale~$\Lambda$ using the one-loop RG equations implemented in {\tt DsixTools~2.0}~\cite{Fuentes-Martin:2020zaz}, thereby resumming the logarithmically enhanced contributions associated with the scale separation between $m_t$ and~$\Lambda$. At LL accuracy, this procedure is equivalent to evaluating the Wilson coefficients at a characteristic scale of the process, such as the invariant mass $M$ of the~$t\bar t$~system, and then evolving them to~$\Lambda$ using the one-loop RG equations. This equivalence follows from RG invariance, which ensures that the dependence on the intermediate scale cancels at LL accuracy up to subleading logarithmic corrections.

For the SM predictions and their theoretical uncertainties, we employ NNLO QCD calculations obtained with {\tt MATRIX}~\cite{Catani:2019iny,Catani:2019hip}, normalized to the inclusive $t\bar t$ production cross section computed with {\tt TOP++}~\cite{Czakon:2011xx}. The latter is accurate at NNLO in QCD and includes the resummation of soft-gluon contributions at NNLL accuracy. By comparing these predictions with our own LO QCD calculation of $t\bar t$ production, we find that the scale choice $\mu_R=\mu_F=m_t$ reproduces the shapes of all $t\bar t$ observables relevant for our analysis remarkably well. The higher-order QCD corrections are largely captured by approximately flat $K$-factors, providing additional support for the scale choice adopted in the SMEFT calculation. We do not, however, apply these SM $K$-factors to the SMEFT contributions. Such a procedure would not be theoretically justified, since the QCD corrections to the SM and SMEFT amplitudes are in general different. In particular, the SMEFT predictions involve operator renormalization and RG-induced mixing effects that have no counterpart in the SM and therefore cannot be captured by a simple multiplicative rescaling. As a result, the use of SM $K$-factors would amount to an uncontrolled approximation whose impact is difficult to quantify reliably.

\subsection{Four-top quark production}
\label{app:4top}

The dependence of the total $pp \to t\bar t t\bar t$ cross section on the Wilson coefficients is taken from~Table~3 of Appendix~B in~\cite{DiNoi:2025uhu}.\footnote{We have independently cross-checked the quoted coefficients using {\tt MadGraph5\_aMC@NLO}~\cite{Alwall:2014hca} together with {\tt SMEFT@NLO}~\cite{Degrande:2020evl}, finding complete agreement within theoretical uncertainties.} As shown in~\cite{Aoude:2022deh}, the ${\cal O}(\Lambda^{-2})$ interference terms between the tree-level SMEFT amplitudes involving third-generation four-quark operators and the tree-level SM EW amplitudes mediated by gauge-boson exchange are numerically relevant. These contributions are therefore retained both in~\cite{DiNoi:2025uhu} and in our analysis. The~predictions for the total $pp \to t\bar t t\bar t$ cross section reported in that work are obtained with the scale choice $\mu_R=\mu_F=2\hspace{0.125mm}m_t$. To derive constraints on the Wilson coefficients, we evolve them from this scale to~$\Lambda$ using the one-loop RG equations implemented in~{\tt DsixTools~2.0}, thereby resumming the logarithmically enhanced contributions associated with the scale separation. For the SM prediction of the total $pp \to t\bar t t\bar t$ cross section and its theoretical uncertainty, we employ the NLO QCD plus EW calculation supplemented by NLL soft-gluon resummation presented in~\cite{vanBeekveld:2022hty}. With this setup, we are able to reproduce the individual $95\%$~CL bounds on the Wilson coefficients renormalized at the scale $2\hspace{0.125mm}m_t$ obtained from $pp \to t\bar t t\bar t$ production and reported in~Table~5 of the article~\cite{DiNoi:2025uhu}.

\section{LHC data}
\label{app:LHC}

\begin{figure}[t!]
\centering
\includegraphics[width=0.85\linewidth]{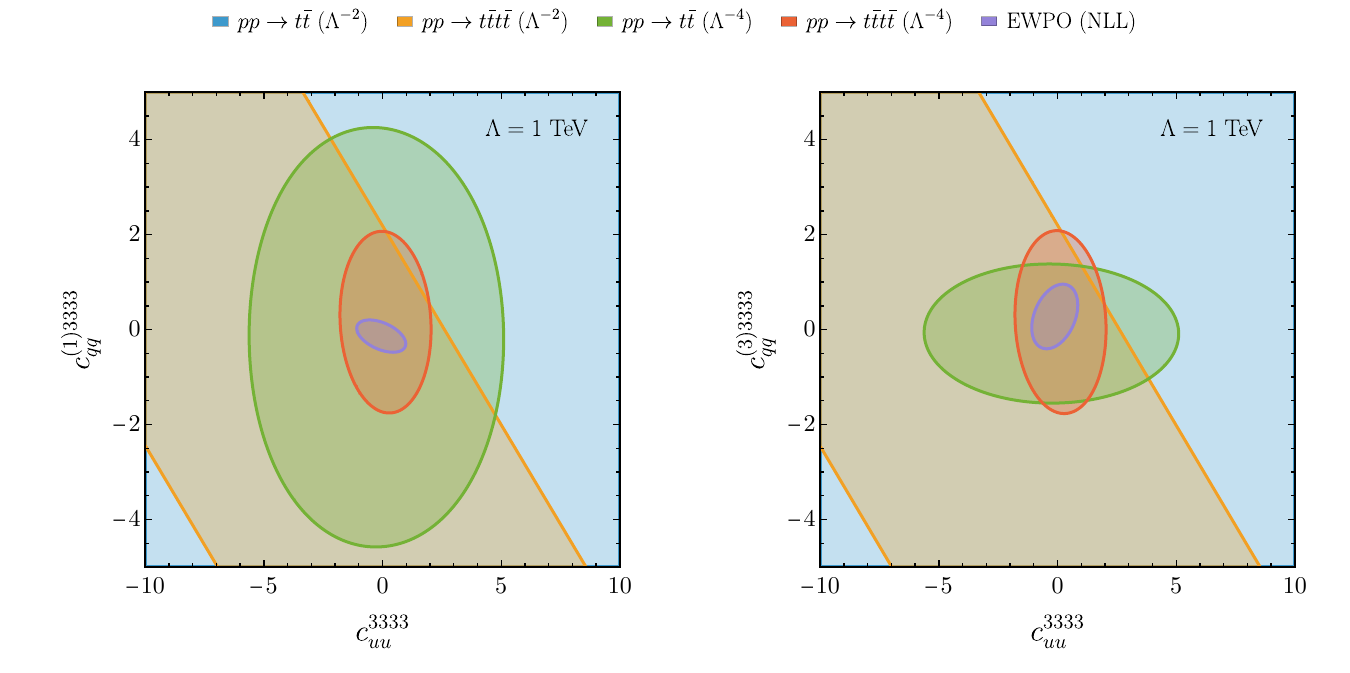}
 
\includegraphics[width=0.85\linewidth]{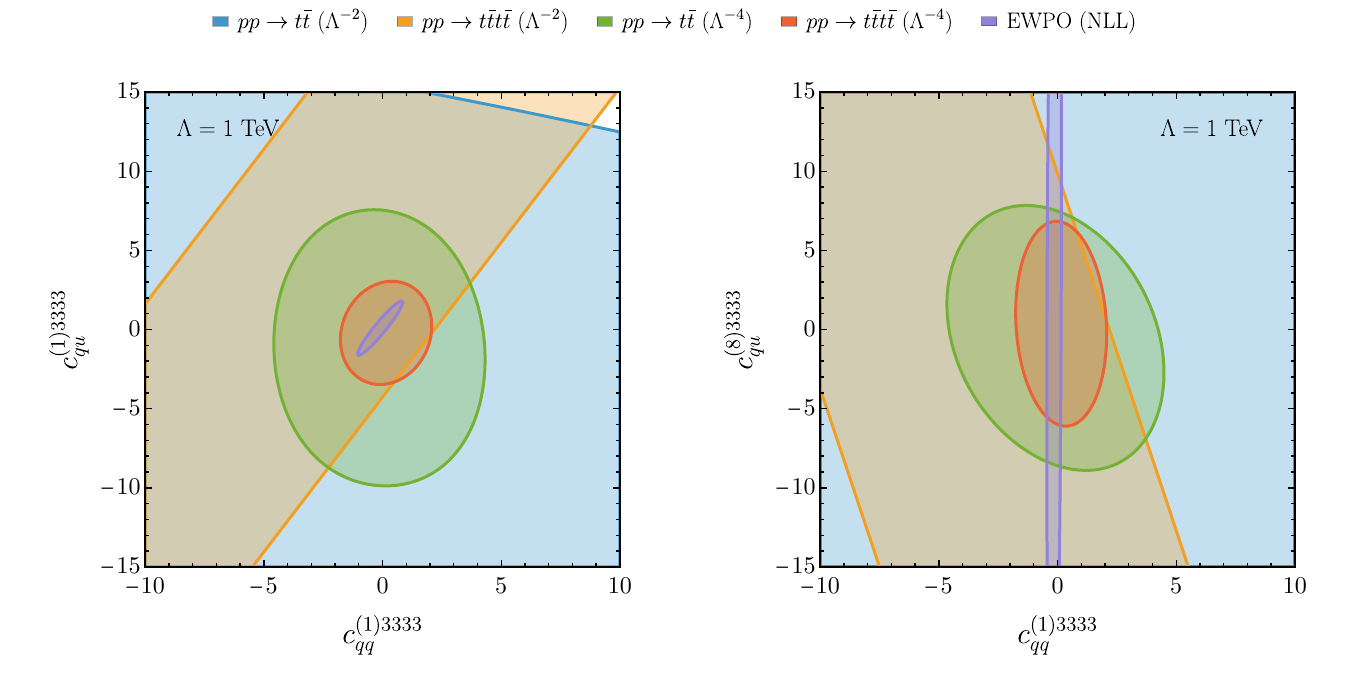}
 
\includegraphics[width=0.85\linewidth]{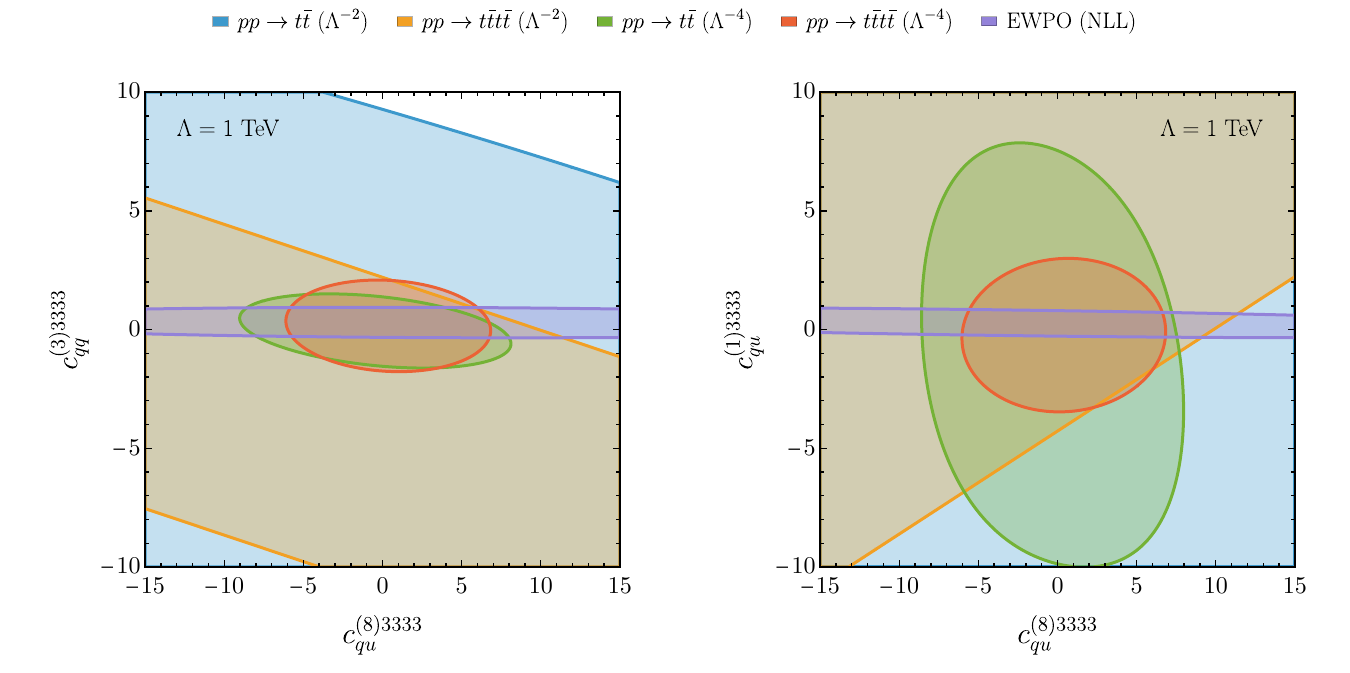}
\vspace{2mm}
\caption{As in~Figure~\ref{fig:qu_vs_uu}, but showing the $95\%$~CL constraints in the two-dimensional Wilson coefficient planes indicated by the axis labels.}
\label{fig:4topplotsapp}
\end{figure}

The current $pp\to t\bar t$ fit results presented in Section~\ref{sec:Zpolefits} are based on the parton-level differential cross sections reported in Figure~19 of the CMS analysis~\cite{CMS:2021vhb}, corresponding to an integrated luminosity of $137\,{\rm fb}^{-1}$ collected in LHC collisions at $\sqrt{S}=13\,{\rm TeV}$. Our fit includes the distributions in the invariant mass $m_{t\bar t}$, the variable $\cos\theta^\ast$, where $\theta^\ast$ denotes the angle between the top quark and the boost direction of the $t\bar t$ system evaluated in the $t\bar t$ rest frame, and the absolute rapidity $|y_{t\bar t}|$ of the $t\bar t$ system. In contrast, the $pp\to t\bar t t\bar t$ analysis is based on the latest ATLAS and CMS measurements of the inclusive cross section~\cite{CMS:2023ftu,ATLAS:2023ajo}, obtained from data sets corresponding to integrated luminosities of approximately $140\,{\rm fb}^{-1}$ collected at $\sqrt{S}=13\,{\rm TeV}$. For the Higgs observables, we consider the signal strengths~$\mu_{\rm prod}^{\rm decay}$ resolved by both production and decay mode, treating each production-decay combination as an independent observable. The corresponding SMEFT predictions are evaluated at linear order in the EFT expansion, i.e.~at ${\cal O}(\Lambda^{-2})$, using the results of~\cite{Falkowski:2019hvp} as implemented in {\tt flavio}~\cite{Giani:2023gfq}. On the experimental side, we employ the most recent ATLAS and CMS measurements of the Higgs signal strengths together with the associated correlation matrices provided by the collaborations~\cite{ATLAS-CONF-2025-006,CMS:2026nce}.

Our HL-LHC projections follow the systematic scenario S2 of~\cite{Cepeda:2019klc,ATLAS:2025eii}, in which statistical uncertainties are rescaled according to the projected integrated luminosity of up to~$3\,{\rm ab}^{-1}$. Systematic uncertainties are treated in a conservative but realistic manner, where most experimental systematics are kept unchanged or only mildly reduced where justified, theoretical systematic uncertainties are reduced by a factor of two where feasible, and contributions from limited Monte Carlo statistics are assumed to be negligible. This~setup corresponds to an intermediate extrapolation scenario in which improved detector performance and analysis strategies yield moderate reductions in systematics alongside the gain in statistical precision from the full HL-LHC data set.

\section{Additional two-dimensional plots}
\label{app:addplots}

Figure~\ref{fig:4topplotsapp} presents additional constraints on the high-scale Wilson coefficients of the third-generation four-quark operators in~\eqref{eq:2loop4top}, again adopting the reference value $\Lambda=1\,{\rm TeV}$. The results further support the conclusion reached in Section~\ref{sec:4HQfits}, namely that the existing LEP and SLC measurements of the~EWPO already provide stronger constraints than current LHC top-quark data for most operators, even when quadratic ${\cal O}(\Lambda^{-4})$ terms are included in the latter. The main exception is $Q_{qu}^{(8)\hspace{0.25mm}3333}$, whose mixing into the operators relevant for~EWPO first occurs at NLL order through numerically suppressed two-loop contributions. As a result,~EWPO provide only weak constraints on this operator, which manifests itself in the pronounced flat directions visible in the last three panels of~Figure~\ref{fig:4topplotsapp}.

\end{appendix}


%

\end{document}